\documentclass[
twocolumn,
english,
aps,
pra,
longbibliography,
superscriptaddress,
amsmath,
amssymb,
floatfix,
nofootinbib,
]{revtex4-2}
\usepackage{amsthm}
\usepackage{amsfonts}
\usepackage{siunitx}
\usepackage{amsmath}
\usepackage{amssymb}
\usepackage{graphicx}
\usepackage{verbatim}
\usepackage{enumitem}
\usepackage[colorlinks]{hyperref}
\usepackage{tikz}
\usepackage{pgfplots}
\usepackage{braket}
\usepackage{xcolor}
\usepackage{amssymb} 
\usepackage{graphicx}
\usepackage{dcolumn}
\usepackage{bm}
\usepackage{mathtools}
\usepackage{hyperref}
\usepackage{mathrsfs}
\usepackage{booktabs,tabularx}

\definecolor{linkcolor}{RGB}{0,83,166}
\hypersetup{
  colorlinks = true,
  allcolors = {linkcolor}
}

\begin{document}

\title{Evaluating the Limits of QAOA Parameter Transfer at High-Rounds on Sparse Ising Models With Geometrically Local Cubic Terms}

\author{Elijah Pelofske}
\email[]{epelofske@lanl.gov}
\affiliation{Information Systems \& Modeling, Los Alamos National Laboratory}

\author{Marek M. Rams}
\affiliation{Institute of Theoretical Physics, Jagiellonian University, Łojasiewicza 11, 30-348 Kraków, Poland}
\affiliation{Mark Kac Complex Systems Research Center, Jagiellonian University, Łojasiewicza 11, 30-348 Kraków, Poland}

\author{Andreas Bärtschi}
\affiliation{Information Sciences, Los Alamos National Laboratory}

\author{Piotr Czarnik}
\affiliation{Institute of Theoretical Physics, Jagiellonian University, Łojasiewicza 11, 30-348 Kraków, Poland}
\affiliation{Mark Kac Complex Systems Research Center, Jagiellonian University, Łojasiewicza 11, 30-348 Kraków, Poland}

\author{Paolo Braccia}
\affiliation{Quantum \& Condensed Matter Physics, Los Alamos National Laboratory}

\author{Lukasz Cincio}
\affiliation{Quantum \& Condensed Matter Physics, Los Alamos National Laboratory}

\author{Stephan Eidenbenz}
\affiliation{Information Sciences, Los Alamos National Laboratory}

\begin{abstract}

The emergent practical applicability of the Quantum Approximate Optimization Algorithm (QAOA) for approximate combinatorial optimization is a subject of considerable interest. One of the primary limitations of QAOA is the task of finding a set of good parameters, which is usually done using a variational optimization loop. Parameter transfer, or parameter concentration, is a phenomenon where QAOA angles trained on problem instances that are self-similar tend to perform well for other problem instances from that similar class. This suggests a potentially highly efficient and scalable non-variational learning method for QAOA angle finding. 
In this work, we systematically study QAOA parameter transferability from small problem sizes (16 and 27 decision variables) onto large problem instances (up to 156 qubits) for heavy-hex graph Ising models with geometrically local higher order terms using the Julia based QAOA simulation tool \texttt{JuliQAOA} to perform classical angle finding for up to $49$ QAOA layers ($p$). Parameter transfer of the fixed angles is validated using a combination of full statevector, Projected Entangled Pair States (PEPS), Matrix Product State (MPS), and LOWESA numerical simulations. 
We find that the QAOA parameter transfer from single instances applied to other (unseen) problem instances does not in general provide monotonically improving performance as a function of $p$ - there are many cases where the performance temporarily decreases as a function of $p$ - but despite this the transferred angles have a general trend of improved expectation value as the QAOA depth increases, in many cases converging close to the true ground-state energy of the $100+$ qubit instances. 
We also sample the hardware-compatible Ising models using the ensemble of transfer-learned QAOA parameters on several superconducting qubit IBM Quantum processors with 127, 133, and 156 qubits. We find continuous solution quality improvement of the hardware-compatible QAOA circuits run on the IBM NISQ processors up to $p=5$ on \texttt{ibm\_fez}, up to $p=9$ on \texttt{ibm\_torino}, and up to $p=10$ on \texttt{ibm\_pittsburgh}. 

\end{abstract}

\maketitle

\section{Introduction}
\label{section:introduction}

The Quantum Approximate Optimization Algorithm (QAOA)~\cite{QAOA, farhi2015quantum, Hadfield_2019, farhi2020quantumapproximateoptimizationalgorithm1, farhi2020quantumapproximateoptimizationalgorithm2} is designed to sample good solutions of combinatorial optimization problems. QAOA is relatively NISQ (Noisy Intermediate-Scale Quantum)~\cite{Preskill_2018} computer friendly in that it is feasible to implement the algorithm at the scale of near term noisy quantum processors (short depth, small qubit counts). There have been a considerable number of studies on the properties of QAOA using both numerical simulations and current quantum processors, in particular in comparison to existing classical algorithms that solve combinatorial optimization problems~\cite{https://doi.org/10.48550/arxiv.1602.07674, Shaydulin_2024, Harrigan_2021, pelofske2023high, shaydulin2023qaoancdotpgeq200, Weidenfeller2022scalingofquantum, Golden_QAOA_Unstructured, boulebnane2024solvingbooleansatisfiabilityproblems}. One of the central challenges with the algorithmic study of QAOA is scaling sampling analysis with respect to number of variables and QAOA depth. This scaling requires either somewhat small scale numerical simulations (limited to tens of qubits for exact statevector computations, or up to hundreds of qubits using tensor network methods or similar types of approximate numerical techniques) or quantum hardware simulations which are subject to a variety of sources of error. QAOA requires hyper-parameter selection that in turn generally requires significant compute time, especially if this process is performed on the quantum device. However, the QAOA parameter learning is distinct from many other types of variational quantum algorithms (VQAs)~\cite{Cerezo_2021} because the number of trainable parameters does not scale, per layer, with the number of variables in the problem, meaning that parameter consistency among problems with varying problem sizes, or coefficient distributions, can be examined. It turns out that QAOA parameters do indeed exhibit parameter consistency for problem instances that share some similarity, such as similar connection structure. This property has been studied and utilized for efficient QAOA parameter finding in numerous studies \footnote{The QAOA parameter similarity property comes by several different terms, including \emph{parameter transfer}, \emph{problem instance independent QAOA parameters}, \emph{fixed angles}, \emph{parameter concentration}, \emph{parameter fixing}, \emph{size invariant QAOA parameters}, \emph{fixed control parameters}, \emph{fixed point QAOA}, and \emph{transfer learning}}~\cite{PhysRevA.104.L010401, 9605323, LIPICS.TQC.2022.7, Farhi_2022, boulebnane2021predictingparametersquantumapproximate, Shaydulin_2023, Sureshbabu_2024, PhysRevA.104.052419, PhysRevA.103.042612, Streif_2020, montanezbarrera2025transferlearningoptimalqaoa, pelofske2024scaling, brandao2018fixedcontrolparametersquantum, sakai2024linearlysimplifiedqaoaparameters, katial2025instance, PhysRevResearch.6.023171, chernyavskiy2025improvingqaoaapproximatequbo, galda2023similaritybasedparametertransferabilityquantum, galda2021transferabilityoptimalqaoaparameters, Lyngfelt_2025}. Many of these types of parameter transfer studies are able to condense QAOA angle usage into fixed angle formulas, at least for average cases of certain problem types and typically at fairly low depth~\cite{Ozaeta_2022}. If QAOA can be used in a non-variational learning manner, then the prospects of using QAOA for meaningfully sized problem instances on large scale error-corrected quantum computers are good -- the variational learning overhead of QAOA can be quite significant. Parameter setting heuristics, namely from parameter concentration, will enable this approach to use QAOA at much larger scale~\cite{he2024parametersettingheuristicsmake}. There do exist other non-variational QAOA approaches, such as using linear-ramp adiabatic schedules~\cite{Sack_2021, Monta_ez_Barrera_2025}, which at sufficiently large $p$ guarantees optimality. 

Here we show through extensive numerical simulation some of the favorable characteristics of parameter transfer, and some of the limitations. More precisely, we use QAOA parameters trained on small, statevector simulation-addressable problem instances to execute full scale QAOA runs on several IBM quantum processors. Finally, we validate the QAOA angle performance using extensive tensor network simulations, in particular PEPS (Projected Entangled Pair States). The QAOA circuits and optimization problems that we use in this study are carefully designed to be extremely hardware compatible to the heavy-hex graph and are therefore very NISQ-computer-friendly see refs~\cite{pelofske2023short, pelofske2023qavsqaoa}. The sparsity of the variable connections in these optimization problem instances makes them an excellent test case for studying fundamental algorithmic engineering questions about QAOA. In this case, we are focusing on \emph{high QAOA depth parameter transfer}.

Previous studies have performed NISQ-hardware simulations of relatively small-scale optimization problems using approximate digital quantum optimization algorithms. Ref.~\cite{romero2024biasfielddigitizedcounterdiabaticquantum} used variational digitized counterdiabatic quantum optimization to sample problems with higher order terms, on a $156$ qubit IBM quantum processor. Refs.~\cite{pelofske2023short, pelofske2023qavsqaoa} used the quantum approximate optimization algorithm to sample $127$ variable problems with higher order terms on a $127$ IBM superconducting qubit processor -- and Ref.~\cite{pelofske2024scaling} then extended this up to a $414$ qubit IBM QPU (Quantum Processing Unit). Ref.~\cite{pelofske2023high} simulated QAOA for sampling MAX-3-SAT problems for up to $22$ variables up to depth $p=2$, and $10$ variables for up to depth $p=20$ using trapped ion quantum processors. Ref.~\cite{Sack_2024} sampled maximum cut on random 3-regular graphs for up to $p=2$ using the IBM QPUs. Ref.~\cite{shaydulin2023qaoancdotpgeq200} sampled up to $32$ variable maximum cut problems on random 3-regular graphs, for up to $p=11$, using a trapped ion quantum processor. In regards specifically to applying QAOA using parameter transfer, the largest scale previous study was ref.~\cite{pelofske2024scaling}, which was able to provide high-quality angles for up to QAOA depth $5$ on $127$-qubit sparse Ising model instances -- in this study, we go well beyond that up to depth $49$ and $156$-qubit instances, using a broader ensemble of QAOA angles.

\begin{table*}[th!]
\begin{center}
\setlength{\tabcolsep}{11pt}
\begin{tabularx}{\textwidth}{@{}llllX@{}}
\toprule
Processor & \#Qubits & \#2q-Gates & Basis gates & QPU names \\ 
\midrule
Eagle r3 & 127 & 144 & ECR, ID, RZ, SX, X & 
\texttt{ibm\_sherbrooke}, \texttt{ibm\_kyiv}, \texttt{ibm\_brisbane}, \texttt{ibm\_kyoto}, \texttt{ibm\_nazca}, \texttt{ibm\_brussels}, \texttt{ibm\_strasbourg} \\ 
Heron r1 & 133 & 150 & CZ, ID, RZ, SX, X & 
\texttt{ibm\_torino} \\ 
Heron r2 & 156 & 176 & CZ, ID, RZ, SX, X & 
\texttt{ibm\_fez}, \texttt{ibm\_marrakesh}, \texttt{ibm\_kingston}, \texttt{ibm\_aachen} \\ 
Heron r3 & 156 & 176 & CZ, ID, RZ, SX, X & 
\texttt{ibm\_pittsburgh}\\ 
\bottomrule
\end{tabularx}
\setlength{\tabcolsep}{6pt}
\end{center}
\caption{Summary of the 12 IBM Quantum Processing Units (QPUs) used in our QAOA hardware experiments. The given basis gates are the gates that the QAOA circuits are compiled to (locally) before executing on the device -- in some cases, such as on \texttt{ibm\_marrakesh}, there are now additional gates available which were not used for the QAOA circuit compilation in this study.}
\label{table:IBM_hardware}
\end{table*}

In this study, the high-depth scaling of QAOA parameters is characterized using an ensemble of learned QAOA angles on small (e.g., less than $30$ qubits) problem instances whose parameters are then transferred to be used on much larger problem instances (up to $156$ qubits). The central question that is examined is how far parameter transfer QAOA angles can be used, with respect to $p$. This is an important question because although parameter transfer seems to be incredibly successful for finding good performing QAOA angles, there are limits to direct parameter training for parameter transfer -- namely, for the task of transferring parameters up to larger problem sizes, better angles cannot be learned past an approximation ratio of $1$, and higher QAOA depths are generally required as problem sizes increase. It is not clear how well parameter transfer can be applied to various combinatorial optimization problem instances in general, but it is crucial to understand where transferred QAOA angles fail to perform, or perhaps work well in certain cases. Remarkably one of the findings of this study is that the intensely optimized QAOA angles, using the classical numerical statevector simulation tool \texttt{JuliQAOA}, recover the smooth-schedule properties of discretized quantum annealing (adiabatic)-type protocols. 

Using an ensemble of numerically trained small problem instance QAOA angles allows very large QAOA circuits to be executed on $127$, $133$, and $156$ qubit IBM quantum processors, using no on-device variational learning. These QAOA executions show clear signal of improved sampling at higher depth circuits, in particular having a consistently improving approximation ratio starting at $p=1$, up to $p=9$ on \texttt{ibm\_torino}, up to $p=10$ on \texttt{ibm\_pittsburgh}, and up to $p=7$ on both \texttt{ibm\_aachen} and \texttt{ibm\_kingston}, and up to $p=3$ on \texttt{ibm\_kyiv} (larger depths of continued improvement are limited by hardware noise on these devices). This is the largest reported continual QAOA performance improvement, as depth increases, on a quantum computer to date. Moreover, this study demonstrates, primarily by performance validation using tensor network approximation methods, that parameter transfer is a highly efficient technique for high-depth QAOA circuits to be executed, with no variational learning.

\section{Methods}
\label{section:methods}

Here we first briefly summarize QAOA. Given a combinatorial optimization problem defined by a cost function $C(z)$ on $n$-spin states $z\in\{+1,-1\}^n$, QAOA is defined by the following components:
\begin{itemize}[noitemsep]
    \item   An~initial state $\ket{\psi}$,
            in this study we use the standard $\ket{\psi} := \ket{+^n}$,
    \item   a~cost Hamiltonian $H_C$ which represents the cost values,
            here $H_C := \sum_z C(z)\ket{z}\bra{z}$,
    \item   a~phase separating Hamiltonian $H_P$
            which is used to phase basis states $z$ by a parametrized function of their cost value,
            here $H_P := H_C$
    \item   a~mixing Hamiltonian $H_M$,
            which is used to give parameterized interference between basis states,
            here we use the standard transverse field mixer $H_M = \sum_{j=1}^n X_j$.
\end{itemize}
QAOA then prepares a parameterized quantum state 
\begin{equation} \label{eq:QAOAstate}
\ket{\vec{\beta},\vec{\gamma}} = e^{-i\beta_p H_M} e^{-i\gamma_p H_P} \cdots e^{-i\beta_1 H_M} e^{-i\gamma_1 H_P}\ket{\psi},
\end{equation}
where we have the following parameters:
\begin{itemize}[noitemsep]
    \item   An~integer $p\geq1$ which describes the number of rounds~\footnote{The QAOA parameter $p$ can be found under several different terms in the literature, including \emph{layers}, \emph{rounds}, \emph{depth} (importantly not the same quantity as circuit depth), and \emph{levels}, in this study we will typically refer to this parameter as either rounds or depth. } to run the algorithm,
    \item   a~vector of reals $\vec{\beta} = (\beta_1,...,\beta_p)$ of length $p$, that parameterizes the layer-wise time evolution of $H_M$,
    \item   a~vector of reals $\vec{\gamma} = (\gamma_1,...,\gamma_p)$ of length $p$, that parameterizes the layer-wise time evolution of $H_P$.
\end{itemize}
Measuring the state $\ket{\vec{\beta},\vec{\gamma}}$ in the computational basis returns a sample solution $z$ of cost value $C(z)$ with probability $\lvert \smash{\braket{z\vert\vec{\beta},\vec{\gamma}}} \rvert^2$, with a mean expectation value of $\bra{\vec{\beta},\vec{\gamma}}H_C\ket{\vec{\beta},\vec{\gamma}}$.
Combined, the real-valued parameters are usually called \emph{angles} -- and these are the parameters that are typically tuned using a black-box optimizer algorithm run on a classical processor. Typically, these parameters are tuned such that the mean expectation value is optimized, which is the training performed in the pre-training stage described next.

The QAOA angle finding in this study is carried out using the Julia~\cite{bezanson2017julia} based software package \texttt{JuliQAOA}~\cite{Golden_2023}, where the angle finding is performed using basin hopping optimization, which in turn uses gradient descent, and all angle optimization begins with uniformly random guesses at $p=1$. \texttt{JuliQAOA} is an efficient numerical simulation angle finding tool that can heuristically generate high quality QAOA angles that are not necessarily globally optimal, given a specific set of $C(z)$ cost values for all feasible solutions. As far as we are aware, finding globally optimal QAOA parameters is a very hard computational problem; we are not aware of tools or techniques that would allow us to verify global optimality of QAOA parameters (e.g., up to some precision), even for relatively small optimization problem instance sizes. \texttt{JuliQAOA} has been shown to be a highly effective QAOA angle learning tool~\cite{Golden_QAOA_Unstructured, pelofske2023high, Golden_2021_threshold, Golden_2023_sat}. One basin hopping iteration is used for each QAOA round, and $p$ is incremented in step sizes of $1$. \emph{Angle extrapolation} is also used within each \texttt{JuliQAOA} $p$ angle optimization pass -- where $p+1$ angles are initialized using the best angles found at the preceding $p$ step. Note that similar ideas have been used in previous studies to compute good QAOA angles at $p>1$~\cite{Pagano_2020}. At each step of $p$ past $p=1$, the angle learning is repeated until the mean expectation value has improved over the previous step of $p$ -- each additional angle learning step (e.g., at $p+1$) uses random angle initialization and since the learning is not necessarily optimal, different initializations can result in different learned QAOA angle quality. Similar sets of heuristic rules used in \texttt{JuliQAOA} have been developed in other studies, motivated by similar ideas of iterative or recursive exploration of the $p+1$ energy landscape using the information gained from the $p$ energy landscape~\cite{Sack_2023, apte2025iterativeinterpolationschedulesquantum, Pagano_2020, lee2022depthprogressiveinitializationstrategyquantum}. Importantly, this \texttt{JuliQAOA} simulation is performing full state vector simulation, meaning the training procedure has at least one advantage of not being hindered by shot noise (e.g., finite sampling effects). However, because of the size of the systems, in particular the $27$ qubit Ising models, this simulation is quite memory and compute time intensive. The angle training used both High Performance Computing (HPC) resources and many weeks of elapsed wall clock time (including re-attempts at optimization where the expectation value does not improve monotonically). The aim of this study is to make use of this pre-computation of the QAOA angles at small system sizes, where such state-vector based training is at least tractable, and apply these parameters to much larger QAOA circuits. We have found that this particular QAOA angle learning tool is the most effective tool that we are aware of; other optimization techniques, such as gradient-free optimization methods, fail systematically at high QAOA depths.

The combinatorial optimization problems are all Ising models, which makes them unconstrained minimization problems. This class of Ising models was introduced as a very hardware-compatible Ising model for the IBM quantum hardware heavy-hex graph structure~\cite{PhysRevX.10.011022, benito2025comparativestudyquantumerror, PRXQuantum.5.040334} by Refs.~\cite{pelofske2023short, pelofske2023qavsqaoa}. These Ising models include geometrically local cubic terms ($ZZZ$ terms) which make the problems computationally harder, but the phase separating Hamiltonian circuit implementation is highly optimized to have strictly a two qubit gate depth of $6$ per $p$ (notably, using no additional two-qubit gates to handle the higher order term compared to the heavy-hex quadratic interactions)~\cite{pelofske2023short, pelofske2023qavsqaoa}. This hardware compatibility makes these circuits incredibly NISQ computer friendly, and both these Ising models and QAOA circuits to sample these Ising models have also been used in several prior studies~\cite{pelofske2024scaling, barron2023provableboundsnoisefreeexpectation, oleary2025efficientonlinequantumcircuit}. Importantly, these Ising models are designed with the aim of being NISQ-friendly test cases to evaluate QAOA properties, namely for running on near term noisy (heavy-hex graph) IBM quantum processors, not with the aim of being extremely computationally challenging. These Ising models, primarily because of the sparsity of spin interactions, lack of long-range interactions, and because of the relatively small number of variables used (for example in this study the largest problem instance we consider is $156$ decision variables), can be solved with classical optimizers such as simulated annealing or provable optimality solvers such as Gurobi extremely quickly (on the order of seconds or faster)~\cite{pelofske2024classicalcombinatorialoptimizationscaling, pelofske2023short, pelofske2023qavsqaoa, pelofske2024scaling}. This class of optimization problems is defined by the following cost function $C(z)$ which simultaneously can be understood as a classical Hamiltonian,
\begin{equation}
    H_C = \sum_i d_i Z_i + \sum_{\langle i,j \rangle} d^{(2)}_{i,j} Z_i Z_j + \sum_i d^{(3)}_{i,j,k}  Z_i Z_{j} Z_{k}. 
    \label{equation:problem_instance}
\end{equation}
The heavy-hex graph $G=(V,E)$ is defined by a set of vertices $V$ and edges $E$. The Hamiltonian is defined by single-spin terms indexed by $i \in V$, the spin-spin interactions are indexed by the edges $(i, j) \in E$, and lastly the local cubic terms are indexed by $(i, j, k)$. The cost function takes as input a vector of spins $z = (z_1, z_2, \ldots, z_n) \in \{ +1,-1\}^n$. Every heavy-hex lattice is bipartite and the bipartition can be defined as $V=V_2 \sqcup V_3$ with $E\subset V_2\times V_3$, where $V_3$ consists of vertices of maximum degree $3$, and $V_2$ consists of vertices of maximum degree $2$. $W$ is the set of vertices in $V_2$ where each node has degree $2$, and the spins in this set $W$ define the center spin of each of the geometrically local cubic terms in Eq.~\eqref{equation:problem_instance}, indexed as $(i, j, k)$. All variable states are spins, where the states are either $\{+1, -1\}$, and the polynomial coefficients, $d_i, d^{(2)}, d^{(3)}$ are randomly (and independently) chosen to be either $+1$ or $-1$, making these a type of hardware-compatible disordered $\pm J$ models. The complexity class, for exact optimal solution computation, that the Ising models of the form Eq.~\ref{equation:problem_instance} is not known~\cite{pelofske2024classicalcombinatorialoptimizationscaling}. The QAOA circuits use a direct spin to qubit mapping; each qubit represents a single spin in the logical Ising model.

Fig.~\ref{fig:QAOA_circuit} in Appendix~\ref{section:appendix_QAOA_circuit} gives a circuit diagram translation from the hardware-compatible Ising model to the corresponding QAOA circuit diagram, where the most expensive component of the circuit in terms of gate complexity is the phase separator. The QAOA circuit simulations run on the IBM hardware do not use any dynamical decoupling, and do not use any type of quantum error mitigation. The QAOA circuits rely on computing a 3-edge-coloring of the heavy-hex hardware graph -- the 3-edge-coloring used in all of the hardware simulations is randomly generated using minimum edge coloring heuristic algorithms from NetworkX~\cite{hagberg2008exploring}. That fixed edge coloring is then used in all $p$ layers (this is recalculated for each Ising model instance). In principle, these edge colorings could be randomized to generate circuit invariants with slightly different noise profiles when executed on the hardware, but we leave this for future study. Additionally, these edge colorings could be optimized to have distinct (average) gate-times in each layer, thus reducing the overall circuit time~\cite{kotil2025quantumapproximatemultiobjectiveoptimization}. During compilation testing, we did attempt this type of 3-edge-coloring partitioning. However, in practice, the two qubit gate times of the IBM processors used in this study have very nearly uniform times, which means that no significant reduction of circuit execution time can be gained, and randomization of the two qubit 3-edge-coloring does not yield variance in total circuit durations. In principle, if a quantum computer had two qubit gate timings that were highly variable, an optimized staggered edge-coloring (for example, 3 colors which correspond to short, median, and long gate times) could be used to compile circuits such that the total circuit duration was minimized. In other words, if the two qubit gate timings effectively replicated a 3-edge-coloring via distinct timing bins, then compiling to that edge coloring could yield substantial improvements in total circuit duration.  

All of the quantum hardware simulations are performed using IBM Quantum programmable fixed-frequency superconducting transmon qubit~\cite{PhysRevLett.107.080502} devices, which are summarized in Table~\ref{table:IBM_hardware}. These devices have the heavy-hex two-qubit connectivity graph~\cite{PhysRevX.10.011022}. The quantum circuits are defined and executed using the Qiskit Python 3 library~\cite{Qiskit, javadiabhari2024quantumcomputingqiskit}. The circuits are transpiled using a circuit optimization level of $3$ before being sent to the quantum computer. Qiskit versions \texttt{1.2.2},  \texttt{0.45.0}, and \texttt{2.1.2} were used across all of the circuit executions on the IBM QPUs. The circuits are run on the IBM quantum computer in the form of discrete \emph{jobs} -- for all devices the shot count per job is $1{,}000{,}000$, with the exception of \texttt{ibm\_torino} where because of device usage limitations only $5{,}000$ shots were run per job (thus requiring more jobs to be run to get an equivalent total shot count to the other devices). One of the consequences of this is higher QPU compute time used in total for \texttt{ibm\_torino} because there is a time cost associated with running many more individual jobs. Occasionally, for some of the IBM QPUs a job would fail to execute due to an unknown internal error -- in these cases, the job was simply re-submitted to the backend (failed jobs are not counted towards reported aggregate statistics such as QPU time).

One way to quantify the QAOA solution quality, besides the energy of the Ising model Hamiltonian, is the approximation ratio, which given a sample energy $e$ is defined as
\begin{equation}
\text{Approximation Ratio} = \frac{C_\text{Max} - C_e}{C_\text{Max} - C_\text{Min}}
\label{equation:approximation_ratio}
\end{equation}

For the Ising model where $C_\text{Max}$ is the global maximum possible energy and $C_\text{Min}$ is the global minimum possible energy. $C_e$ is the cost function $C(z)$, Eq.~\eqref{equation:problem_instance}, also called the energy, of a single spin configuration. This approximation ratio formula is maximized for minimum energies, and minimized for maximum energies, meaning that this is intended to be applied to minimization problems.

All binary solutions obtained by the various solvers used in this study are adapted to spins (variable states of $\{+1, -1\}$) via the mapping of $1 \mapsto 1$, $0 \mapsto -1$. Optimal solutions of the Ising models considered in this study, specifically the optimal energy (both minimum and maximum cost function value) is deterministically computed using the commercial mathematical optimization software packages CPLEX and Gurobi~\cite{cplex, gurobi}. The random $J$ coupling of these models means that these can be viewed as a type of sparse spin glass model (e.g., a heavy-hex spin glass model), in particular the $\pm J$ distribution leads to some competing frustration throughout the model, and therefore fairly high ground-state degeneracy. However, we do not investigate any degeneracy properties, such as degeneracy lifting, or frustration properties, of these models in this study. 

\begin{figure*}[th!]
    \centering
    \includegraphics[width=0.495\textwidth]{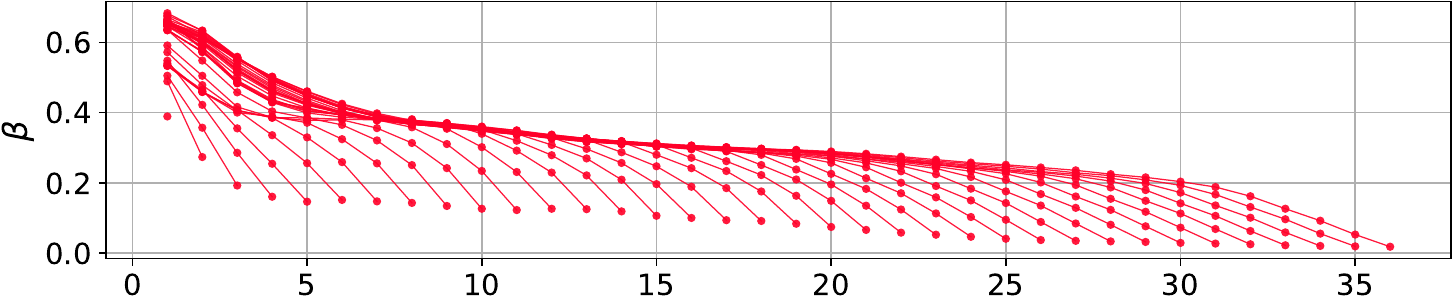}
    \includegraphics[width=0.495\textwidth]{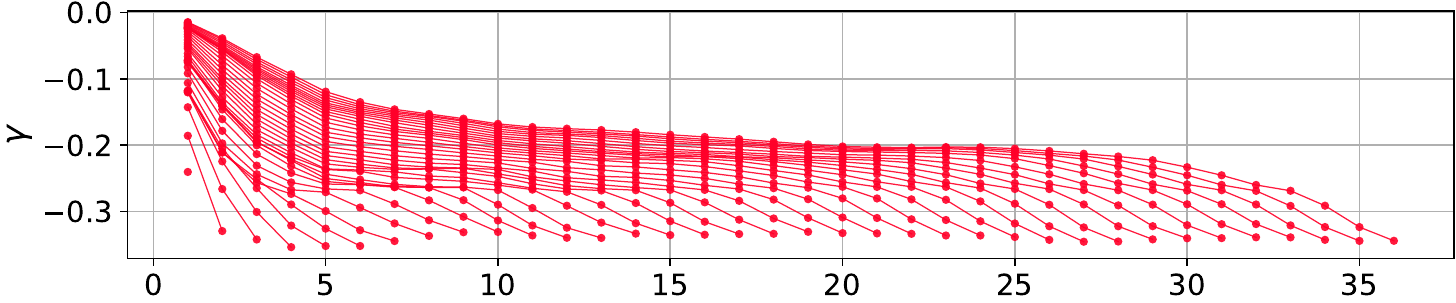}
    \includegraphics[width=0.495\textwidth]{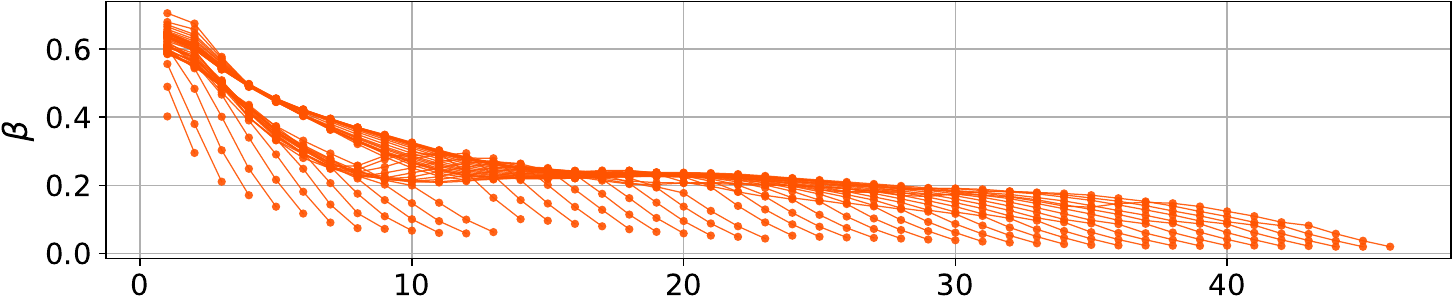}
    \includegraphics[width=0.495\textwidth]{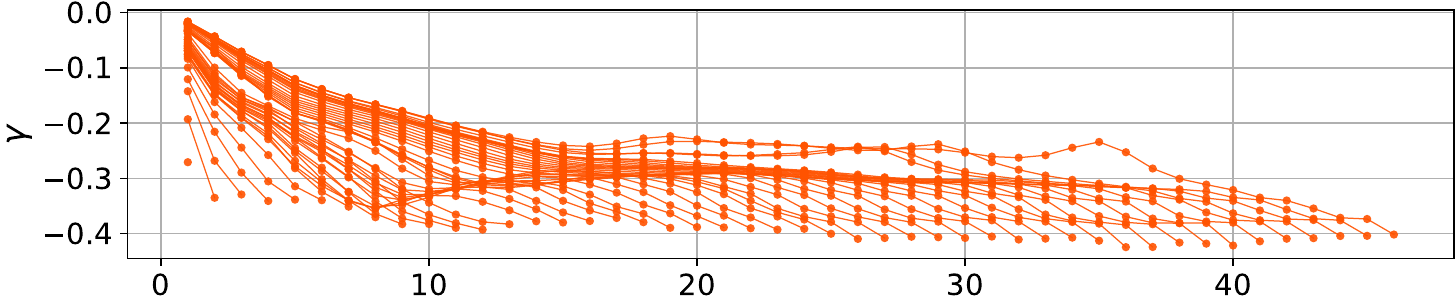}
    \includegraphics[width=0.495\textwidth]{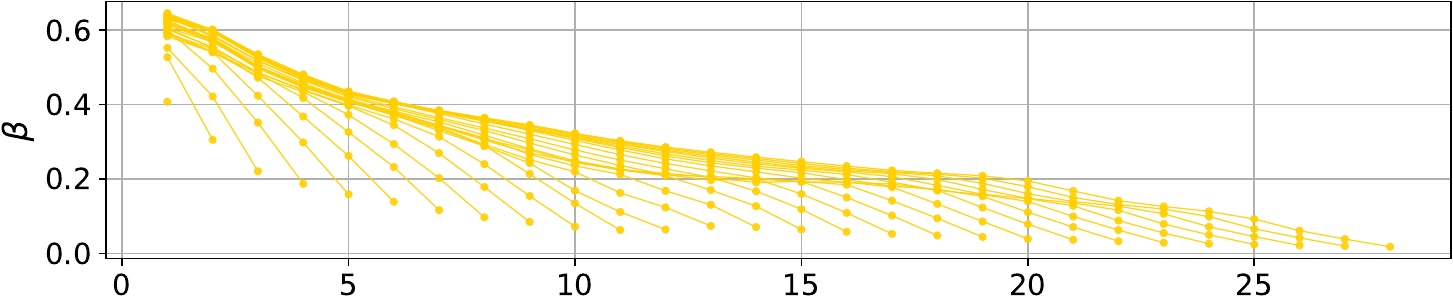}
    \includegraphics[width=0.495\textwidth]{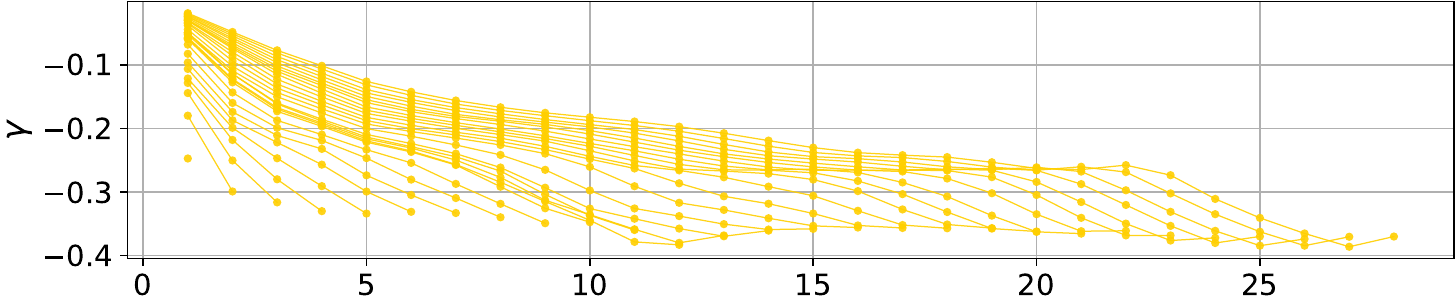}
    \includegraphics[width=0.495\textwidth]{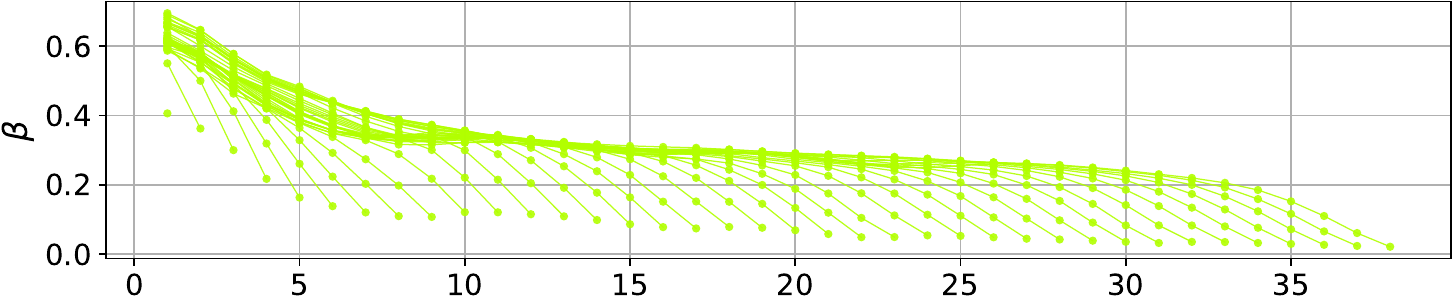}
    \includegraphics[width=0.495\textwidth]{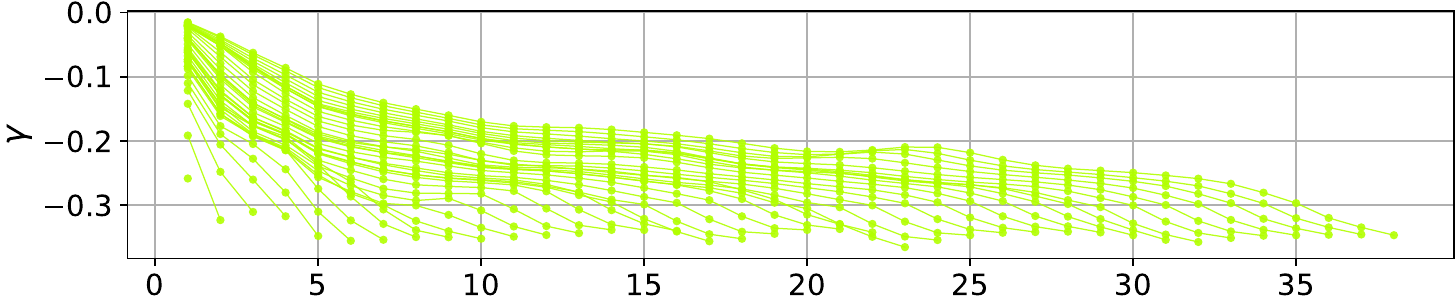}
    \includegraphics[width=0.495\textwidth]{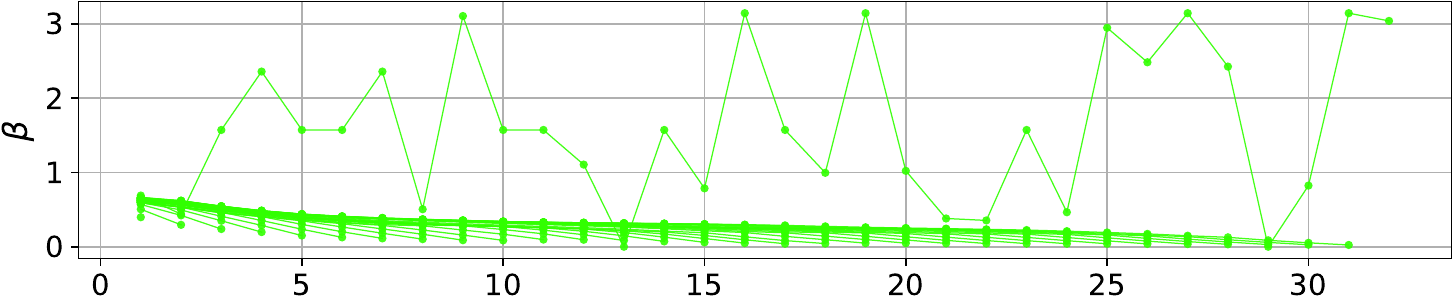}
    \includegraphics[width=0.495\textwidth]{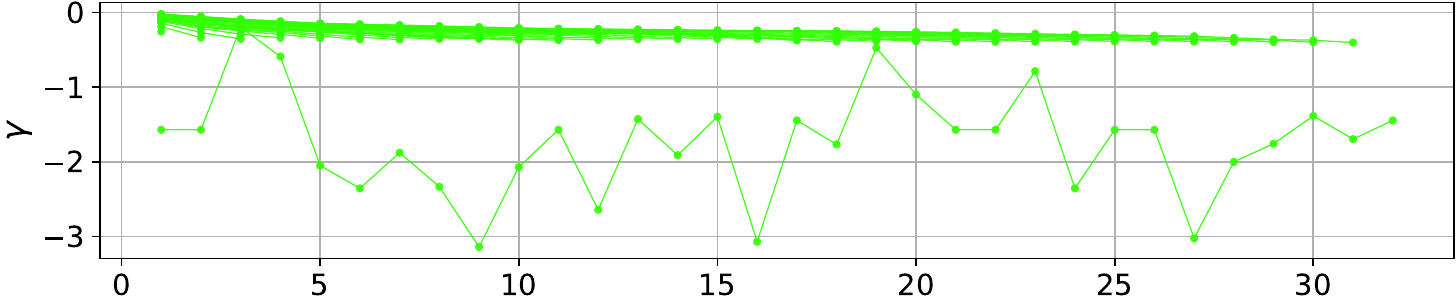}
    \includegraphics[width=0.495\textwidth]{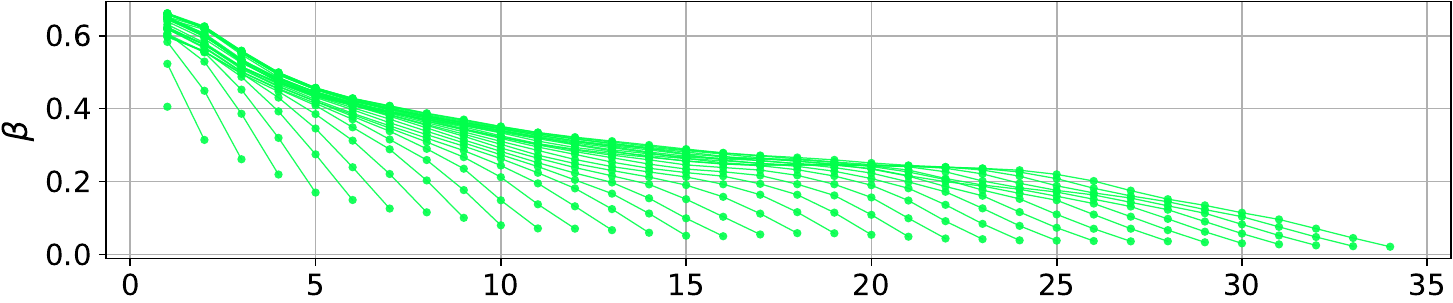}
    \includegraphics[width=0.495\textwidth]{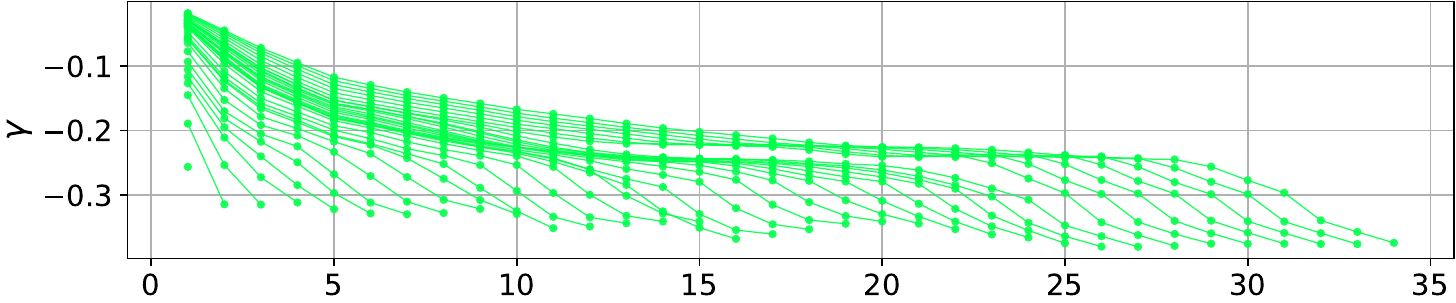}
    \includegraphics[width=0.495\textwidth]{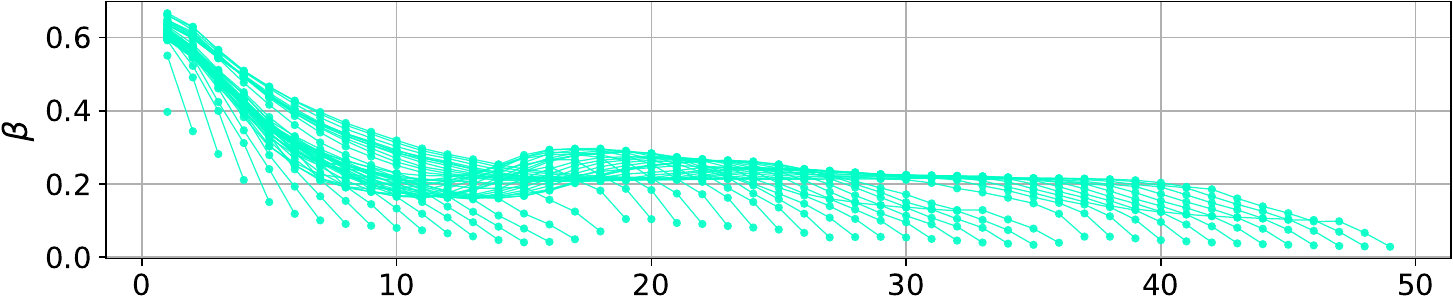}
    \includegraphics[width=0.495\textwidth]{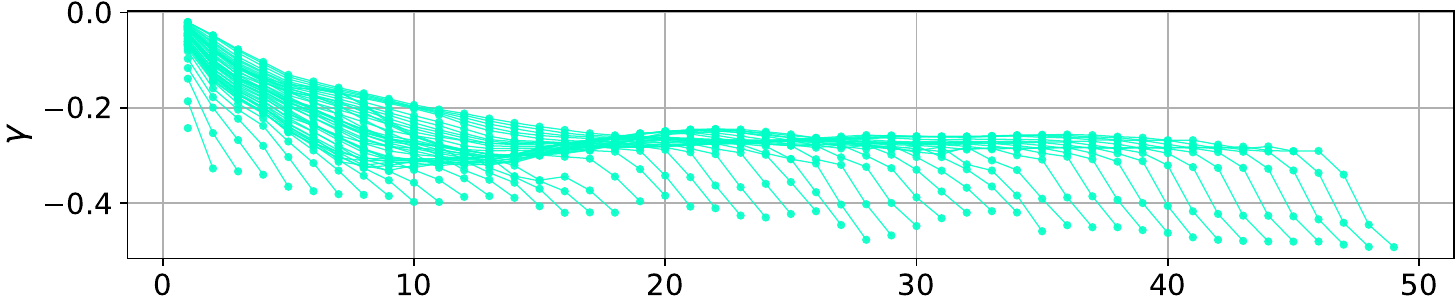}
    \includegraphics[width=0.495\textwidth]{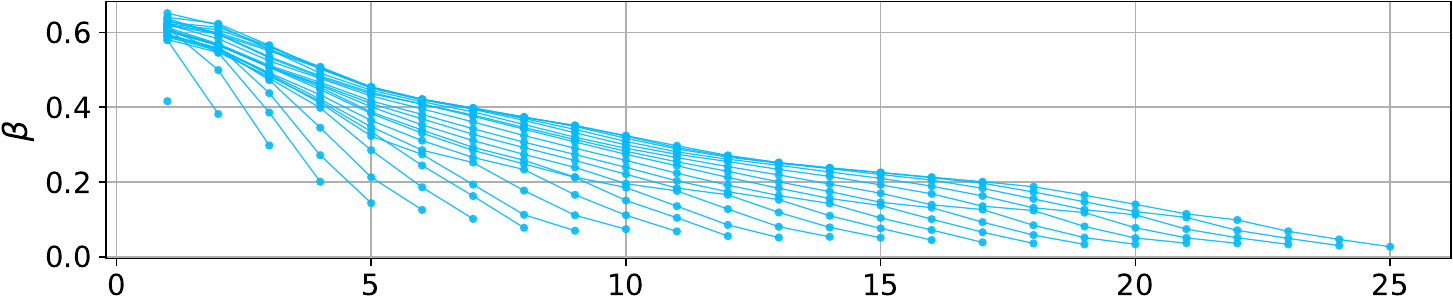}
    \includegraphics[width=0.495\textwidth]{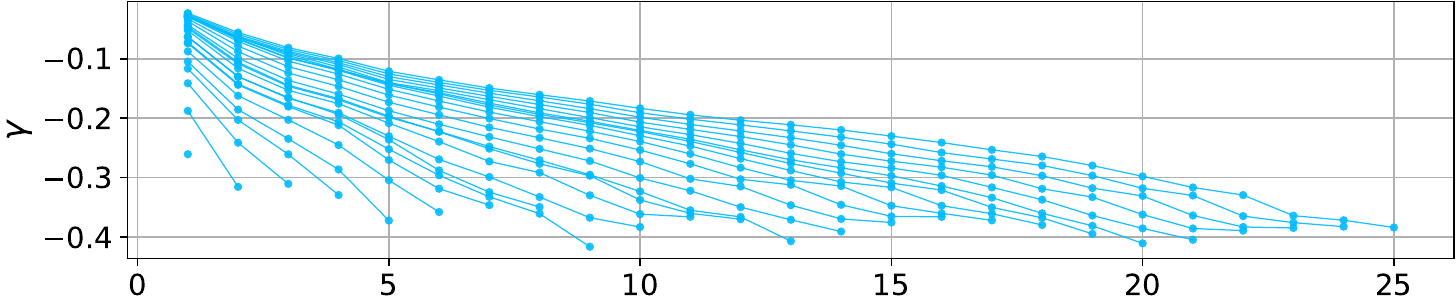}
    \includegraphics[width=0.495\textwidth]{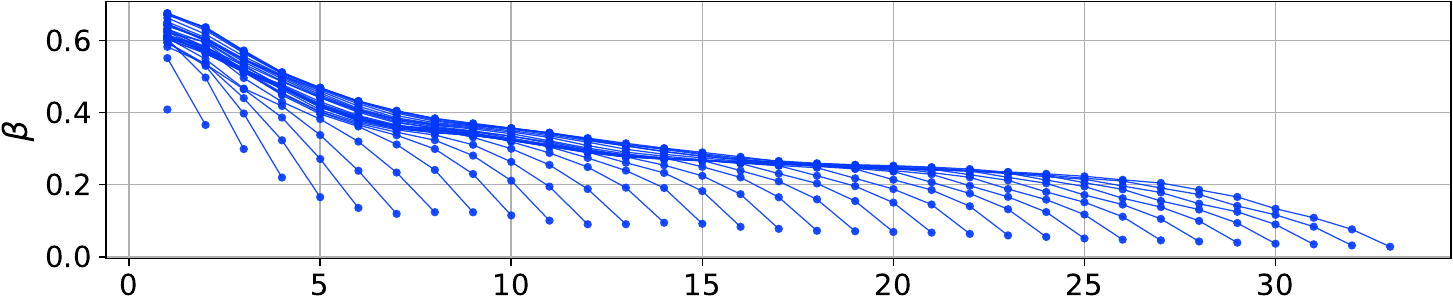}
    \includegraphics[width=0.495\textwidth]{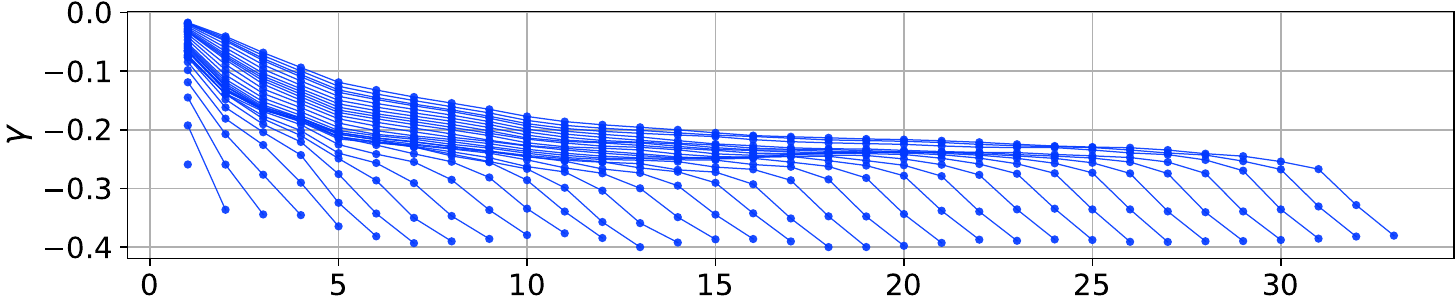}
    \includegraphics[width=0.495\textwidth]{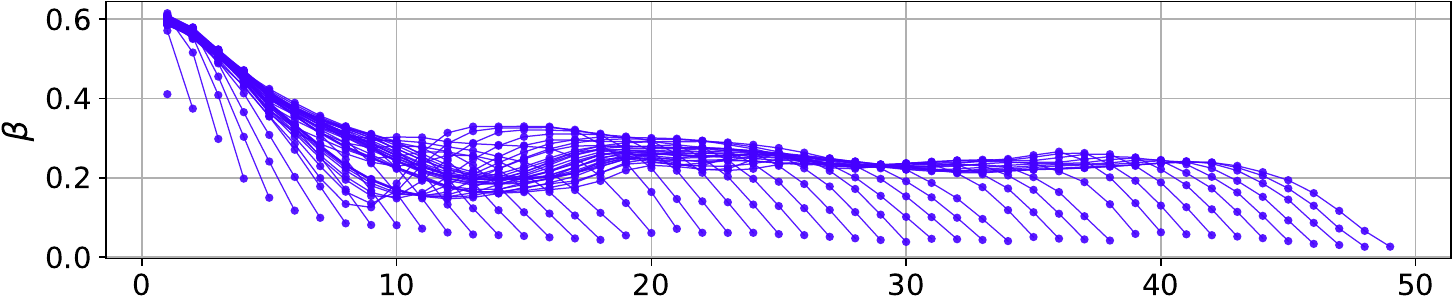}
    \includegraphics[width=0.495\textwidth]{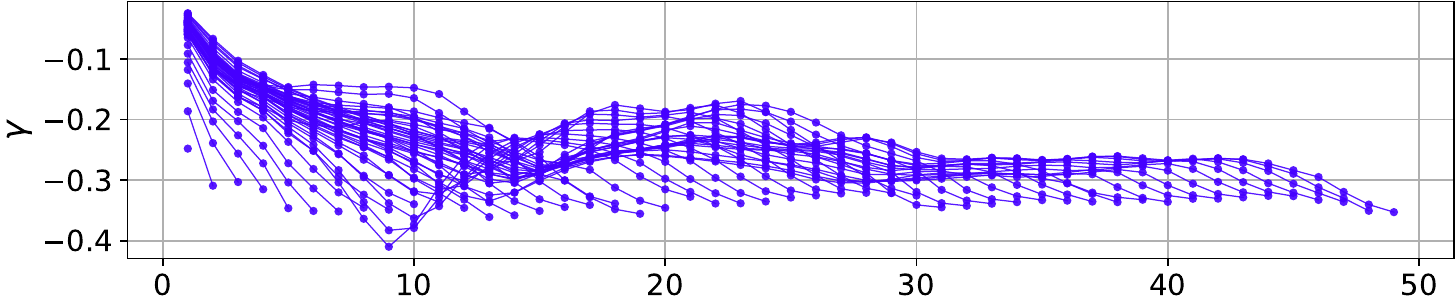}
    \includegraphics[width=0.495\textwidth]{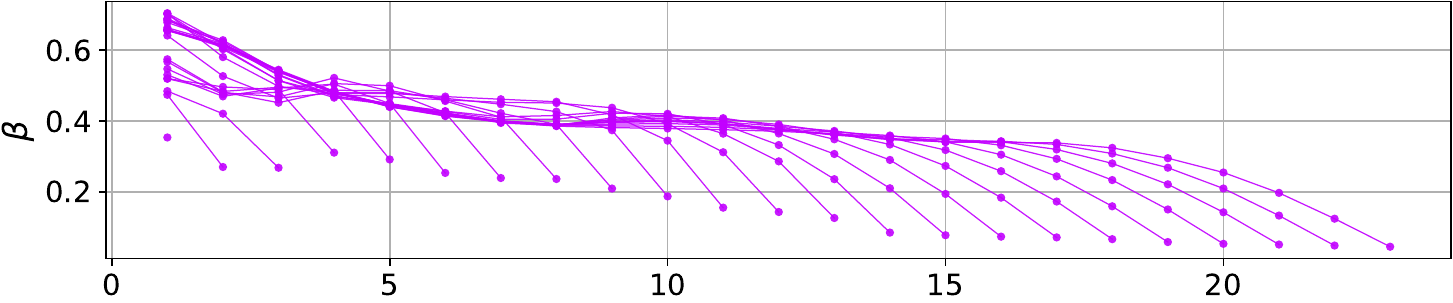}
    \includegraphics[width=0.495\textwidth]{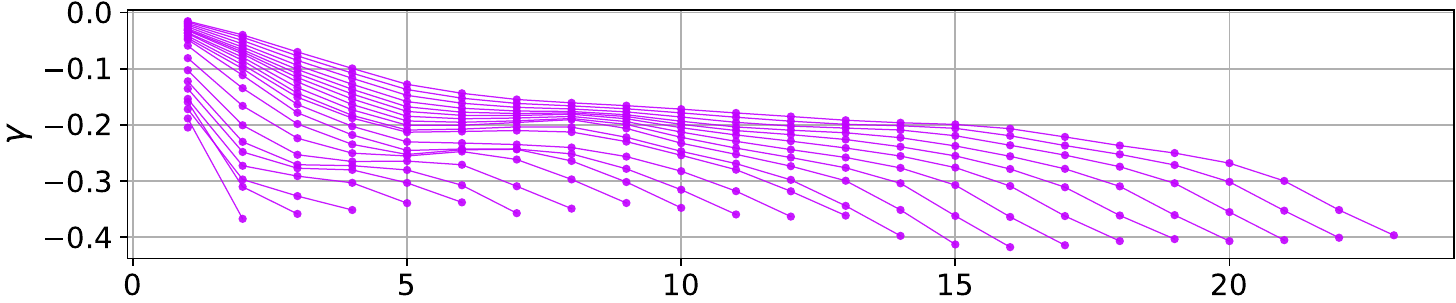}
    \includegraphics[width=0.495\textwidth]{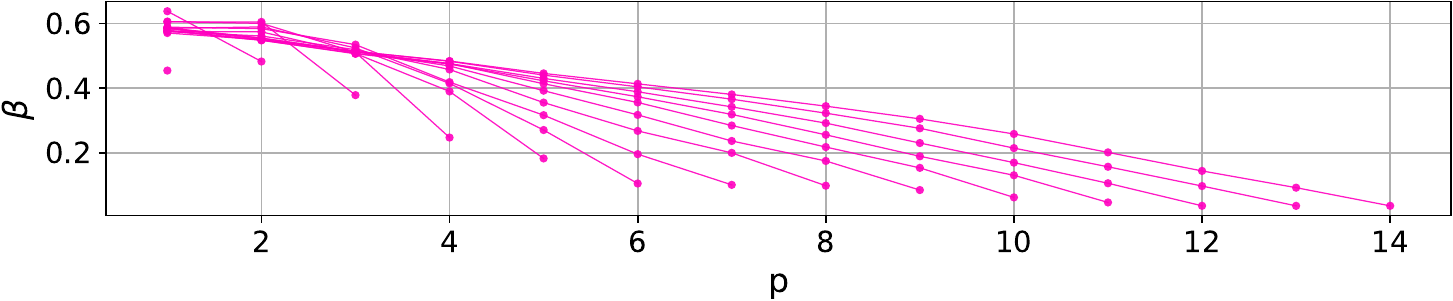}
    \includegraphics[width=0.495\textwidth]{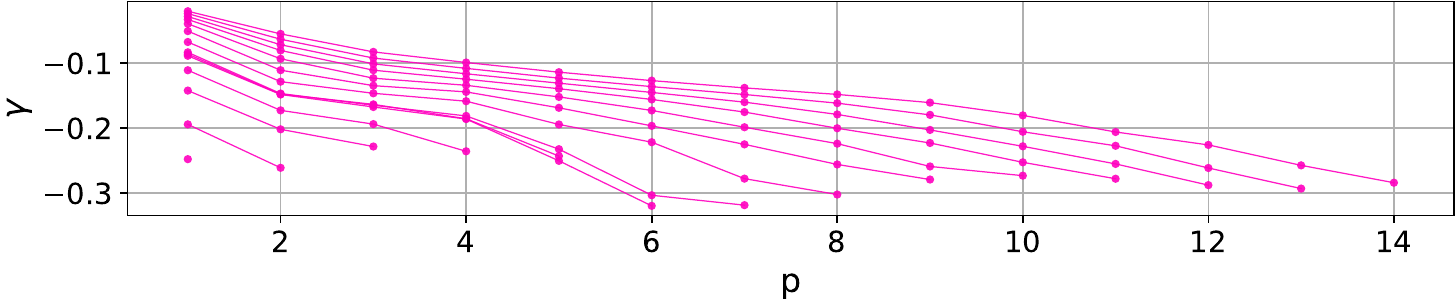}
    \caption{$16$ qubit instance QAOA angles learned up to $p=49$ or close to the ground-state energy. Each row corresponds to a different problem instance (first 10 are random-coefficient, bottom two are uniform positive and negative coefficients).  }
    \label{fig:learned_angles_16_qubits}
\end{figure*}

\begin{figure*}[ht!]
    \centering
    \includegraphics[width=0.495\textwidth]{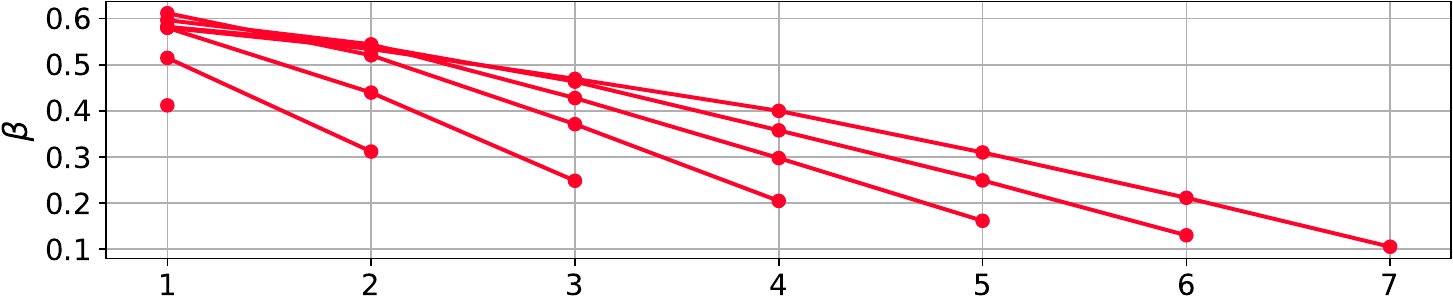}
    \includegraphics[width=0.495\textwidth]{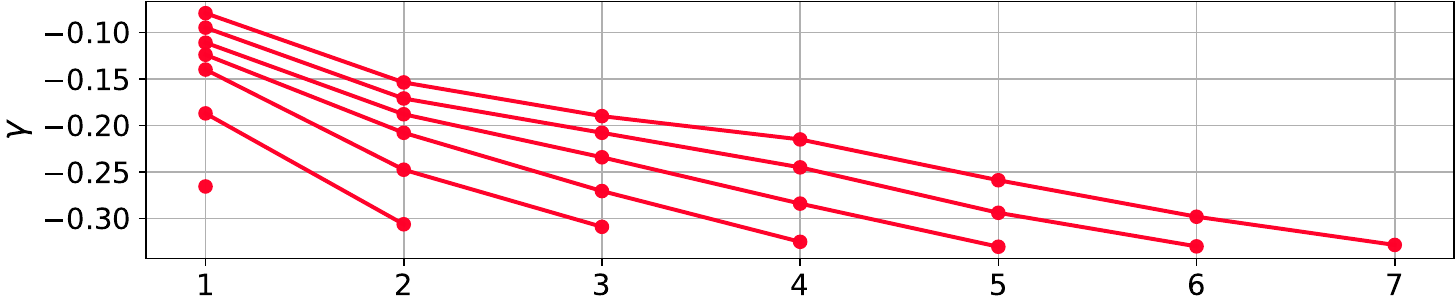}
    \includegraphics[width=0.495\textwidth]{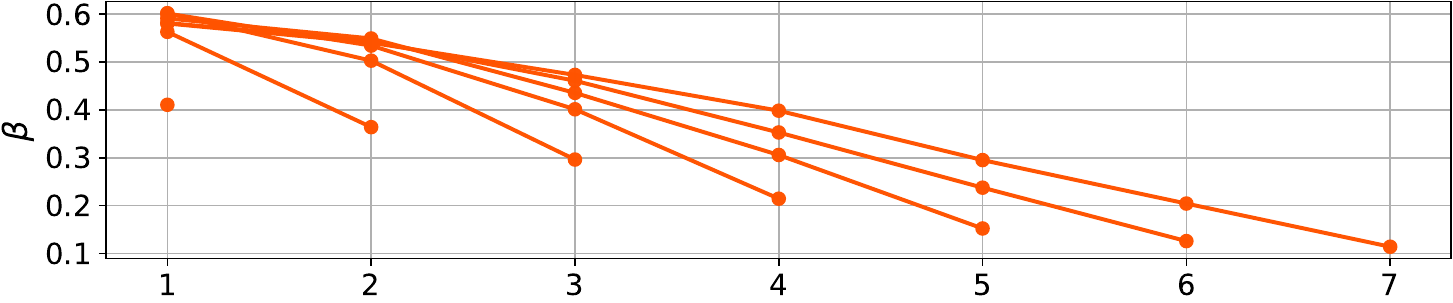}
    \includegraphics[width=0.495\textwidth]{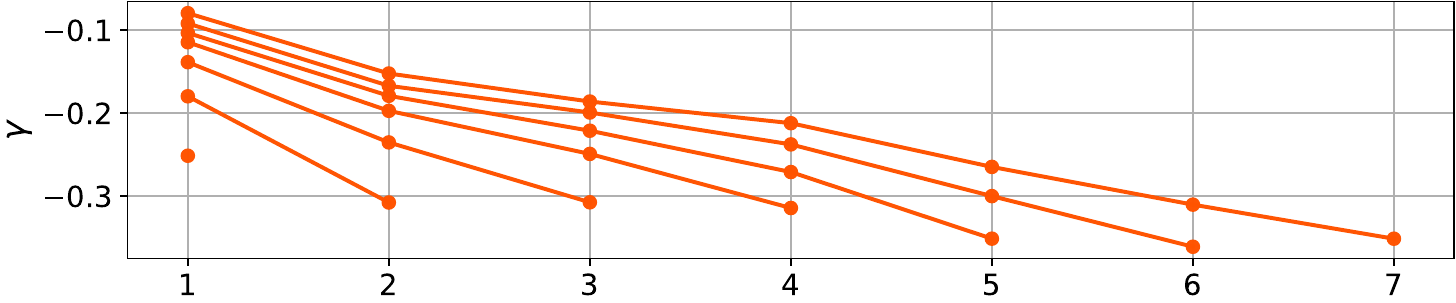}
    \includegraphics[width=0.495\textwidth]{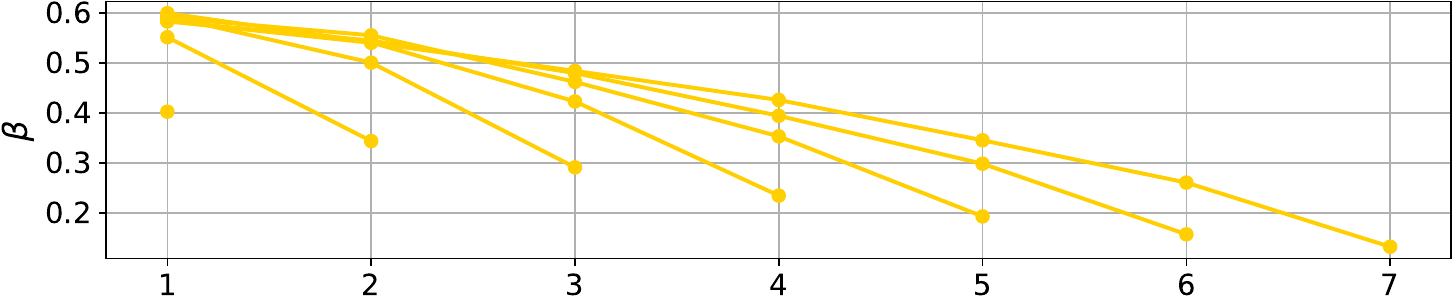}
    \includegraphics[width=0.495\textwidth]{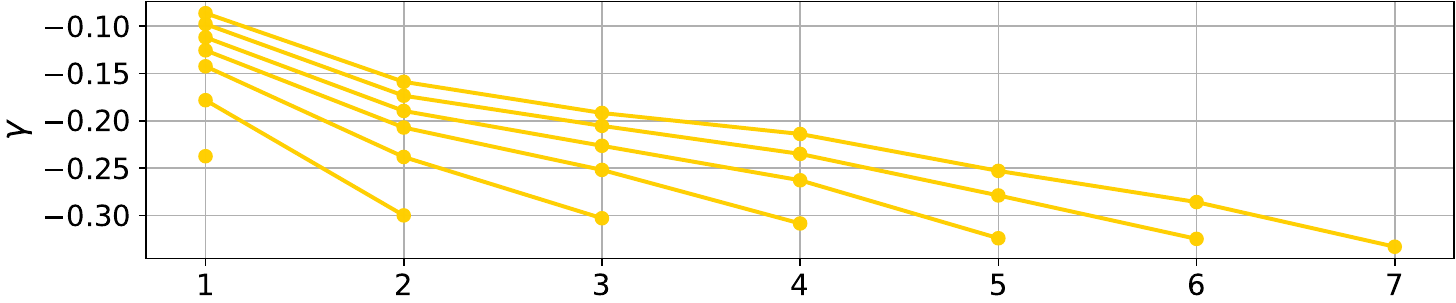}
    \includegraphics[width=0.495\textwidth]{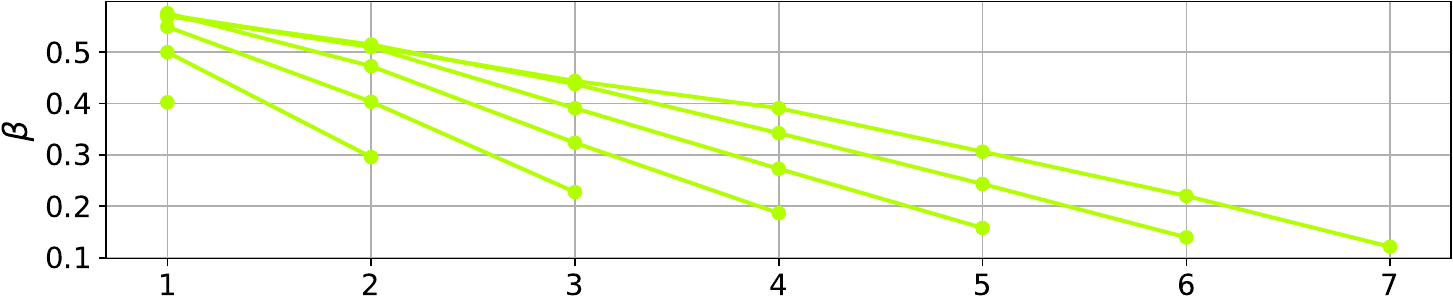}
    \includegraphics[width=0.495\textwidth]{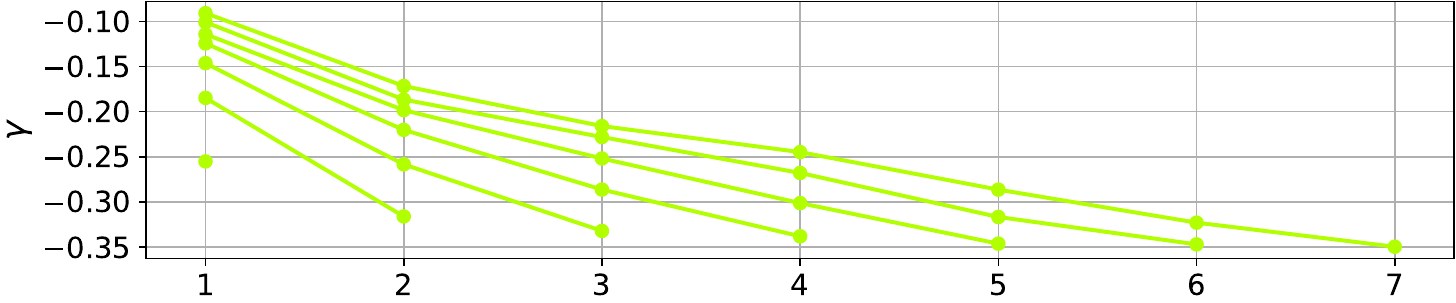}
    \includegraphics[width=0.495\textwidth]{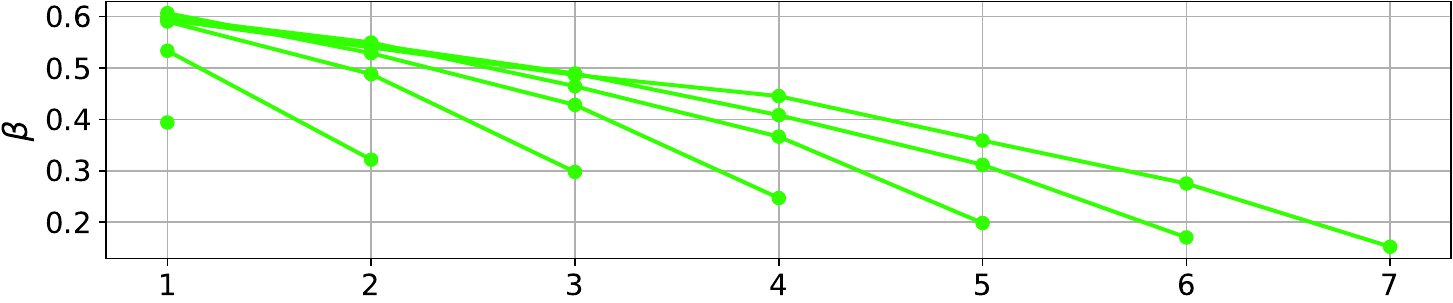}
    \includegraphics[width=0.495\textwidth]{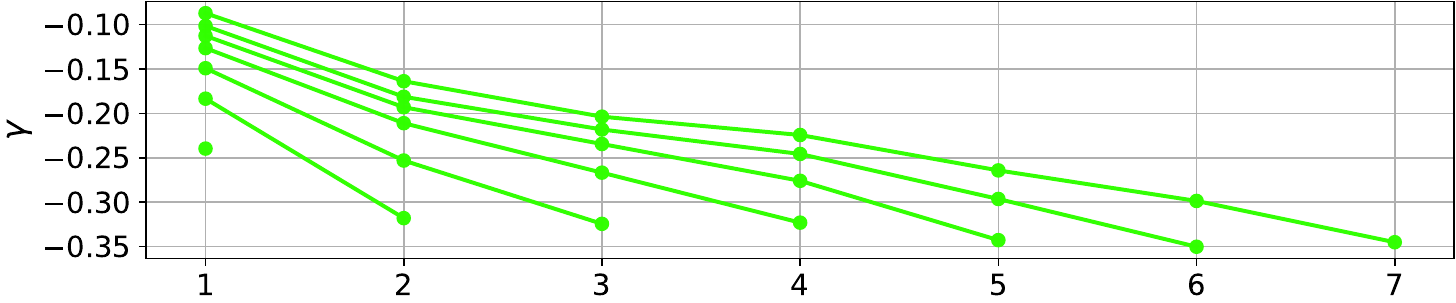}
    \includegraphics[width=0.495\textwidth]{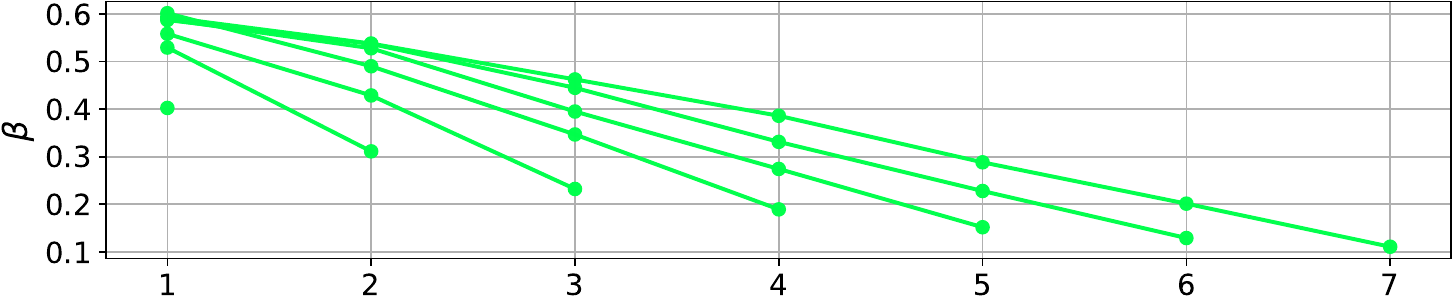}
    \includegraphics[width=0.495\textwidth]{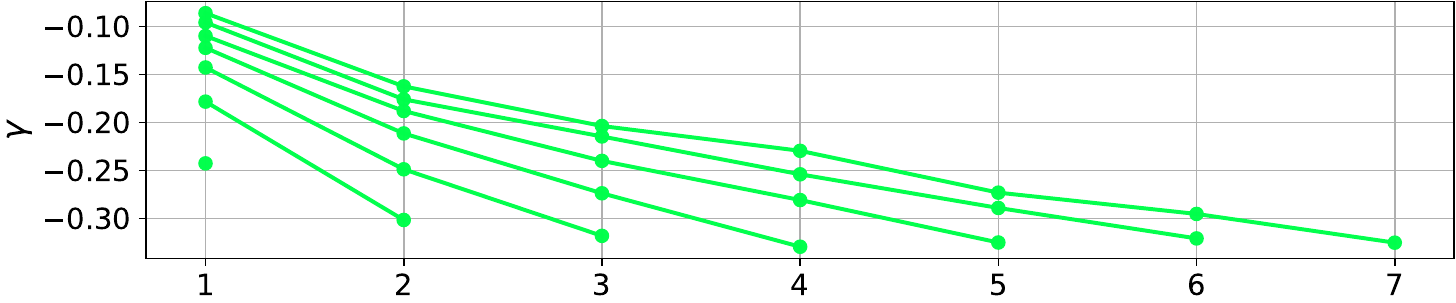}
    \includegraphics[width=0.495\textwidth]{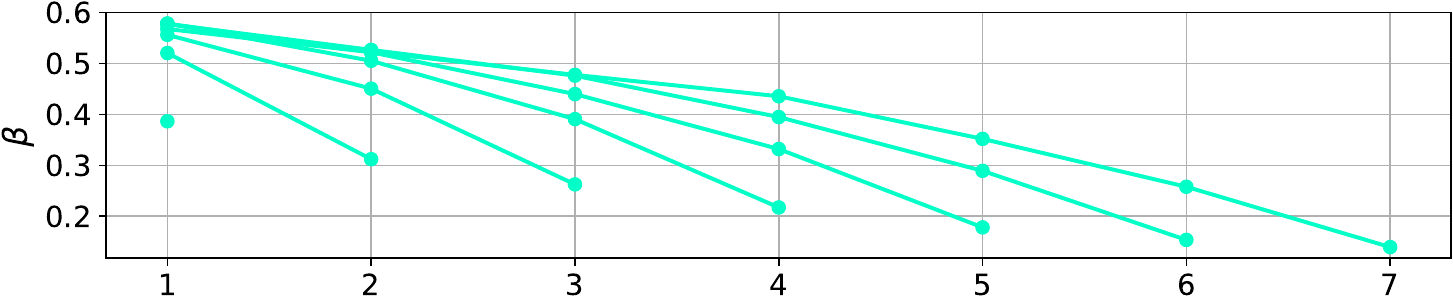}
    \includegraphics[width=0.495\textwidth]{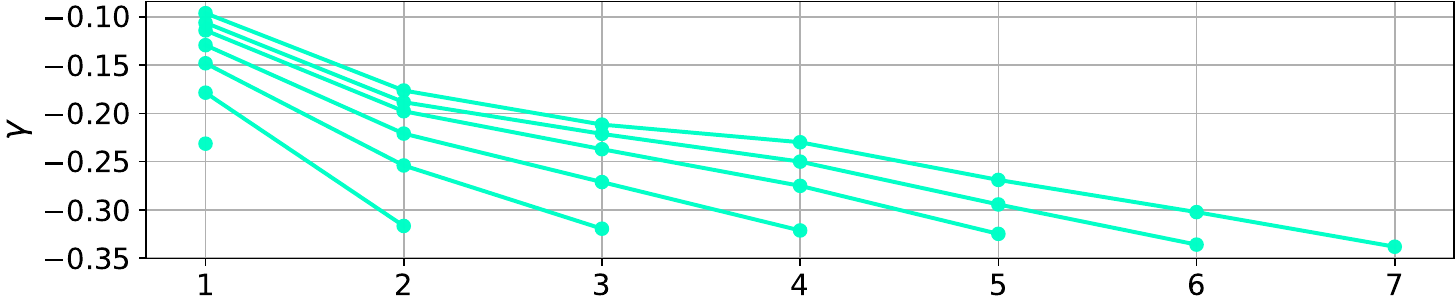}
    \includegraphics[width=0.495\textwidth]{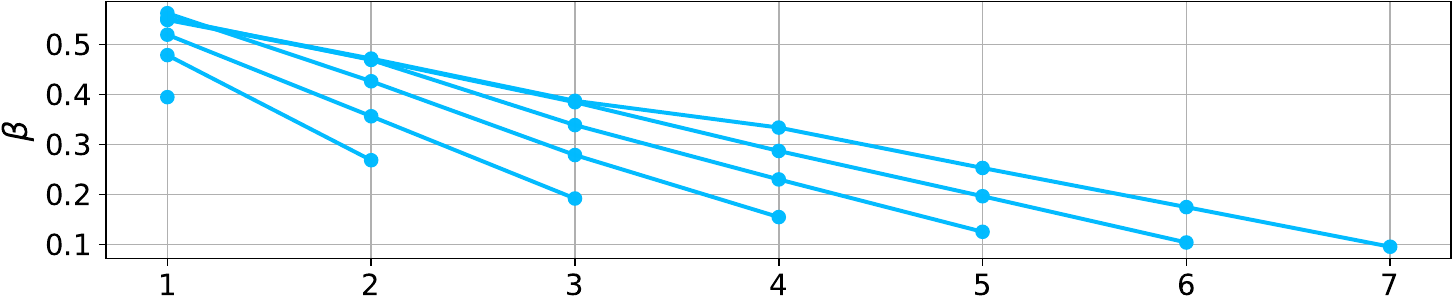}
    \includegraphics[width=0.495\textwidth]{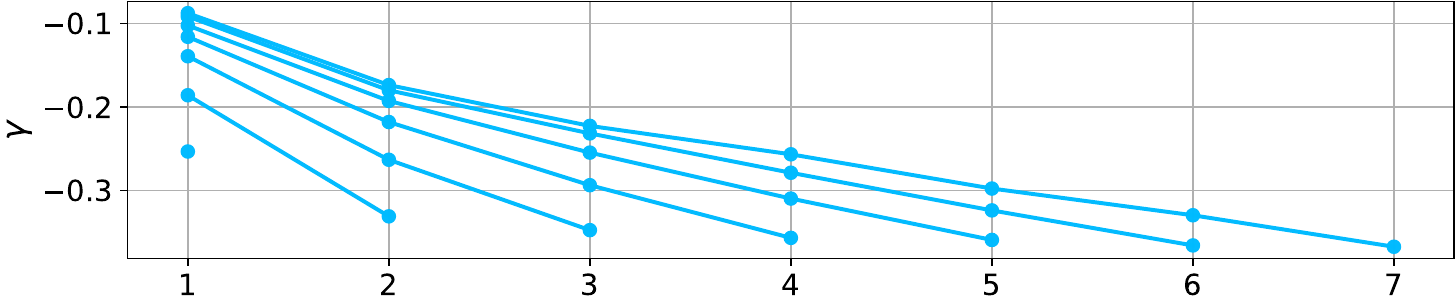}
    \includegraphics[width=0.495\textwidth]{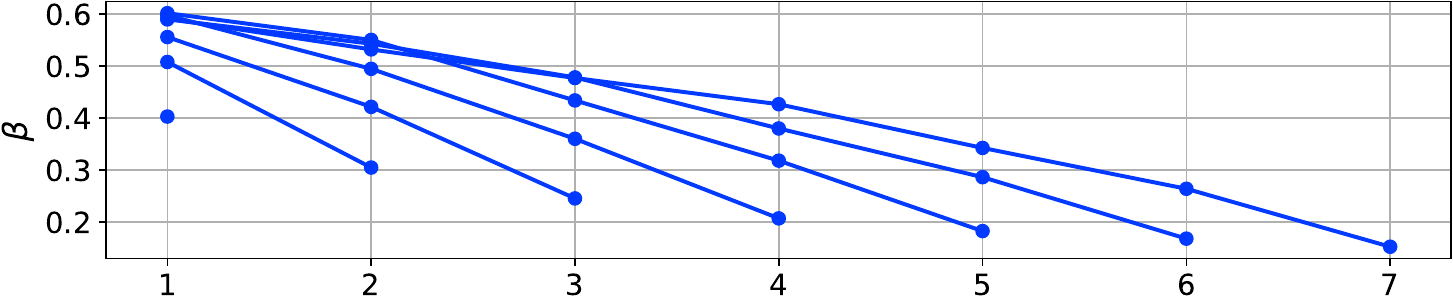}
    \includegraphics[width=0.495\textwidth]{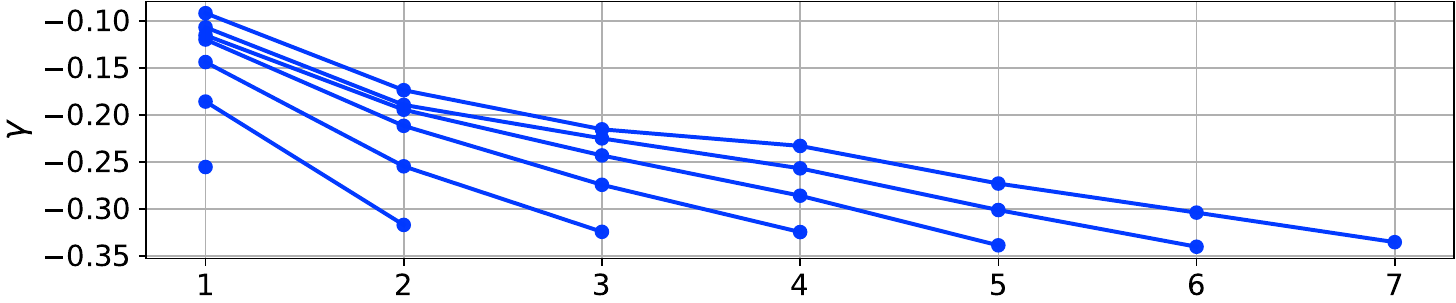}
    \includegraphics[width=0.495\textwidth]{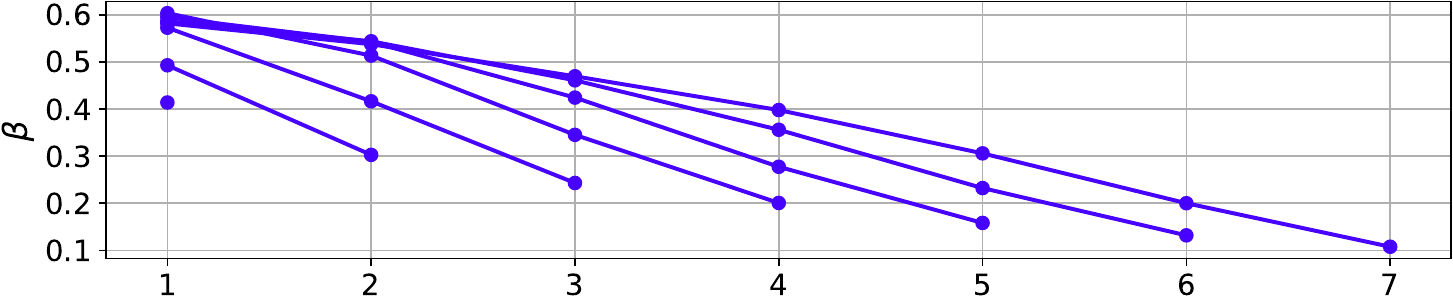}
    \includegraphics[width=0.495\textwidth]{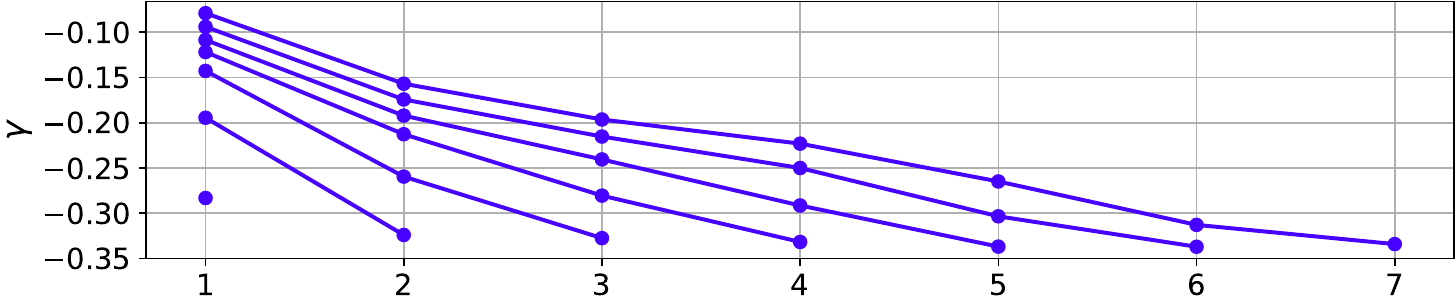}
    \includegraphics[width=0.495\textwidth]{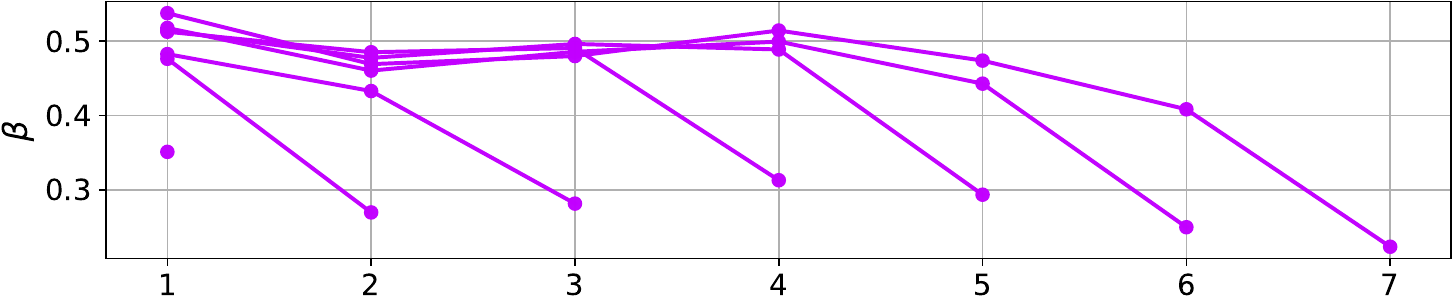}
    \includegraphics[width=0.495\textwidth]{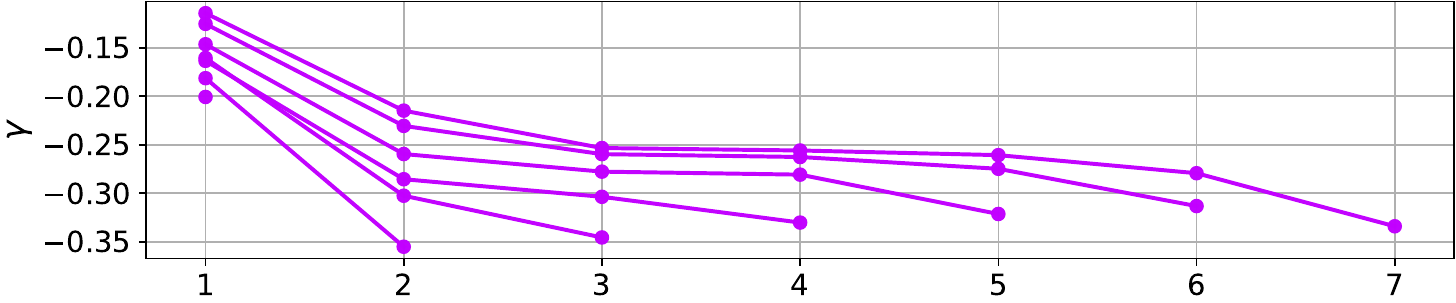}
    \includegraphics[width=0.495\textwidth]{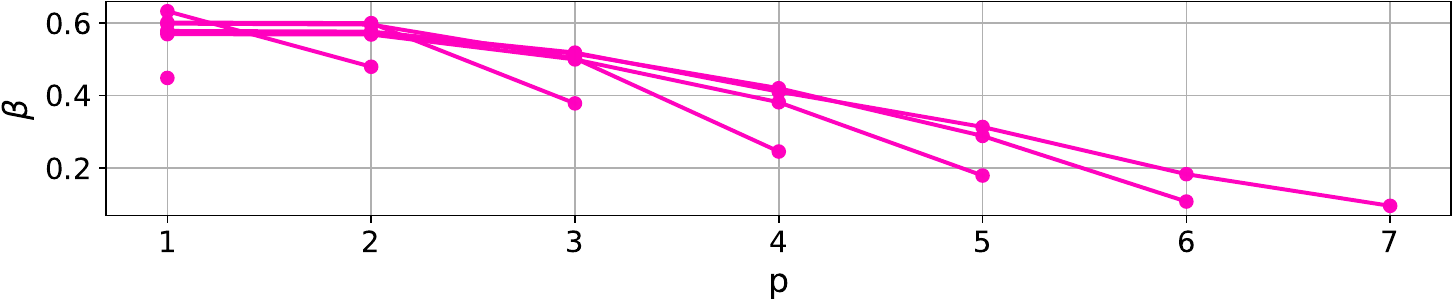}
    \includegraphics[width=0.495\textwidth]{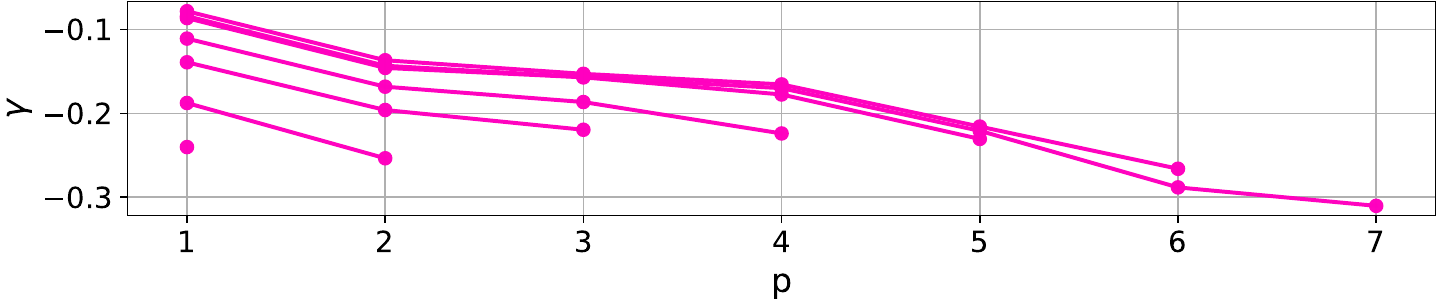}
    \caption{$27$ qubit instance QAOA angles learned up to $p=7$ for all instances. Each row corresponds to a different problem instance (first 10 are random-coefficient, bottom two are uniform positive and negative coefficients).  }
    \label{fig:learned_angles_27_qubits}
\end{figure*}

\subsection{Projected Entangled Pair States Simulation Methods}
\label{section:methods_PEPS}

The scale of the quantum simulations here means that it is intractable to perform full state-vector simulations. However, there exist approximation tensor network-based methods that can in many instances perform high quality simulations of these types of quantum circuits, namely on the order of 100's of qubits and with entangling graph structure (in this case defined by the IBM heavy-hex hardware graphs), which are relatively sparse. To this end, it would be useful to validate how well these QAOA angles work on very large instances, using simulations that have lower noise than the IBM QPUs themselves. For this we turn to PEPS simulations, complemented by MPS and LOWESA simulations at low depth to cross-verify that these simulations agree. Tensor network based approximation of low-entanglement digital quantum circuits is a versatile field~\cite{Vidal_2003, Shi_2006, jozsa2006simulationquantumcircuits} with a number of different tools and allows us in this case to verify, with some reasonable level of numerical approximation, the expected performance of QAOA on this large set of parameter-transferred angles. 

Appendix~\ref{section:appendix_PEPS_contraction_methods_comparison} compares different tensor network contraction methods, and we find that belief propagation (BP) offers a very good convergence of local expectation values, while being more efficient than alternative methods. 
Therefore, we use BP-based simple update~\cite{alkabetz2021beliefpropagation, tindall2023beliefpropagation} (which is equivalent to the original simple update algorithm~\cite{jiang2008simpleupdate}) to simulate the time evolution of the PEPS representation of the quantum system, and BP to perform the final contraction of the tensor network for calculation of the expectation values or sampling (unless specifically stated otherwise). This is corroborated by Refs.~\cite{tindall2024IBM, begusic2024IBM, siddhartha2024IBM, rudolph2025simulatingsamplingquantumcircuits}, who found similar properties of simple update and BP specifically for digital circuits defined on the heavy-hex connectivity graph.

Our PEPS simulations up to bond dimension of $D=64$, including sampling, have been performed using the YASTN tensor network package~\cite{YASTN}, and at $D=128$ were performed using a slightly different in-house implementation built on top of ITensors~\cite{itensor, itensor-r0.3}. The YASTN-based QAOA circuit simulation implementation is publicly available on Github~\cite{marek_github}. The difference between these two implementations is in how the geometrically local cubic terms are handled, resulting in a tradeoff between computational cost and accuracy at any fixed bond dimension value. 
In the YASTN-based implementation, we assign each PEPS tensor to a group of few qubits, so that each cubic term results in an update of two nearest-neighbor PEPS tensors. In the in-house implementation, each PEPS tensor corresponds to a single qubit and each cubic term results in an update of three PEPS tensors. 
Thanks to the grouping of spins, the YASTN-based implementation approximates the exact evolution of the PEPS state more accurately for a given bond dimension $D$. However, it incurs an extra numerical cost due to having to deal with a larger physical space. In contrast, the in-house implementation compensates for its lower accuracy at fixed $D$ by allowing for larger values of the latter, thanks to its lighter computational cost.

\subsection{Matrix Product State Simulation Methods}
\label{section:methods_MPS}

There exist other types of approximate simulation methods which have been able to approximate larger scale quantum circuits, specifically in the case of heavy-hex connectivity circuits, namely LOWESA~\cite{fontana2025classical,rudolph2023classical} and MPS~\cite{tindall2024IBM}. For this study, we focus on using PEPS simulations because PEPS is particularly suited for 2D system simulations. However, we do evaluate how well MPS and LOWESA perform on simulating these QAOA circuits for two purposes. First, at low circuit depth, we see very good agreement between all three methods, thus validating the accuracy of all three simulators in the low-entanglement/low-$p$ regime. But, at higher circuit depths, we see breakdown of numerical stability and convergence, in particular for LOWESA, thus highlighting the necessity of validating the QAOA simulations with PEPS. Appendix~\ref{section:appendix_LOWESA} details these LOWESA simulations.

Here, we detail the 1-dimensional tensor network simulations that we perform on these QAOA circuits, in order to additionally validate the performance of the PEPS simulations. In particular, we expect that if the circuits generate sufficiently low entanglement, we should observe very good agreement between PEPS and MPS at sufficiently large bond dimensions. 

Here, we represent the state $\ket{\vec{\beta},\vec{\gamma}}$ in Eq.~\eqref{eq:QAOAstate} with MPS. The action of $e^{-i\beta_j H_M}$ and $e^{-i\gamma_j H_C}$ is simulated with a version of time-evolving block decimation~\cite{vidal2003efficient}. MPS uses a refinement parameter (called bond dimension and denoted by $\chi$ here) that controls the accuracy of the simulation. The Hamiltonian $H_C$ in Eq.~\eqref{equation:problem_instance} contains a significant number of non-local (from MPS point of view) interactions and therefore accurate description of $\ket{\vec{\beta},\vec{\gamma}}$ may require large bond dimension.

Our simulation algorithm collects the terms in $H_C$ acting on the same qubits and decomposes $e^{-i\gamma_j H_C}$ into a product of three-qubit unitaries. It turns out that those unitaries can be exactly represented by a Matrix Product Operator with bond dimension 2. It can be done by performing SVD of a properly reshaped unitary~\cite{orus2014practical}. This way of treating non-local interactions improves the efficiency of our MPS simulation algorithm. Our algorithm collects terms in the Hamiltonian $H_C$ into layers of non-overlapping unitaries. Unitaries $e^{-i\gamma Z_iZ_jZ_k}$ and $e^{-i\gamma Z_{i'}Z_{j'}Z_{k'}}$ belong to the same layer if $i \geq k'$ or $i' \geq k$. MPS is compressed after all unitary gates in a given layer are applied, possibly resulting in a loss of fidelity. The compression errors (expressed as the sum of discarded eigenvalues of the density matrix of the half of the system) are tracked and found to decay rapidly with~$\chi$. The application of $e^{-i\beta_j H_M}$ to an MPS is trivial and does not require compression.

\begin{figure*}[th!]
    \centering
    \includegraphics[width=0.495\textwidth]{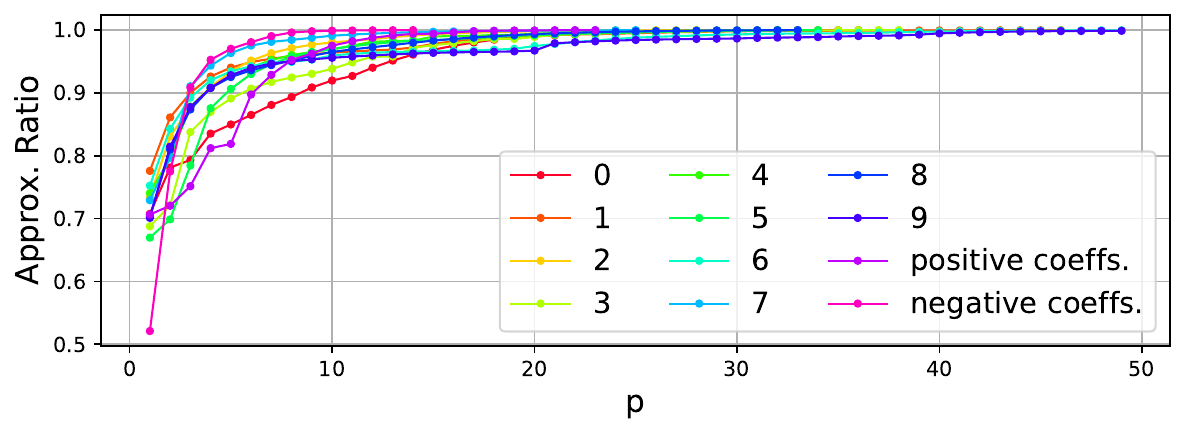}
    \includegraphics[width=0.495\textwidth]{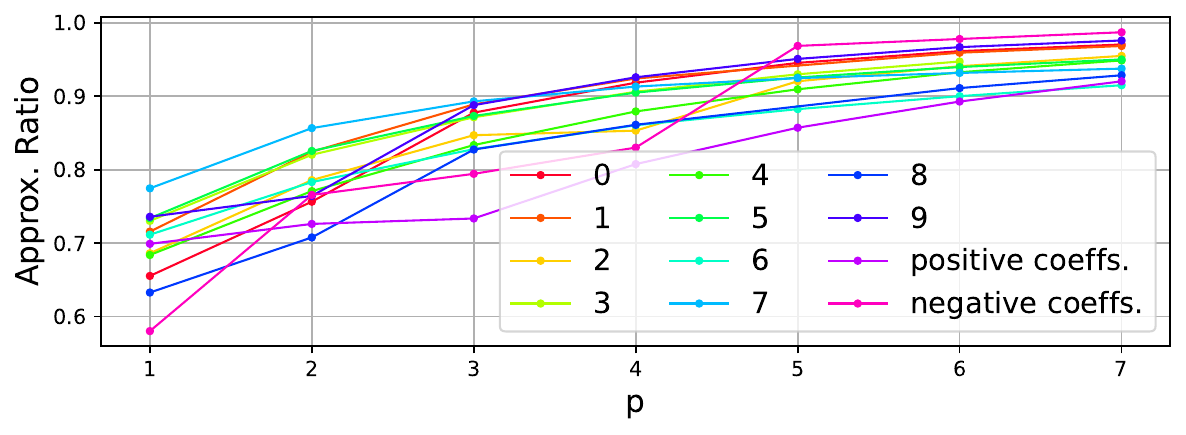}
    \caption{Approximation ratio (y-axis) vs $p$ (x-axis) for the $12$ distinct $16$ qubit Ising model (left) and the $12$ distinct $27$ qubit Ising models (right). The legends show the specific instance type -- $10$ random coefficient models and one model with entirely positive coefficients and one model with entirely negative coefficients. The $27$ qubit instance training is cutoff at $p=7$; note that for all instances the approximation ratio is above $0.90$ at $p=7$. }
    \label{fig:learned_energy_approximation_ratios}
\end{figure*}

\section{Results}
\label{section:results}

The results are organized into sub-sections as follows. First, in Section~\ref{section:results_learning_QAOA_angles} we present the trained QAOA angles, which we will then use for the remainder of the study as fixed QAOA parameters and evaluate how well they work for direct parameter transfer on much larger problem sizes. Next, Section~\ref{section:results_small_scale_parameter_transfer_validation} validates how well some of these fixed parameters work when transferred onto new (unseen) test-problem instances, using exact classical statevector simulations (up to $27$ qubits). 

Section~\ref{section:results_IBM_Quantum_computer_experiments} presents extensive large-scale IBM NISQ computer QAOA executions, using the ensemble of transferred QAOA angles. This section includes extensive QAOA sampling result analysis, including reporting large QAOA volume measures. 

Section~\ref{section:results_PEPS_simulations} presents extensive classical tensor network simulations, specifically PEPS, of the high depth QAOA circuits, validating how well the QAOA parameter transfer works on Ising model problem sizes of up to $156$ spins. This includes ground-state sampling rates, and convergence analysis of the PEPS simulations. Complementing the PEPS simulations, Section~\ref{section:results_MPS_simulations} then further validates PEPS simulation accuracy using MPS tensor network simulations. Combined, Section~\ref{section:results_PEPS_simulations} and Section~\ref{section:results_MPS_simulations} (as well as Appendix~\ref{section:appendix_LOWESA}) show strong numerical evidence that large scale parameter transfer QAOA would work, up to high depth $p$ QAOA, on quantum computers with significantly lower error than the actual quantum computer hardware results in Section~\ref{section:results_IBM_Quantum_computer_experiments}.

\subsection{Classical QAOA Angle Learning with Statevector Simulations}
\label{section:results_learning_QAOA_angles}

Fig.~\ref{fig:learned_angles_16_qubits} plots the numerical quantities of the learned QAOA angles $\beta, \gamma$ for the ensemble of $12$ different $16$-qubit heavy-hex spin glass instances, learned until $p=49$ or the mean (noiseless) energy is at least $0.01$ (energy) below the optimal ground-state energy. During training, the QAOA angles are not restricted or bounded to a certain range. Fig.~\ref{fig:learned_angles_27_qubits} plots the same but for the ensemble of $27$ qubit instances. This set of $24$ QAOA angles are the parameters that are used in the entirety of this study for evaluating how well, or not, parameter transfer works up to these very high depth QAOA circuits.

\begin{figure*}[ht!]
    \centering
    \includegraphics[width=0.495\textwidth]{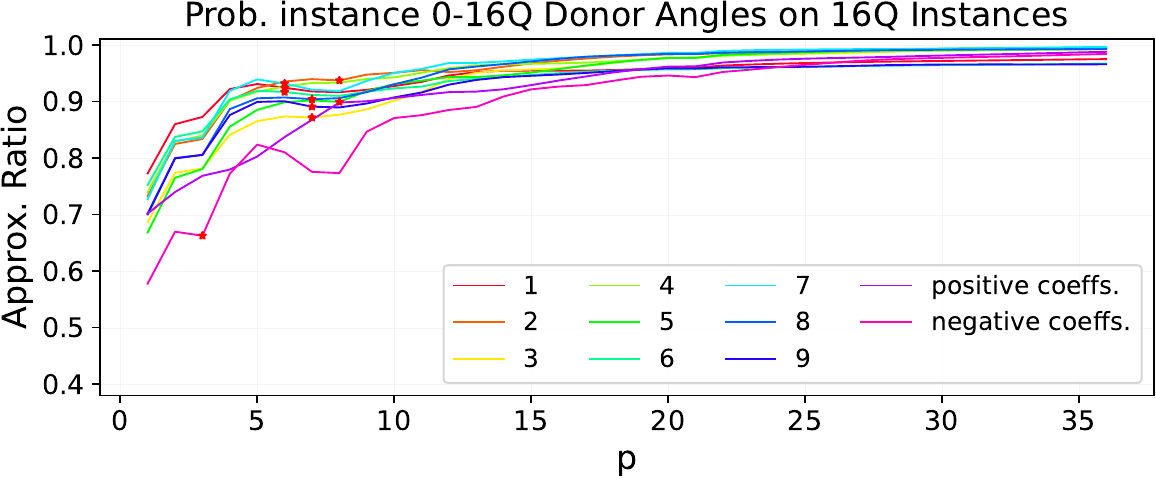}
    \includegraphics[width=0.495\textwidth]{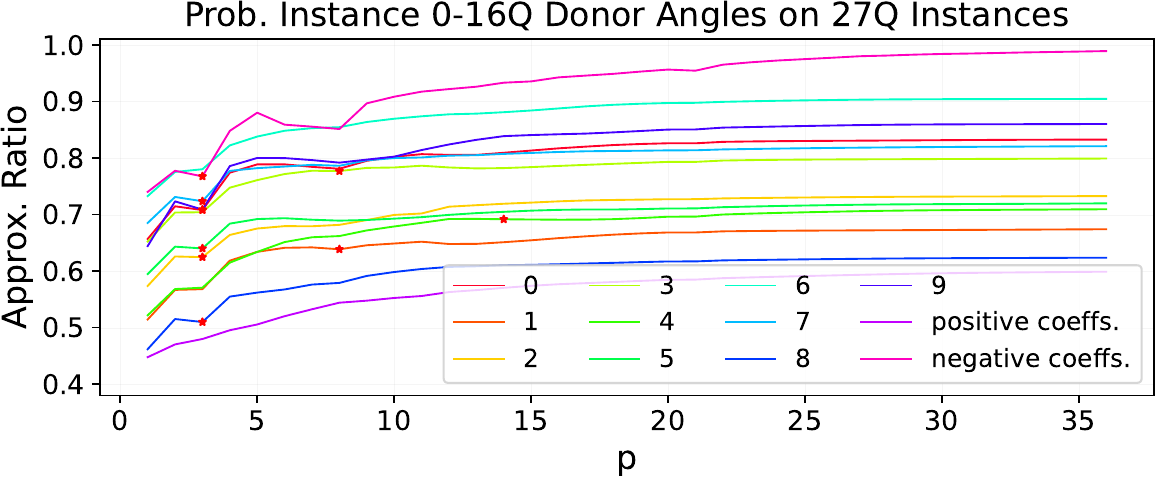}
    \caption{Numerical simulation of parameter transfer of one of the $16$ qubit problem instances onto the other $11$ $16-$qubit problem instances (left) and onto the $12$ $27-$qubit problem instances (right). These simulations are all noiseless (no shot noise). y-axis is the approximation ratio for the given target problem instances, x-axis is $p$ QAOA steps. Each line corresponds to a different problem instance (which were not using in the training of this set of fixed angles), which is indexed in the legend numerically and by the all positive and negative coefficient Ising models. Red asterisk markers denote the $p$ step where the expectation value decreases in quality (increases in energy) for that given problem instance, if that occurs. $8$ out the $11$ instances in the left plot have at least one non-improving $p$ step, and $9$ out of $12$ of the instances in the right plot have at least one non-improving $p$ step. However, there is an overall trend for all problem instances shown in these plots that the higher $p$ angles do give consistently improving energies, even if there are some blips of non-monotonic improvements.  }
    \label{fig:parameter_transfer_16_qubits_instance_0}
\end{figure*}

Our trained angles all lie in the range $[0,2\pi)$, however, there are several periodicities and symmetries that enable us to further transfer the angles into mathematical equivalent angles plotted in Figs.~\ref{fig:learned_angles_16_qubits},~\ref{fig:learned_angles_27_qubits}:
\begin{itemize}[noitemsep]
    \item   The~phase separator $e^{-i\gamma H_P}$ is $\pi$-periodic in $\gamma$, 
            since for each instance defined in Eq.~\eqref{equation:problem_instance}, 
            $H_P=H_C$ has only odd or only even integer eigenvalues,
    \item   the~mixer $e^{-i\beta H_M}$ is $\pi$-periodic up to a global phase, 
            as $e^{-i(\pi+\beta)X} = e^{-i\pi X} e^{-i\beta X} = -\mathit{Id}\cdot e^{-i\beta X}$,
    \item   $H_C$ and $H_M$ are real valued and satisfy time reversal symmetry, hence 
            $\bra{-\vec{\beta},-\vec{\gamma}}H_C\ket{-\vec{\beta},-\vec{\gamma}} = \bra{\vec{\beta},\vec{\gamma}}H_C\ket{\vec{\beta},\vec{\gamma}}$.
\end{itemize}

Flipping signs of all angles for an instance, and/or shifting all $\beta$s or all $\gamma$s by multiples of $\pi$, we plot angles according to their deviation from $0$.
With the angle quantity visualizations having been clarified, and standardized so as to be comparable, using these mathematical transformations, the two sets of angle plots show a number of notable observations in regards to angle trends:

\begin{itemize}[noitemsep]
    \item   The $\vec{\beta}, \vec{\gamma}$ angle schedules start from, and finally approach, particular values,
    \begin{itemize}[noitemsep]
        \item    $\vec{\beta}$ begins at around $\beta_1 \approx 0.6$ and approaches $\beta_p \approx 0$, converging closer to those values as $p$ increases;
        \item   $\vec{\gamma}$ begins at around $\gamma_1 \approx 0$ and approaches $\gamma_p \approx -0.35$ (equivalently $\approx 2.79$ ), converging closer to those values as $p$ increases.
    \end{itemize}
            This convergence is mostly only seen in the 16-qubit instances because of the available data going to much higher depth $p$.
    \item   $\vec{\gamma}$ starting at $\gamma_1 \approx 0$ (and increasing in absolute value), with $\vec{\beta}$ approaching $\beta_p \approx 0$ (by decreasing in absolute value) resemble an adiabatic schedule, but we do not have an explanation of the close agreement of specific values of $\gamma_p, \beta_1$ across different schedules.
    \item   Another quantity we can examine is the absolute range of the values in $\vec{\beta}, \vec{\gamma}$.
            These quantities correspond to the effective simulation time that was taken at each QAOA step. 
            We can see that $\vec{\beta}$ spans a wider range of values than $\vec{\gamma}$.
    \item   The slope of the QAOA schedules begin at low $p$ depth being large, and then as $p$ grows larger the slope becomes flatter, with each subsequent $p$ taking a smaller timestep. 
    \item   The overall shape of the learned QAOA angles show that for each increment of $p$, the QAOA schedule, while similar, does ``drift'' slightly. Still, the consistent trend means that we could perform regression on these learned QAOA schedules (such as in Ref.~\cite{Farhi_2022}) in an attempt to extrapolate good QAOA schedules up to larger $p$.\\
            The limitation with this approach is the convergence behavior mentioned above, which occurs as the expectation value approaches $1$. But for good extrapolation onto unseen larger problem instances the extrapolated QAOA angles would need to \emph{not converge as quickly} as on the smaller instances. 
            QAOA parameter extrapolation -- beyond parameter transfer -- would be a valuable future research question to probe. 
    \item   Lastly, some of the trained QAOA angles have very different behavior that is not seen in the other problem instances -- namely, instance number 5 in Fig.~\ref{fig:learned_angles_16_qubits} shows highly fluctuating angles for the largest (converged) $p$ that was found. This highly fluctuating instance could be related to the convergence to nearly an approximation ratio of $1$. 
\end{itemize}

Notably, these qualitative trend agreements of the QAOA angles between different optimization problem instances has been observed before in several other studies that have applied the QAOA to other types of optimization problems~\cite{crooks2018performancequantumapproximateoptimization, Farhi_2022, Pagano_2020, Sack_2023, PhysRevResearch.7.023165} -- this suggests that good angles can be extrapolated using computationally simple polynomial fits based on a smaller number of trained $p$ steps.

More generally, these optimized QAOA schedules closely follow recent studies that have drawn connections between highly optimized QAOA trajectories, and continuous-time quantum annealing schedules, which typically result in smooth QAOA angle trajectories~\cite{díezvalle2025universalresourcesqaoaquantum, boulebnane2025equivalencequantumapproximateoptimization, apte2025iterativeinterpolationschedulesquantum, Kovalsky_2023, mbeng2019quantumannealingjourneydigitalization} (this connection can be understood as QAOA being a type of discretized continuous-time adiabatic quantum annealing). These highly optimized QAOA angles, shown in Figs.~\ref{fig:learned_angles_16_qubits} and \ref{fig:learned_angles_27_qubits}, therefore provide more evidence for this fundamental connection between continuous time adiabatic evolution and the discretized QAOA algorithm. This smooth-schedule property could be due to the underlying angle extrapolation technique used in the training, however because the ground-truth of optimal QAOA angles are not known, there is no way to know whether these angles are optimal or not and therefore we consider the \texttt{JuliQAOA} angles to be good \emph{heuristic} QAOA control parameters. Whether the QAOA schedule characteristics we observe are due an artifact of angle extrapolation or not, their performance on large-scale problem instances shows that these types of QAOA schedules work very well, supporting similar findings from prior studies~\cite{díezvalle2025universalresourcesqaoaquantum, boulebnane2025equivalencequantumapproximateoptimization, apte2025iterativeinterpolationschedulesquantum}. 

Fig.~\ref{fig:learned_energy_approximation_ratios} plots the mean noiseless approximation ratios for the set of trained QAOA angles shown in Figs.~\ref{fig:learned_angles_16_qubits} and~\ref{fig:learned_angles_27_qubits}, evaluated on the problem instances on which the QAOA parameters were trained. Fig.~\ref{fig:learned_energy_approximation_ratios} shows the rate of energy improvement as a function of $p$ (which quickly plateaus at large $p$ for the $16$-qubit instances), and moreover shows that the trained QAOA parameters give continual solution improvement as a function of $p$.

The $27$ qubit instances of Figs.~\ref{fig:learned_angles_27_qubits}~and~\ref{fig:learned_energy_approximation_ratios}~(right) were trained only up to $p=7$ because of the high classical computational cost required to train these instances up to very high rounds -- the $16$ qubit instances of Figs.~\ref{fig:learned_angles_16_qubits}~and~\ref{fig:learned_energy_approximation_ratios}~(left) were significantly easier to train because the size of the statevector simulation is significantly smaller.

\begin{figure*}[th!]
    \centering
    \includegraphics[width=0.53\textwidth]{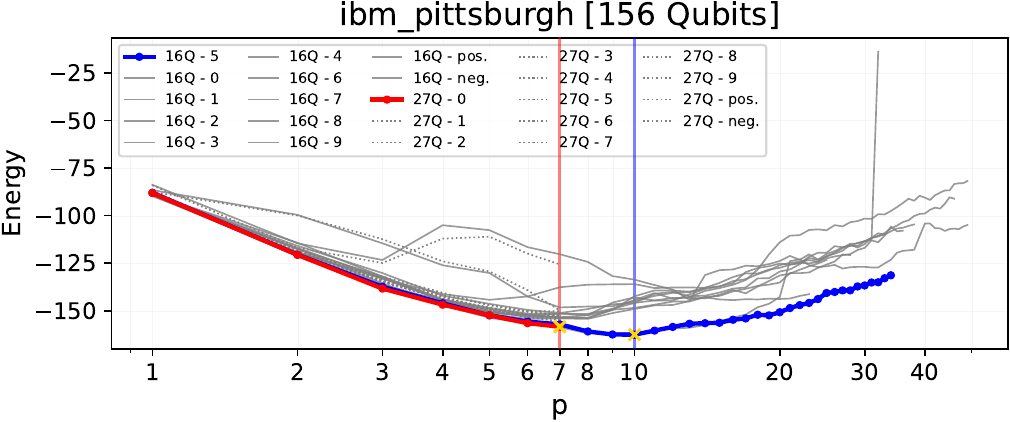}\\
    \includegraphics[width=0.495\textwidth]{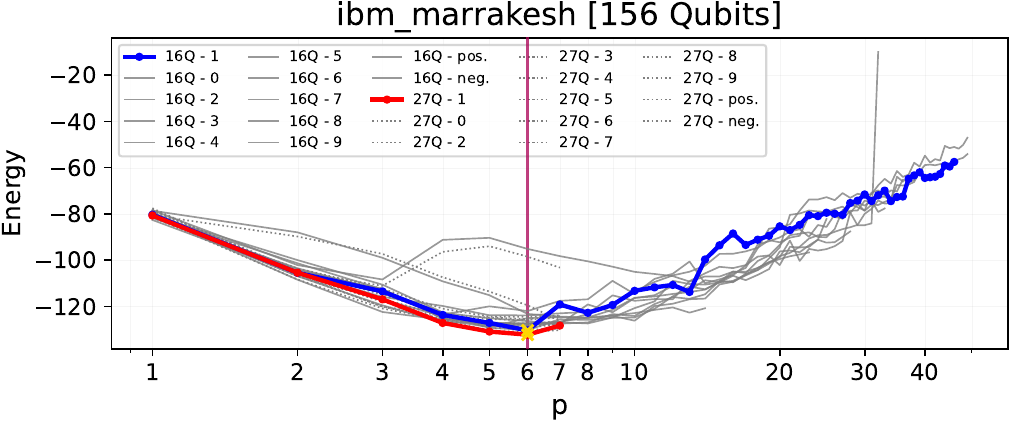}
    \includegraphics[width=0.495\textwidth]{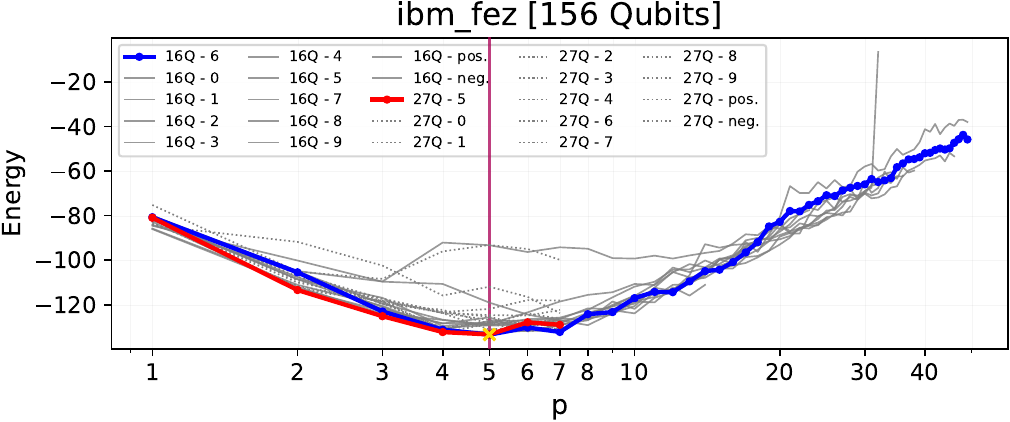}
    \includegraphics[width=0.495\textwidth]{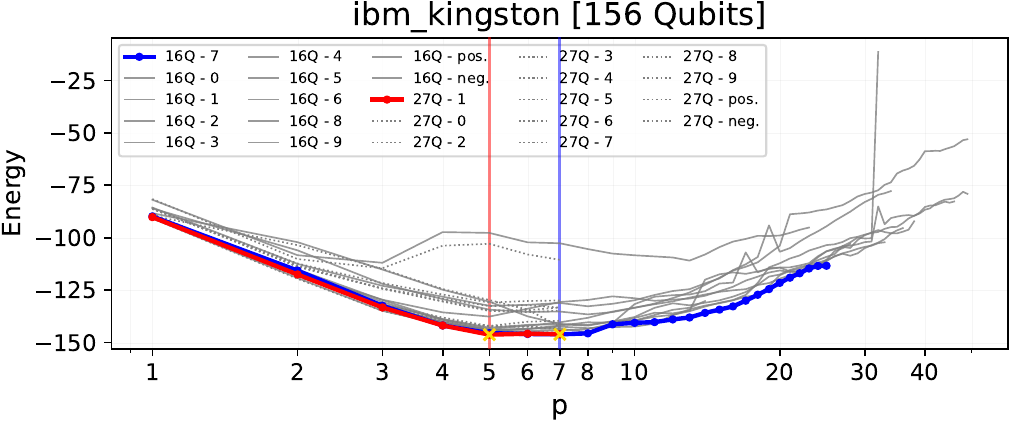}
    \includegraphics[width=0.495\textwidth]{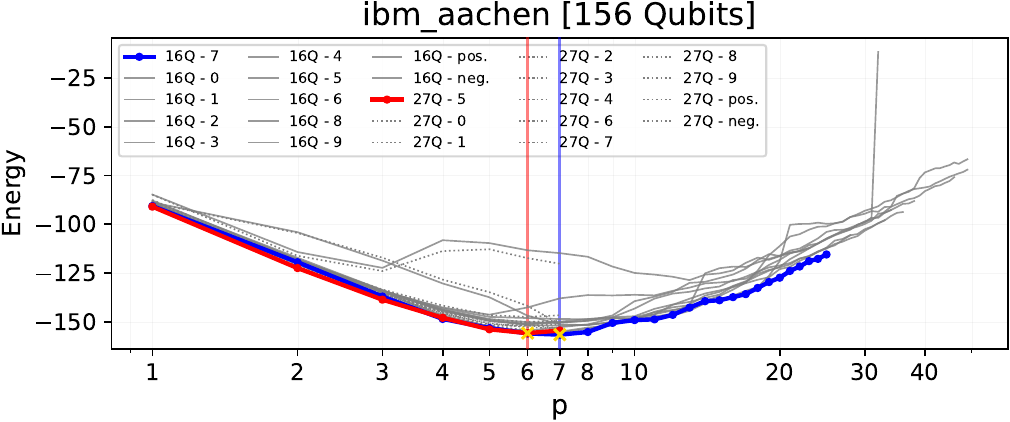}
    \caption{IBM quantum processor QAOA performance results for the $156$ qubit IBM quantum computers in terms of objective function (Ising model) energy on the y-axis as a function of the QAOA depth $p$ on the log scale x-axis. Each processor is sampling a single $156$-spin hardware-defined Ising model instance, whose global minimum energy is $-246$. Each plotted point (defined by a $p$ and a fixed set of transferred QAOA parameters) is showing the mean expectation value estimated using $10^5$ samples, except \texttt{ibm\_fez} where $10^6$ samples are used. The classically trained angles are denoted in the legend by the qubit count of the problem followed by an index integer of the random instance (or all positive or negative coefficients), resulting in a total of $24$ sets of QAOA angles. Within each sub-plot, the red and blue lines each show the two sets of transferred QAOA angles, for $27$ and $16$-qubit source instances, which resulted in the best (lowest) expectation value found on that particular QPU. Although these $5$ sub-plots show sampling results from the same fixed Ising model, the y-axis are not scaled to be the same for each device so as to highlight any small differences between the noisy computations. Note that each figure legend contains multiple lines that are either solid ($16$-qubit training instance) or dashed ($27$-qubit training instance) gray -- these lines are not visually differentiable, and are intended to show the collective sampling rates for the sets of fixed QAOA angles. The minimum (mean) energies, with continuous energy improvement as a function of $p$, are marked with $\times$ symbols and vertical lines at that $p$ index. }
    \label{fig:hardware_results_scaling_p_156_qubits}
\end{figure*}

\subsection{Validating Small-Scale Parameter Transfer with Classical Statevector Simulations}
\label{section:results_small_scale_parameter_transfer_validation}

Fig.~\ref{fig:parameter_transfer_16_qubits_instance_0} shows, as a function of $p$, what the mean energy is when using a fixed set of QAOA angles learned from one of the (random-coefficient) $16$-qubit instances on all of the other $16$ and $27$-qubit problem instances (which, this set of fixed angles was specifically not trained on). These energy curves remarkably show that despite several blips of non-improving energy as a function of $p$, there is a very consistent energy improvement up to very large $p$. Fundamentally, this shows that the limits of parameter transfer, at least for this size of problem instance (in terms of what size of $p$ can still be used effectively) is the saturation of an approximation ratio of $1$ on the training instance -- if you are willing to accept temporary non-ideal blips of non consistent improvement as a function of $p$. We argue that parameter transfer works if the performance (computed expectation value) is monotonically improving as a function of $p$; this is the same requirement we impose on the initial QAOA training. However, if this condition is not met, parameter transferred angles could still work well if the performance on average tends to improve at larger $p$ relative to smaller $p$ despite transitory decreases in performance.

\begin{figure*}[th!]
    \centering
    \includegraphics[width=0.495\textwidth]{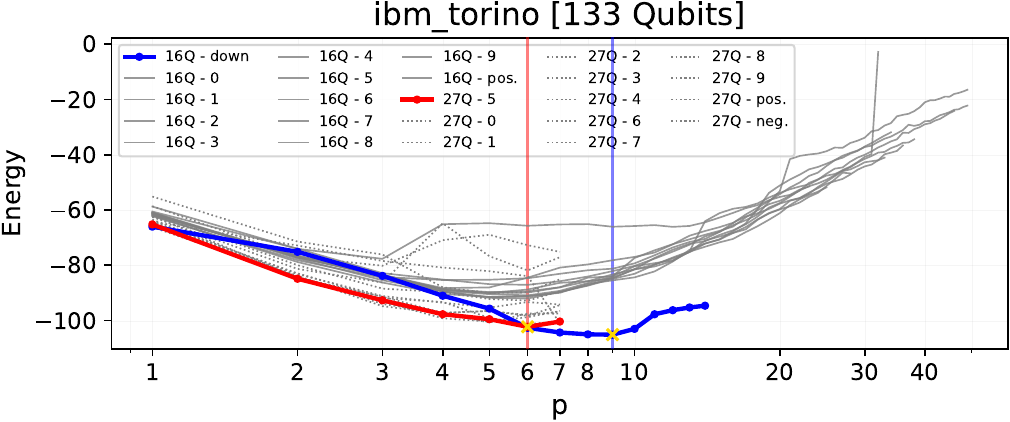}
    \includegraphics[width=0.495\textwidth]{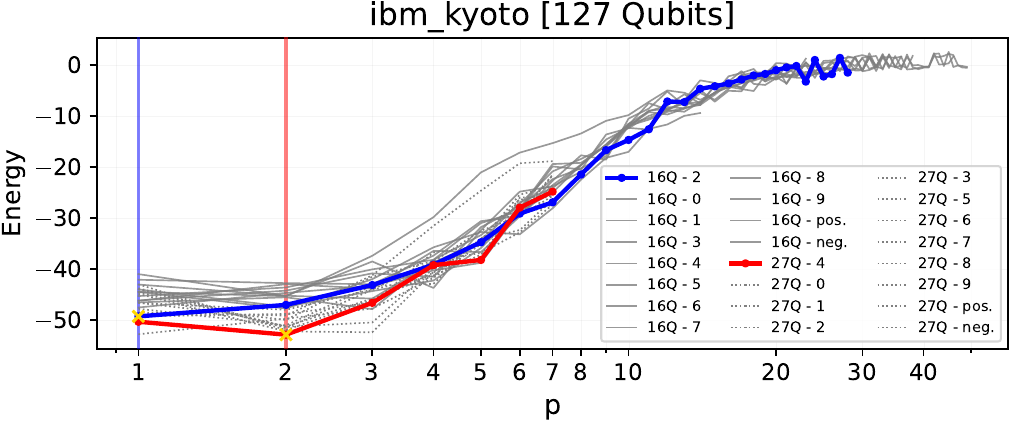}
    \includegraphics[width=0.495\textwidth]{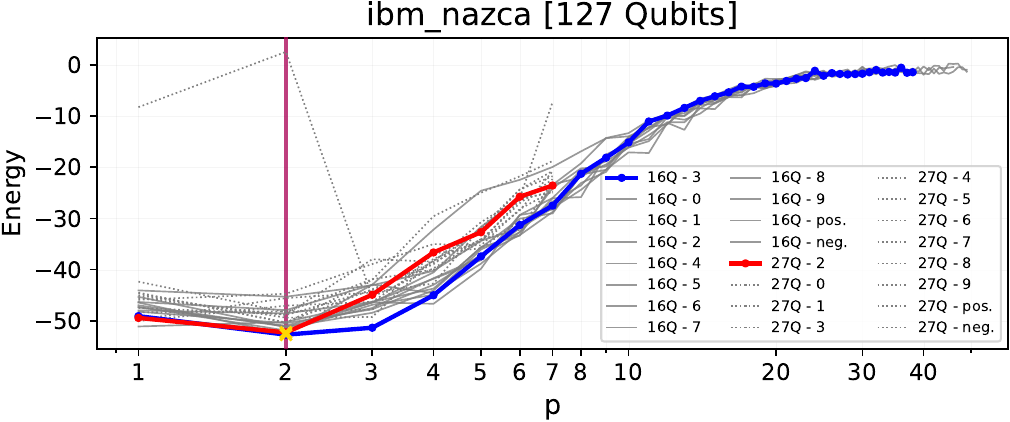}
    \includegraphics[width=0.495\textwidth]{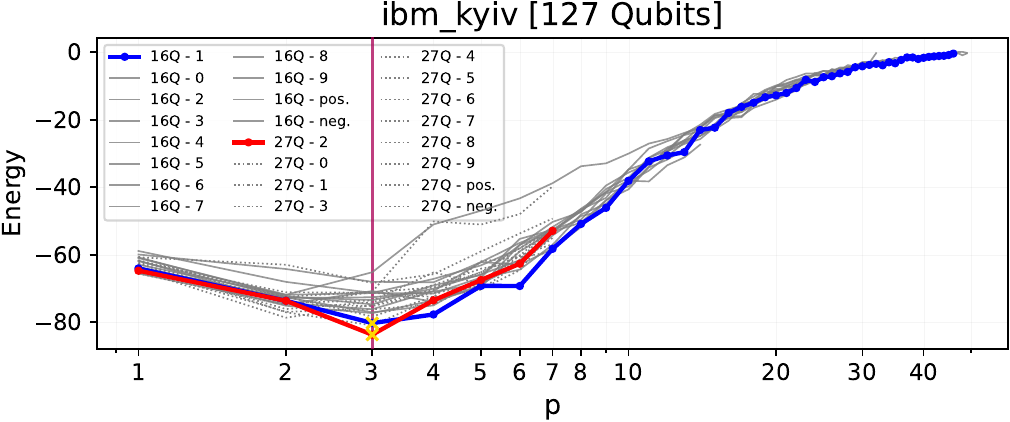}
    \includegraphics[width=0.495\textwidth]{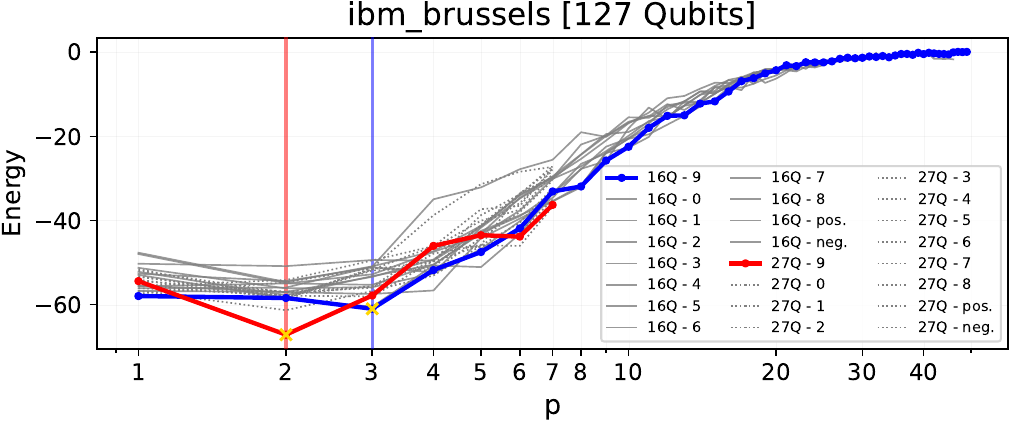}
    \includegraphics[width=0.495\textwidth]{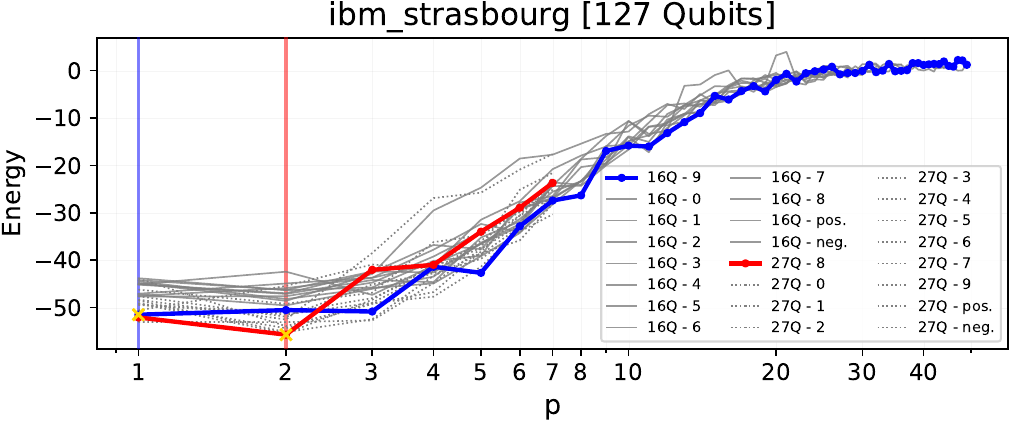}
    \includegraphics[width=0.495\textwidth]{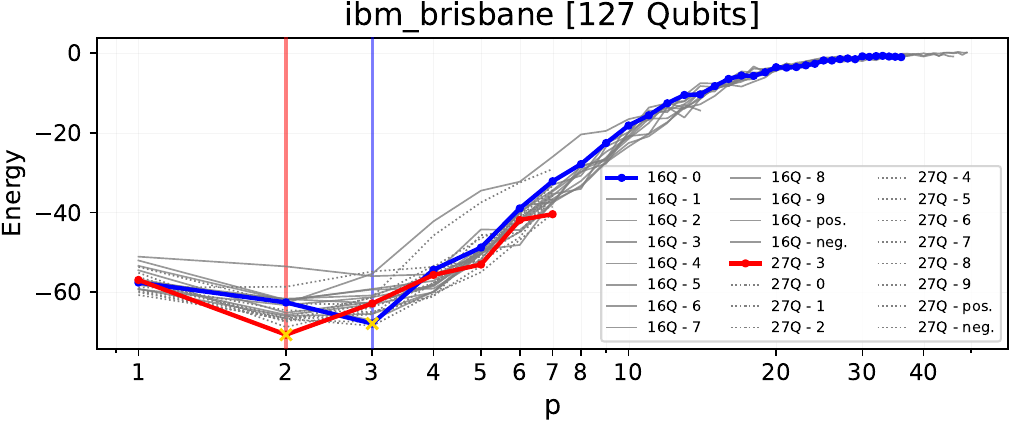}
    \includegraphics[width=0.495\textwidth]{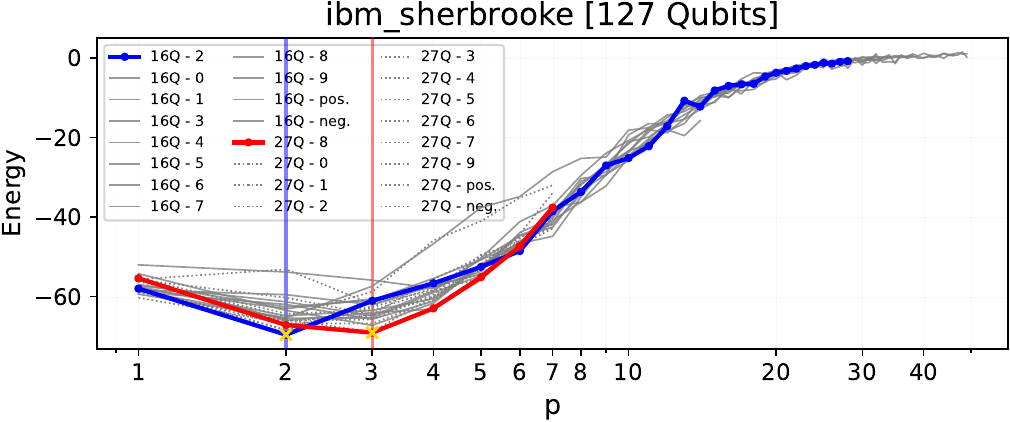}
    \caption{IBM quantum processor QAOA performance results in terms of objective function (Ising model) energy on the y-axis as a function of the QAOA depth $p$ on the log scale x-axis, where the parameters used for each $p$ are the same angles that were learned on the smaller problem instances classically. This is a continuation of Fig.~\ref{fig:hardware_results_scaling_p_156_qubits} using the same format, but for different IBM devices. The classically trained angles are denoted by the qubit count of the problem followed by an index integer of the random instance (or all positive or negative coefficients), resulting in a total of $24$ sets of QAOA angles. The optimization problem is a minimization task, meaning that lower energies are better solutions. Each point (defined by a $p$ and a fixed set of transferred QAOA parameters) is showing the mean energy sampled from $10^6$ measurements, with the exceptions of \texttt{ibm\_torino} where only $10^5$ were measured. The set of angles which resulted in the lowest (best) mean energy for both the $16$-qubit and the $27$-qubit instances are shown as blue and red lines respectively. The minimum (mean) energies, with continuous energy improvement as a function of $p$, are marked with $\times$ symbols and vertical lines at that $p$ index. The optimization problem being solved by each QPU is identical for each QPU size (since the hardware two-qubit entangling gate connectivity structure is exactly the same for each QPU size); the instance is the same for all of the $127$ qubit devices, and there is a single $133$ qubit instance for \texttt{ibm\_torino}. }
    \label{fig:hardware_results_scaling_p}
\end{figure*}

\begin{figure*}[th!]
    \centering
    \includegraphics[width=0.495\textwidth]{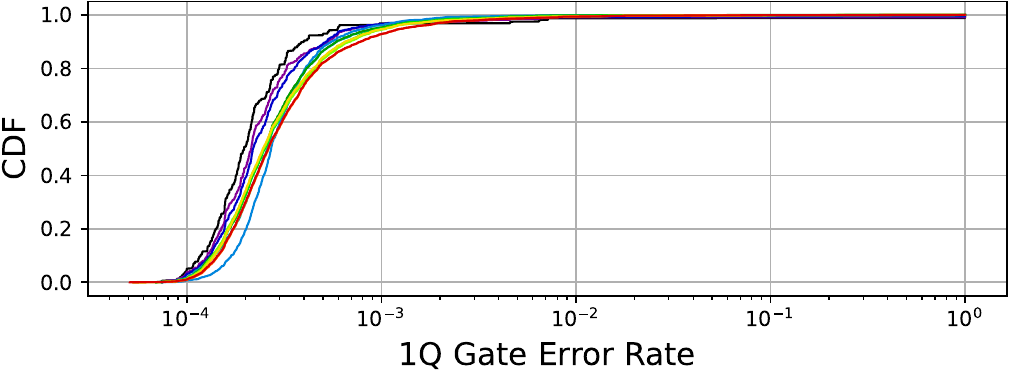}
    \includegraphics[width=0.495\textwidth]{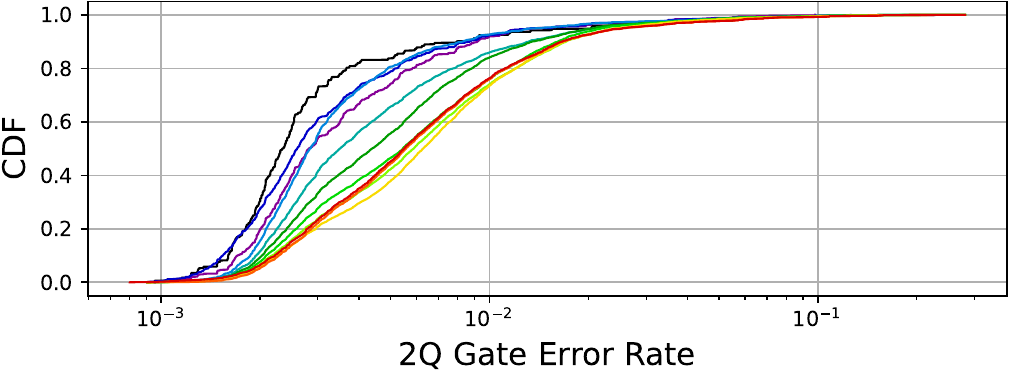}
    \includegraphics[width=0.495\textwidth]{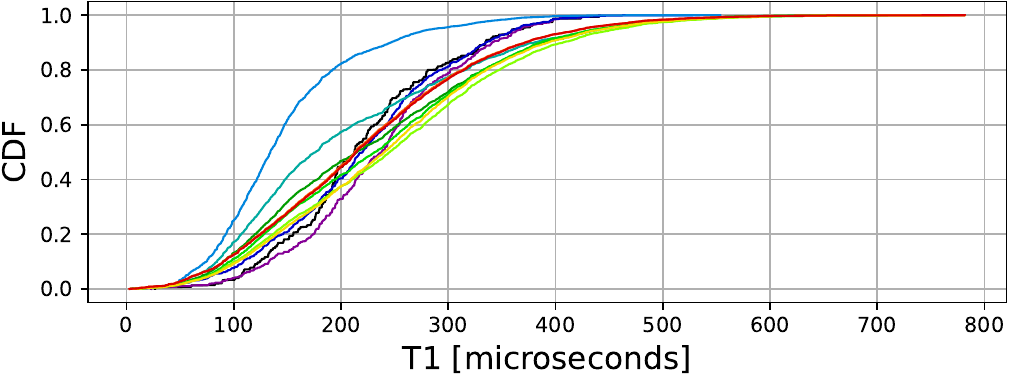}
    \includegraphics[width=0.495\textwidth]{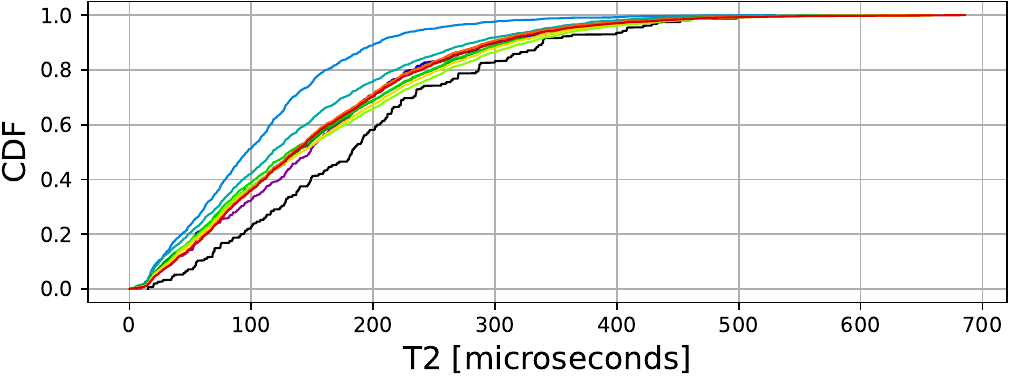}
    \includegraphics[width=0.999\textwidth]{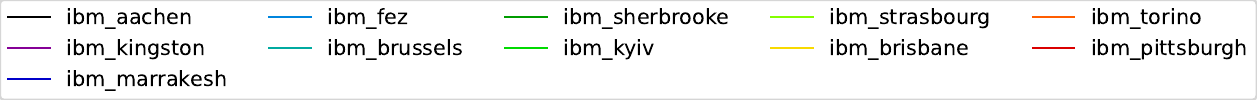}
    \caption{IBM quantum processor hardware error rates, aggregated from the vendor-provided calibration data for all QAOA circuits that were executed and whose results are shown in Fig.~\ref{fig:hardware_results_scaling_p} and Fig.~\ref{fig:hardware_results_scaling_p_156_qubits}. Top row shows the single and two qubit gate randomized benchmarking error rates (log scale x-axis), bottom row shows T1 and T2 coherence times (x-axis is in microseconds). Each plot presents the cumulative distribution function (y-axis) of the combined hardware measured error rates for each device. }
    \label{fig:IBM_hardware_error_rates}
\end{figure*}

Having trained an ensemble of $24$ sets of QAOA ``schedules'', one approach could be to compute average quantities for these learned angles, which could represent a good schedule for an average problem. However, here we utilize all $24$ sets of angles to sample the underlying heavy-hex Ising models in order to obtain a spectrum of QAOA performance -- where some outlier angles perform worse, but potentially others perform better. This is seen very clearly in Figs.~\ref{fig:parameter_transfer_16_qubits_instance_0} where the performance of the fixed angles differs significantly across the different instances. 

Appendix~\ref{section:appendix_all_positive_all_negative_param_transfer} reports small-instance size parameter transfer numerical simulation results in Figs.~\ref{fig:parameter_transfer_16_qubits_instance_negative}, and~\ref{fig:parameter_transfer_16_qubits_instance_positive} where the donor instances are entirely positive and entirely negative coefficient Ising models. The angles from these uniformly weighted models show a similar story where the transferred angles work well, but in some cases and at some $p$ depths the approximation ratio improvement is not strictly monotonic. Moreover, the entirely positive-weighted coefficient Ising model training angles do not perform as well when transferred to other instances compared to the all negative coefficient model.

\subsection{IBM Quantum Computer Sampling Using the Transferred QAOA Angles}
\label{section:results_IBM_Quantum_computer_experiments}

Figs.~\ref{fig:hardware_results_scaling_p_156_qubits} and~\ref{fig:hardware_results_scaling_p} plots the finite-shot estimates of energy expectation value as a function of the (learned) increasing $p$ angles when executed on $127$, $133$, and $156$ qubit IBM QPUs. There are only two differences between the $24$ distinct connected lines in the scaling plots as a function of $p$: i) Time, meaning there is possible time dependent noise drift on the quantum hardware (which does certainly occur on these QPUs~\cite{9259941, dasgupta2021stabilitynoisyquantumcomputing}), and ii) Performance differences of the transferred QAOA angles. The ensemble of learned angles are all unique (although, they are typically quite similar) because they were all trained on distinct random heavy-hex spin glass instances. The underlying combinatorial optimization problem instance that is being optimized is fixed, as is the QPU, as is the structure of the circuit being defined strictly on the heavy-hex graph (only single qubit gate rotations are changed across the different compiled circuits from the different sets of angles). Therefore, the intention is for these plots to show the relative differences between the IBM QPUs across the set of $24$ distinct sets of QAOA parameters (i.e., QAOA schedules) as the QAOA circuit depth increases, up to the largest parameter-transferred $p$-depth. There is not a single source problem instance which resulted in trained QAOA angles that resulted in the best performance on NISQ hardware, instead Figs.~\ref{fig:hardware_results_scaling_p} and~\ref{fig:hardware_results_scaling_p_156_qubits} show that the on-hardware performance is quite similar across the different sets of angles (this is very likely due to the error rates, and likely biased errors, on the QPUs).

Fig.~\ref{fig:hardware_results_scaling_p} shows that there is a clear and consistent loss of signal where the mean expectation value converges to an energy of $0$ for the $127$ qubit processors when $p$ is sufficiently large (around $p=30$). Remarkably however, both the $133$ and $156$ qubit processors (see Fig.~\ref{fig:hardware_results_scaling_p_156_qubits}) show a non-zero signal up to the deepest QAOA circuits tested, which were up to $p=49$. This is an indication that the qubit coherence times are significantly better for the Heron processors compared to the $127$ qubit Eagle processor devices. Although these extremely high depth QAOA circuits at $p=49$ do not show a continual improvement of mean expectation value (because of noise), these are the highest depth QAOA digital circuits executed on noisy quantum processor to date -- and it is notable that there is a nonzero signal generated by some, but not all, of the IBM quantum processors. It is surprising that despite the peak performance of the $156$-qubit IBM Heron r2 and r3 devices being at $p \approx 10$, there is a non-random Hamiltonian expectation found all the way up to $p=49$. For quantum circuit size context, the largest one of these QAOA circuits (both with respect to qubit count and circuit depth) uses $156$ qubits, $17,248$ CZ two-qubit gates, $40,924$ SX single qubit gates, $34,513$ RZ single qubit gates, and $156$ qubit measurement operations at the end of the circuit. This qualitative result is made especially clear by the \texttt{ibm\_pittsburgh} data, where the largest QAOA circuit expectation value at $p=49$ is a lower energy (better) than, or approximately equal to, the $p=1$ results.

\begin{table*}[th!]
\begin{center}
\setlength{\tabcolsep}{8pt}
\begin{tabularx}{\textwidth}{@{}l X c@{\hspace*{8pt}}c r r r@{}}
\toprule
 & & \multicolumn{2}{c}{Energy} & QPU time, & Measurement count, & QAOA volume, \\[-1pt]
\cmidrule(lr){3-4}
QPU size & QPU name & Ising GS. & min sample & \emph{total} [seconds] & \emph{per} parameter~comb. & \emph{best} [$p\cdot N$] \\
\midrule
127 qubits & \texttt{ibm\_sherbrooke} & $-194$ & $-160$ & 192{,}740 & $10^{6}$ & $3 \cdot 127 = 381$ \\
           & \texttt{ibm\_kyiv}       & $-194$ & $-168$ & 415{,}753 & $10^{6}$ & $3 \cdot 127 = 381$ \\
           & \texttt{ibm\_brisbane}   & $-194$ & $-156$ & 181{,}311 & $10^{6}$ & $3 \cdot 127 = 381$ \\
           & \texttt{ibm\_kyoto}      & $-194$ & $-144$ & n/a        & $10^{6}$ & $2 \cdot 127 = 254$ \\
           & \texttt{ibm\_nazca}      & $-194$ & $-142$ & n/a        & $10^{6}$ & $2 \cdot 127 = 254$ \\
           & \texttt{ibm\_brussels}   & $-194$ & $-154$ & 176{,}812 & $10^{6}$ & $3 \cdot 127 = 381$ \\
           & \texttt{ibm\_strasbourg} & $-194$ & $-146$ & 176{,}526 & $10^{6}$ & $2 \cdot 127 = 254$ \\
\midrule
133 qubits & \texttt{ibm\_torino}     & $-196$ & $-178$ & 41{,}331 & $10^{5}$ & $9 \cdot 133 = 1{,}197$ \\
\midrule
156 qubits & \texttt{ibm\_fez}        & $-246$ & $-226$ & 156{,}322 & $10^{6}$ & $5 \cdot 156 = 780$ \\
           & \texttt{ibm\_marrakesh}  & $-246$ & $-218$ & 16{,}723  & $10^{5}$ & $6 \cdot 156 = 936$ \\
           & \texttt{ibm\_kingston}   & $-246$ & $-228$ & 14{,}171  & $10^{5}$ & $7 \cdot 156 = 1{,}092$ \\
           & \texttt{ibm\_aachen}     & $-246$ & $-238$ & 15{,}825  & $10^{5}$ & $7 \cdot 156 = 1{,}092$ \\
           & \texttt{ibm\_pittsburgh}  & $-246$ & $-232$ & 14{,}459  & $10^{5}$ & $10 \cdot 156 = 1{,}560$ \\
\bottomrule
\end{tabularx}
\setlength{\tabcolsep}{6pt}
\caption{IBM NISQ computer QAOA hardware-compatible Ising model sampling summary. The \emph{min sample energy} is the absolute lowest energy found on that QPU, across all QAOA parameters. The corresponding ground state energies are $-194$ (for 127 spins), $-196$ (for 133 spins) and $-246$ (for 156 spins).
The \emph{total QPU time} is the absolute sum of QPU time (server side, not including local compilation and circuit construction) used across all QAOA circuit executions on that QPU (data not available for \texttt{ibm\_kyoto} and \texttt{ibm\_nazca}).
The \emph{measurement count} is the number of shots sampled for each $p$, for each of the $24$ QAOA angle sets. Note that this constitutes different numbers of backend jobs -- for all $127$ qubit processors and \texttt{ibm\_fez}, this is $10$ distinct backend jobs, for \texttt{ibm\_torino} this is $20$ jobs, and for the remaining three $156$-qubit devices, this was a single backend job. 
Finally, the last column denotes the \emph{best QAOA volume} that was found, in terms of the mean expectation value of the best performing QAOA angle with continuous improving performance at higher depth circuits, in the form of $p\cdot N$ where $p$ is the QAOA depth and $N$ is the number of discrete variables.}
\label{table:IBM_NISQ_simulation_summary}
\end{center}
\end{table*}

These QAOA simulations on NISQ computers required a large amount of QPU time, which is summarized in Table~\ref{table:IBM_NISQ_simulation_summary}. Note that these compute times depend heavily on the total number of samples (e.g., circuits executed). These total compute times gives a clear perspective on how efficient the different QPUs are (taking into account the lower shot counts on several of the devices) -- for example that \texttt{ibm\_fez} and \texttt{ibm\_marrakesh} are faster than all of the $127$-qubit devices (i.e., consumes less QPU time per sample).

Notably, none of the NISQ computer QAOA distributions shown in Fig.~\ref{fig:hardware_results_scaling_p} sampled a globally optimal energy spin configuration, despite the usage of a wide range of trained parameter transfer angles, in addition to high measurement counts. Table~\ref{table:IBM_NISQ_simulation_summary} reports the lowest energies sampled on each QPU.

In order to show the component level errors rates on these IBM NISQ processors used in Fig.~\ref{fig:hardware_results_scaling_p_156_qubits} and Fig.~\ref{fig:hardware_results_scaling_p}, Fig.~\ref{fig:IBM_hardware_error_rates} shows aggregate single and two qubit gate error rates, as well as the qubit coherence time metrics of $T1$ and $T2$ times across most of the IBM NISQ computers. $T1$ is the relaxation time and $T2$ is the dephasing time~\cite{clarke2008superconducting} -- these metrics give estimates for how long the qubits remain coherent (the larger these times are, the better the qubits are). These error rates are reported by the device backends, and here we are showing the cumulative distribution function of the distribution of all error rates collected from every job that was submitted to the device (and that successfully executed) whose results are shown in Fig.~\ref{fig:hardware_results_scaling_p} -- this means that the reported distribution technically contains duplicates because the IBM devices are only calibrated once every some set number of hours (dependent on the device), and these metrics are reported by this device calibration on the backend. The goal of showing this distribution in this way is because it shows the \emph{aggregate effective error rates} encountered by the QAOA circuits seen in Fig.~\ref{fig:hardware_results_scaling_p}, and moreover show differences in qubit coherence times on the different hardware platforms. These metrics are measured using randomized benchmarking~\cite{Magesan_2012, Harper_2019} techniques. Note that \texttt{ibm\_nazca} and \texttt{ibm\_kyoto} are not included in this dataset because the hardware level error rates could not be extracted from the Qiskit API post-circuit execution. The gate error rates shown in Fig.~\ref{fig:IBM_hardware_error_rates} are based on recording the error rates of the gateset to which the circuits are compiled to. The error rates of the single qubit gates are recorded based on the gates \texttt{sx} and \texttt{x}, and the error rates of the two qubit gates are recorded based on the gates \texttt{ecr} and \texttt{cz}. Note that the single qubit gate \texttt{rz} is a \emph{virtual} gate which is implemented as an abstract circuit compilation operation~\cite{mckay2017efficient}, which means that the error rate for this gate is always $0$. When retrieving two qubit gate errors rates from the IBM backends, there are many instances of reported error rates of $1.0$. It seems that these reported error rates are indicating either an uncalibrated two qubit gate, or a gate which has been de-activated at the software level. Because these error rates are an erroneous backend related artifact, we have discarded all $1.0$ error rates in the data shown in Fig.~\ref{fig:IBM_hardware_error_rates}.

Fig.~\ref{fig:IBM_hardware_error_rates} shows that the $T1$ and $T2$ times of the newer generations of IBM processors, such as \texttt{ibm\_marrakesh}, are smaller than the previous generation QPUs. However, this does not mean that the newer devices actually have worse performance -- the reason is because the single and two qubit gate times were also reduced, along with the coherence times, on these Heron processors, and the relevant measure therefore becomes circuit depth that can be executed within the coherence times.

One way of quantifying the size of QAOA circuits that have been executed on noisy quantum computers such that there is continual improvement at each step of p is a concept called ``QAOA volume''~\cite{shaydulin2023qaoancdotpgeq200, pelofske2023high} which defines the largest QAOA circuit from $N$ variables and depth $p$ that results in monotonic improvement of solutions up to that depth $p$ (analogous to the quantum volume NISQ computer benchmark~\cite{PhysRevA.100.032328, Baldwin2022reexaminingquantum, Pelofske_2022_QV, Jurcevic_2021}). The QAOA volumes reported in the hardware simulations shown in Figs.~\ref{fig:hardware_results_scaling_p_156_qubits} and~\ref{fig:hardware_results_scaling_p} are summarized in Table~\ref{table:IBM_NISQ_simulation_summary}. These are some of the largest QAOA volumes reported on noisy quantum computers to date, going up to a continual improvement of $p=10$ on a $156$ qubit processor. QAOA volume here shows a benchmark of current IBM NISQ computer capability.

\subsection{Validating Large Scale QAOA Parameter Transfer with PEPS Simulations}
\label{section:results_PEPS_simulations}

\begin{figure*}[th!]
    \centering
    \includegraphics[width=0.325\textwidth]{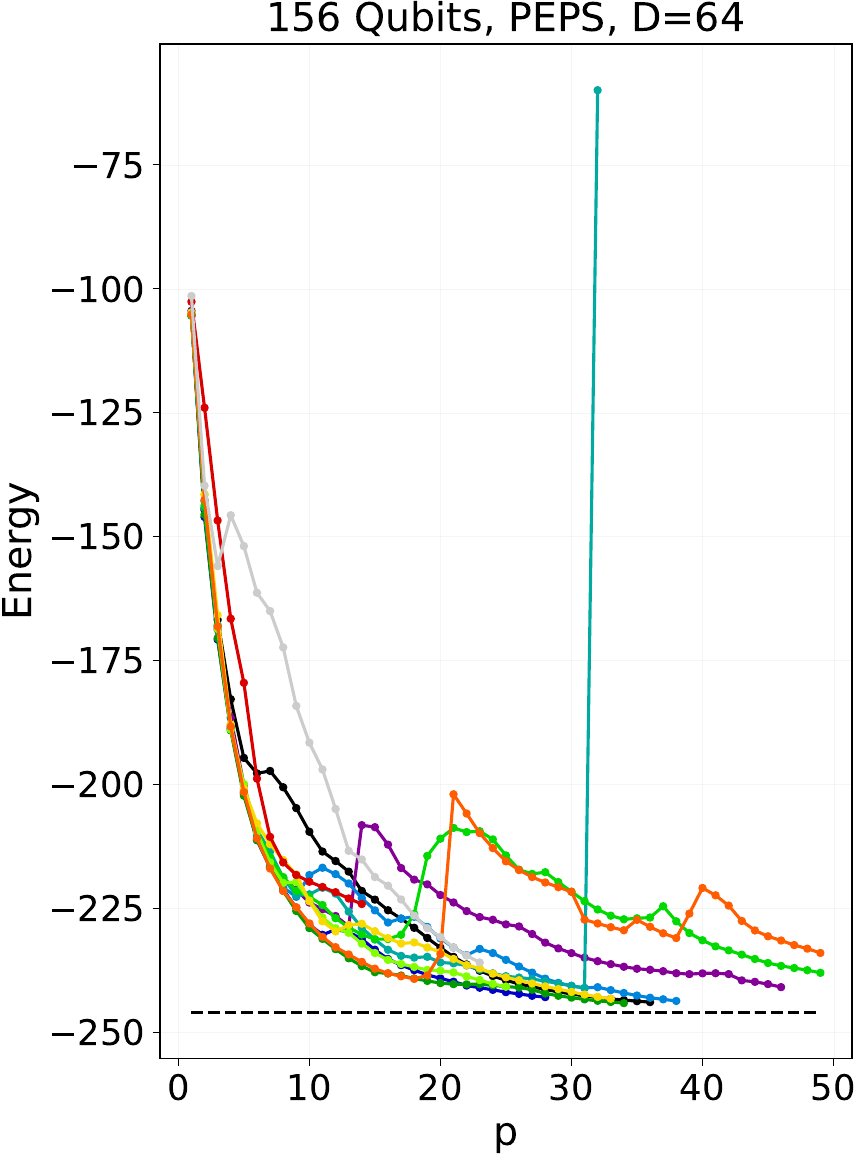}
    \includegraphics[width=0.325\textwidth]{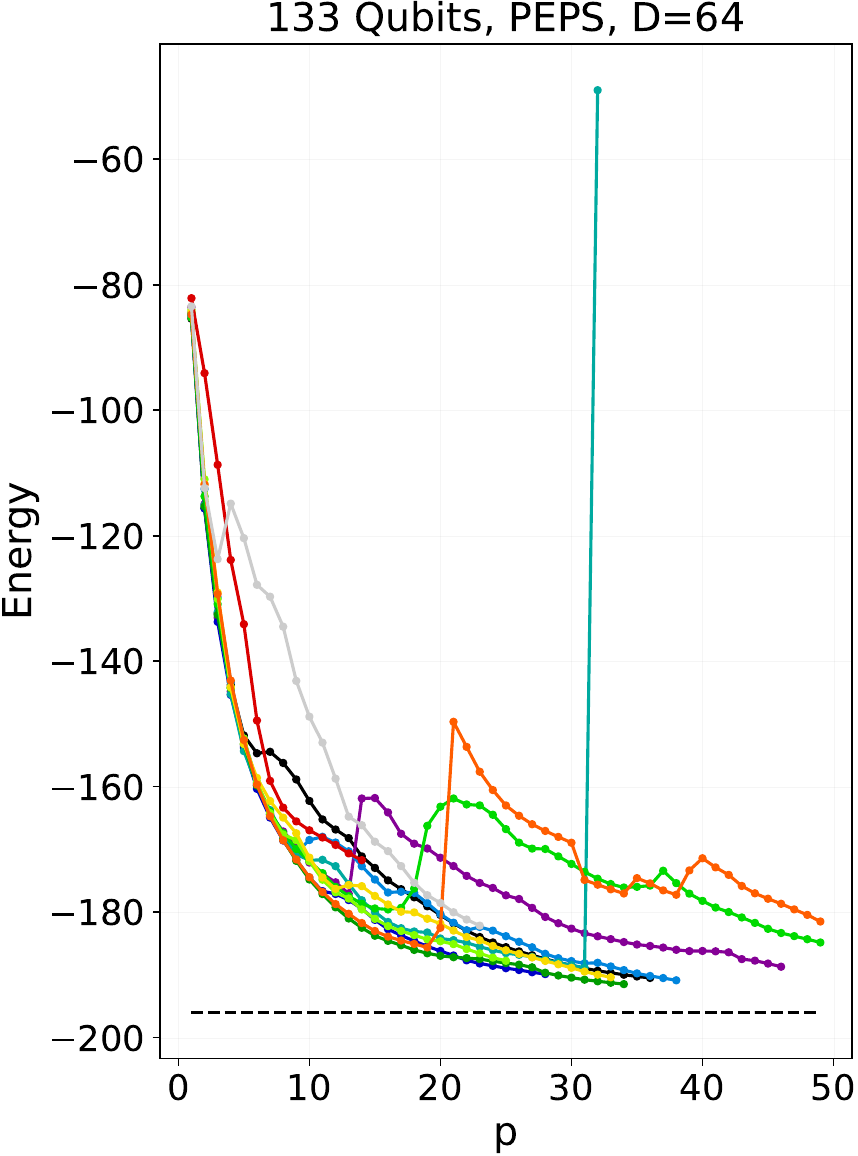}
    \includegraphics[width=0.325\textwidth]{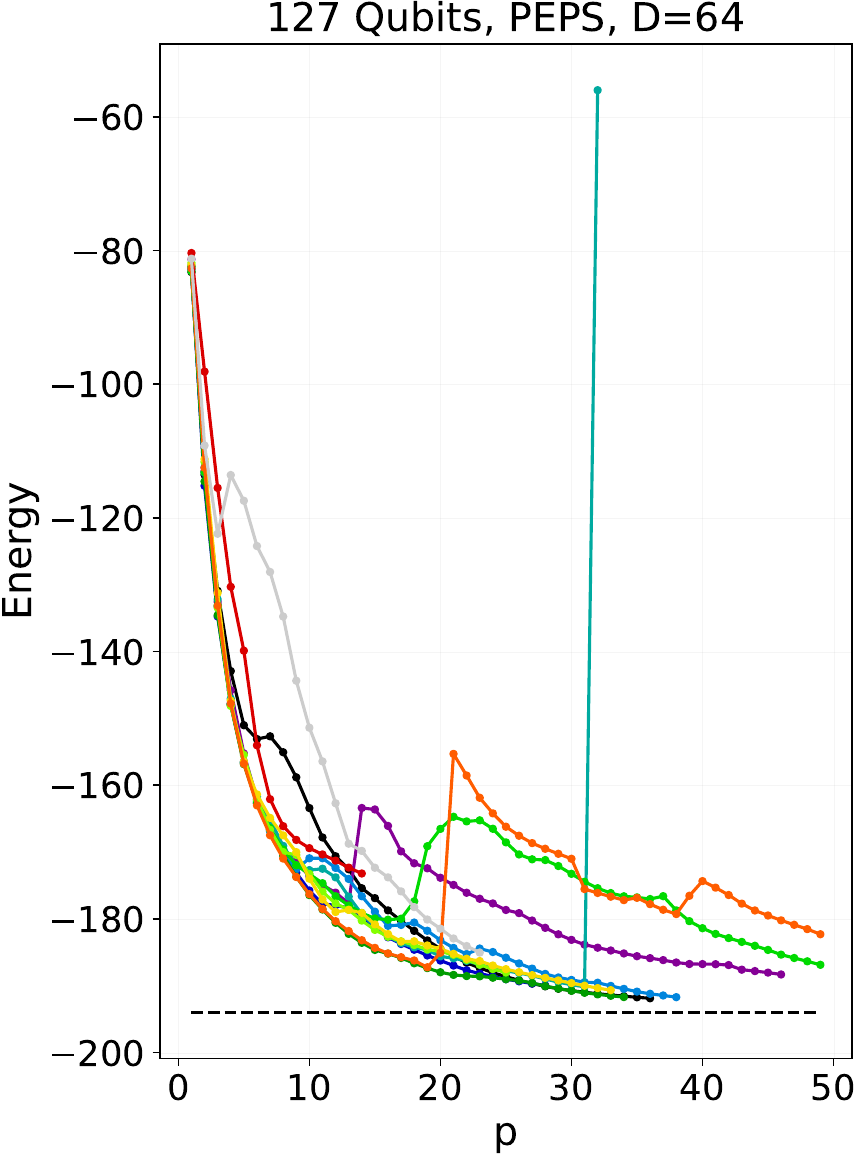}
    \includegraphics[width=1.0\textwidth]{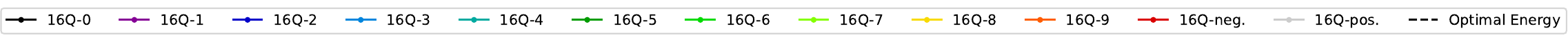}
    \caption{Validating the QAOA angle quality, trained on the set of $12$ $16$-qubit instances, applied to the $156$-qubit (left), $133$-qubit (middle), and $127$-qubit (right) instances, using $D=64$ PEPS simulations with belief propagation for approximate tensor network contraction. The different fixed QAOA angles are denoted by the $12$ different lines shown on the plots, for example \texttt{16Q-pos.}~denotes the set of QAOA angles learned on a $16$ qubit heavy-hex Ising model that has entirely positive coefficients. The true ground-state of each Ising model is denoted by a dashed black line.  }
    \label{fig:PEPS_16Q_angles}
\end{figure*}

Fig.~\ref{fig:PEPS_16Q_angles} plots the Hamiltonian expectation value as a function of $p$, for the complete set of $12$ trained sets of QAOA angles, using PEPS with a bond dimension of $D=64$, for the $127$, $133$, and $156$ qubit Ising model instances. The goal here is to validate whether the parameter transfer continues to work at larger problem sizes. Indeed, we see that for some of the QAOA angles, parameter transfer works remarkably well -- allowing convergence to nearly the true ground-state. However, similarly to the smaller scale statevector simulations of Figs.~\ref{fig:parameter_transfer_16_qubits_instance_0}, \ref{fig:parameter_transfer_16_qubits_instance_negative} and~\ref{fig:parameter_transfer_16_qubits_instance_positive}, we see that for some sets of parameters, there exist non-monotonic fluctuations, dips, and in one case there is a single set of QAOA angles which results in a significantly worse performance (note that this angle is shown explicitly as the highly fluctuating schedule shown in the 5th row from the top of Fig.~\ref{fig:learned_angles_16_qubits}). This result shows a single case where non-smooth QAOA angles can perform well on the small test case problem instance, but then fail to perform well when those parameter are transferred up to larger problem sizes -- we hypothesize that this is likely a rare artifact associated with having converged to close the ground-state energy, although additional investigation is required on this point.

In summary, Fig.~\ref{fig:PEPS_16Q_angles} shows that QAOA parameter transfer can work well across instances which have a parameter-transfer Ising model size difference (between the trained instance and the tested instance) of up to $140$ discrete variables, and in this case allows us to get near to the true ground-state, assuming a noise-less quantum computation (approximated by PEPS with a bond dimension of $D=64$). 

\begin{figure*}[th!]
    \centering
    \includegraphics[width=0.325\textwidth]{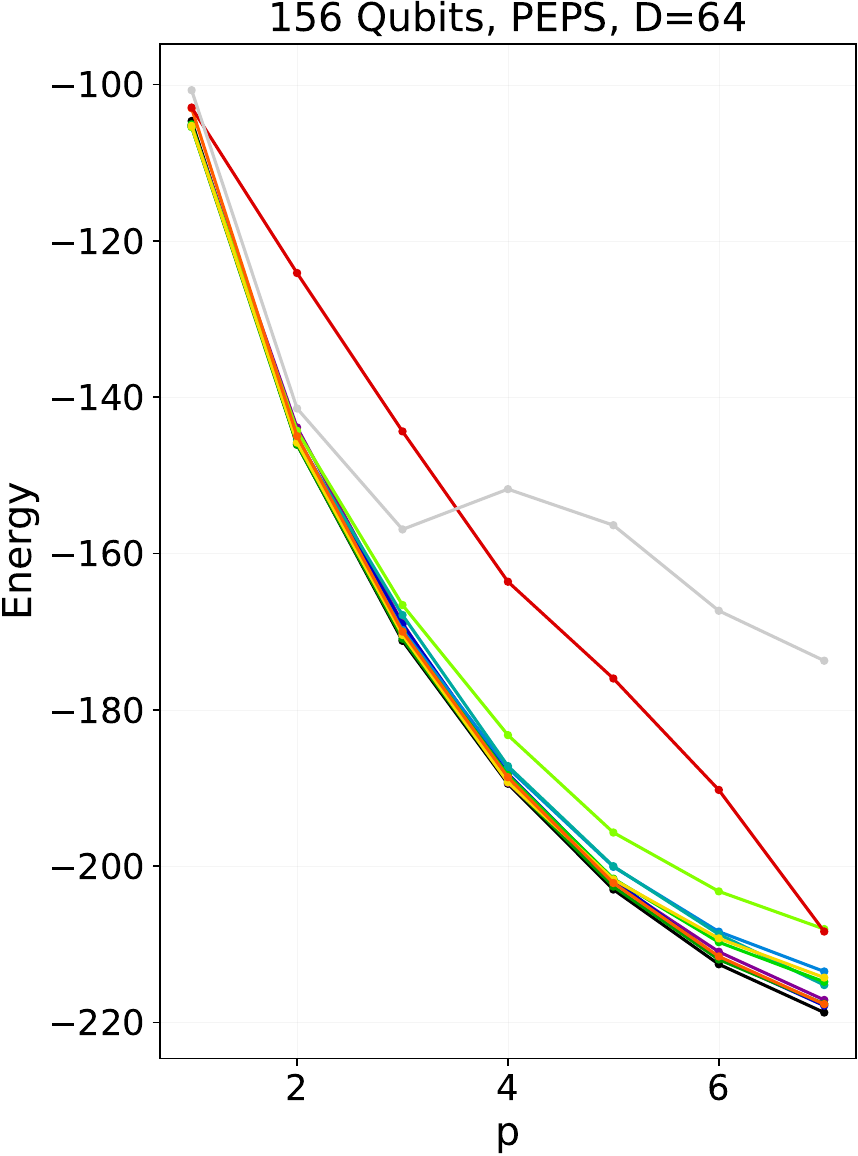}
    \includegraphics[width=0.325\textwidth]{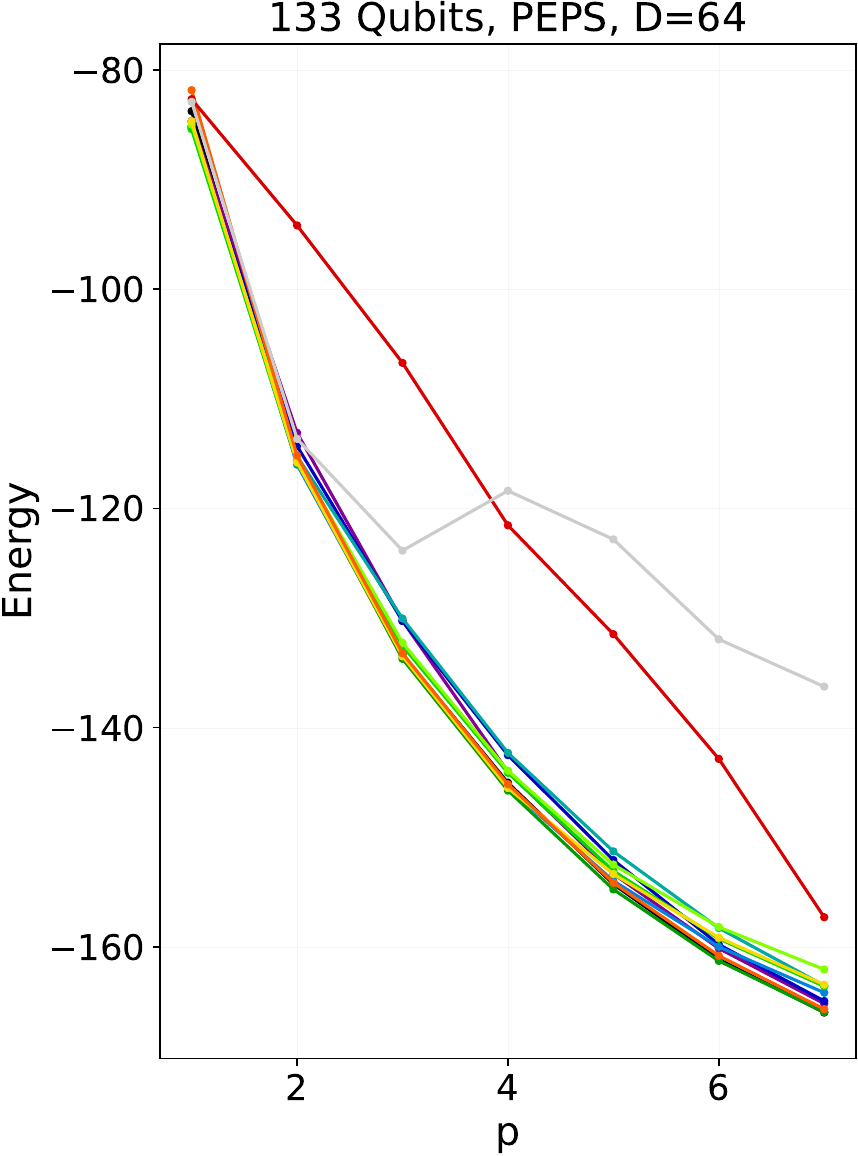}
    \includegraphics[width=0.325\textwidth]{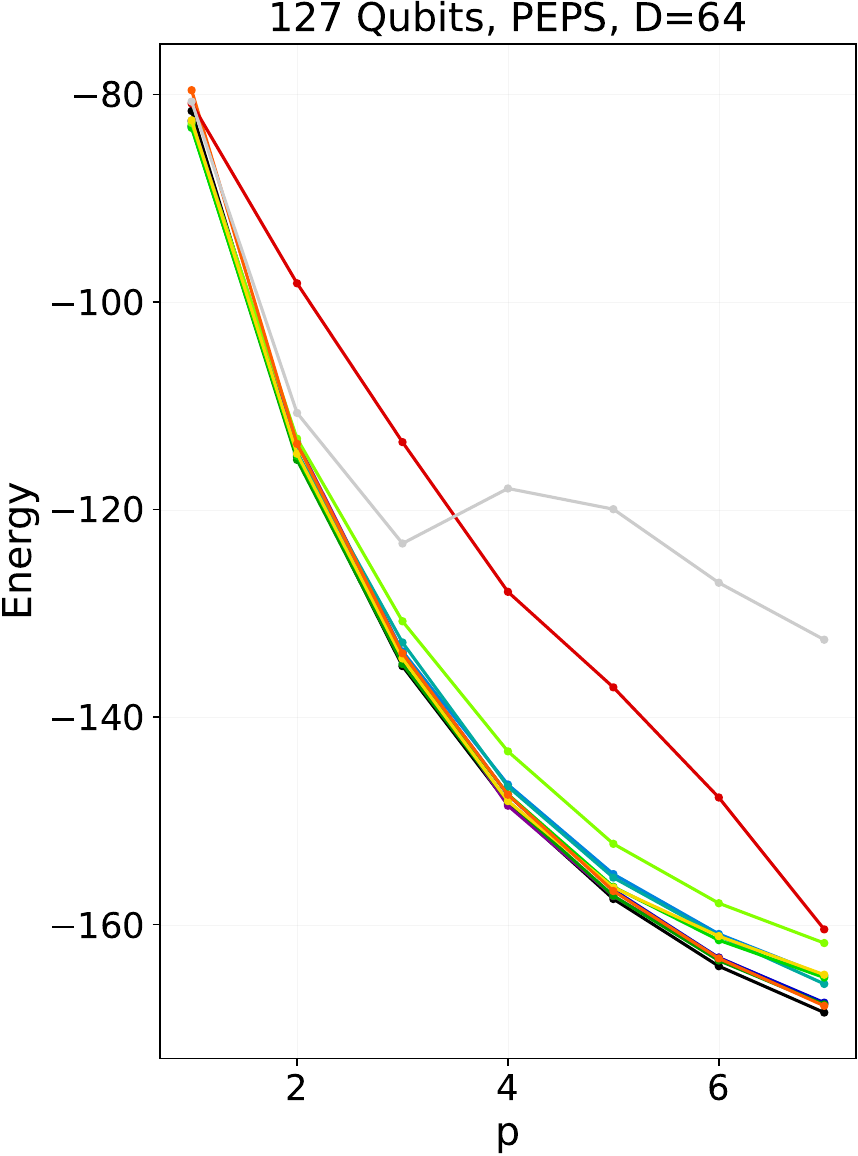}
    \includegraphics[width=1.0\textwidth]{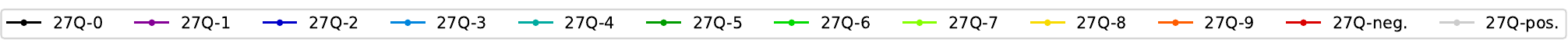}
    \caption{Validating the QAOA angle quality, trained on the set of $12$ distinct $27$-qubit Ising model instances, applied to the $156$-qubit (left), $133$-qubit (middle), and $127$-qubit (right) instances, using $D=64$ PEPS simulations with belief propagation for approximate tensor network contraction. Ground state energy not marked because these expectation values $\leq 7$ do not reach near the ground state, as opposed to Fig.~\ref{fig:PEPS_16Q_angles} where the mean expectation value does approach the true ground state energy. }
    \label{fig:PEPS_27Q_angles}
\end{figure*}

\begin{figure*}[ht!]
    \centering
    \includegraphics[width=0.495\textwidth]{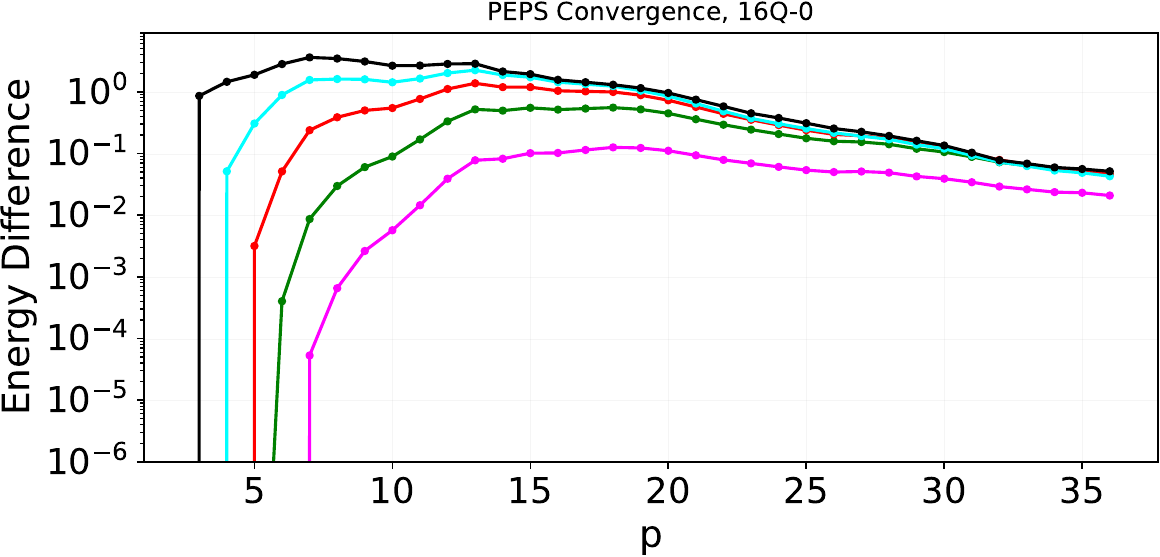}
    \includegraphics[width=0.495\textwidth]{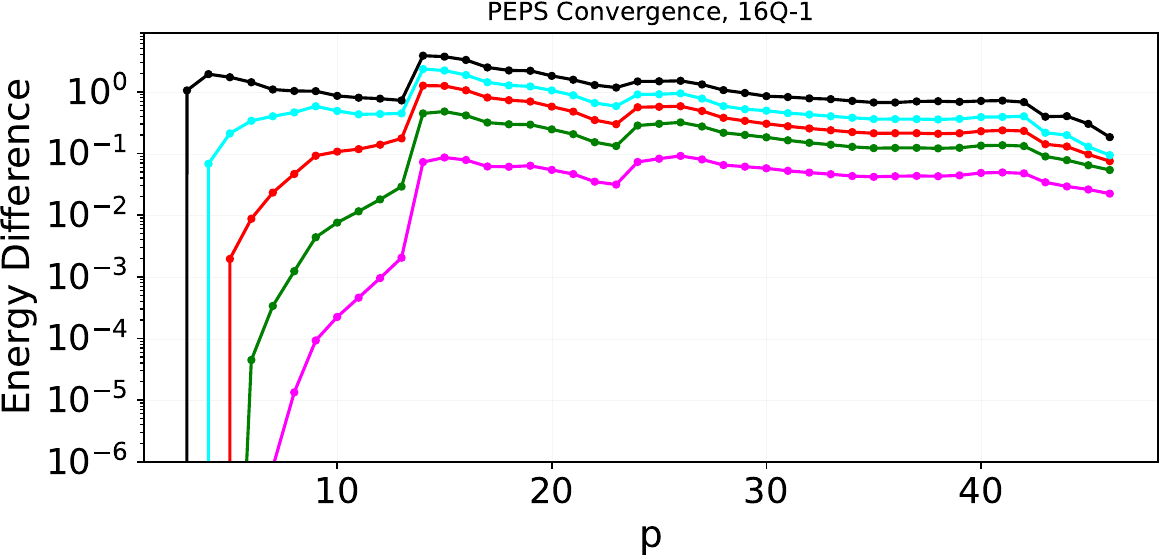}
    \includegraphics[width=0.495\textwidth]{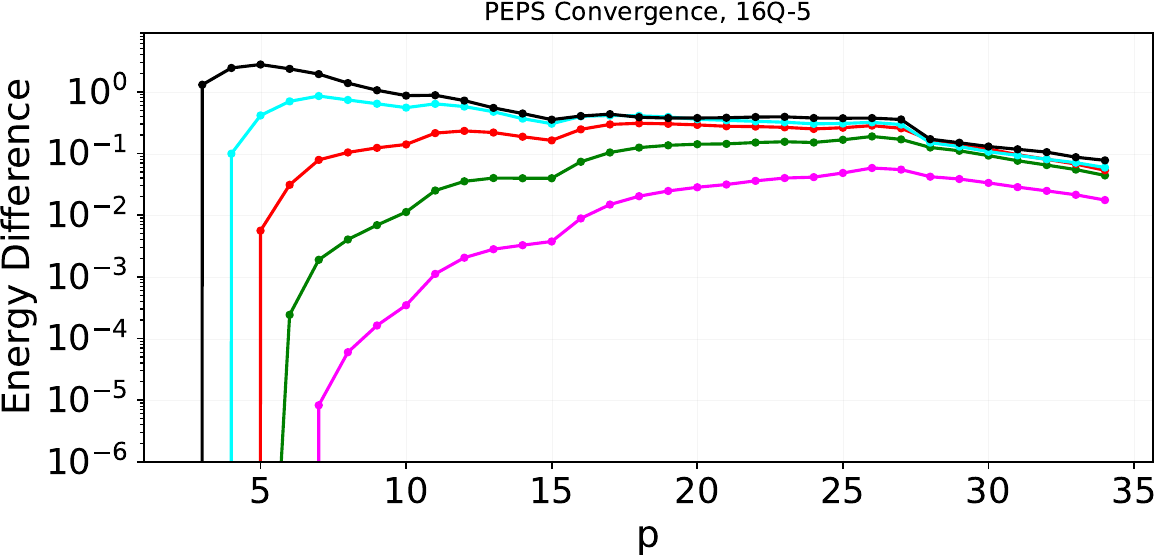}
    \includegraphics[width=0.495\textwidth]{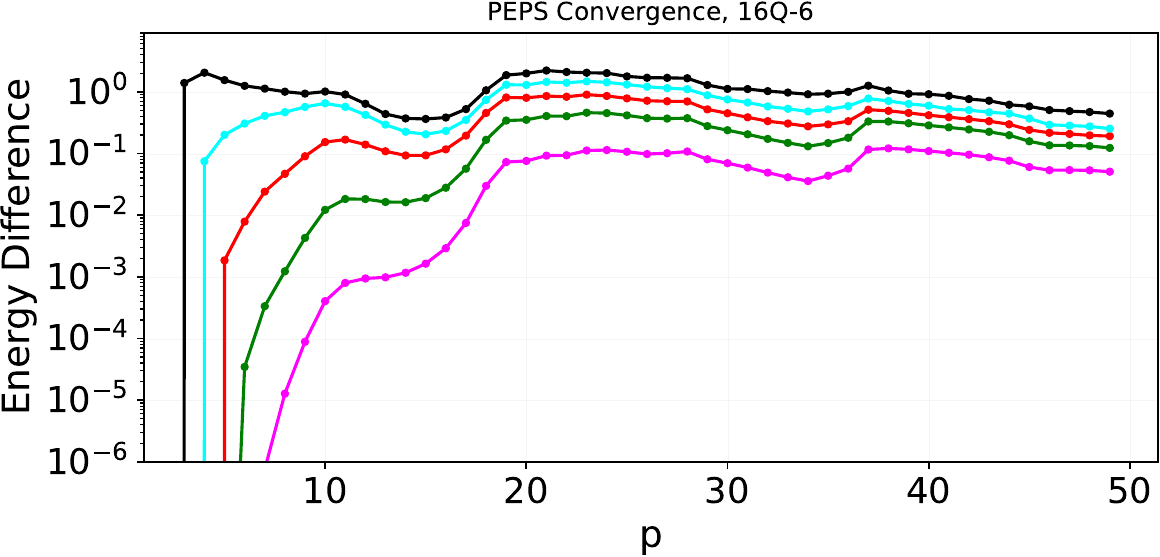}
    \includegraphics[width=0.99\textwidth]{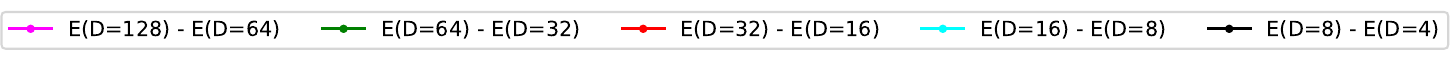}
    \caption{PEPS simulation convergence analysis using four of the sets of parameter transferred QAOA angles (specifically from random-coefficient Ising models) onto the $156$-qubit hardware-compatible Ising model. Tensor network contraction is performed using belief propagation. The log-scale y-axis plots the relative difference between energy estimates obtained with subsequent PEPS bond dimensions. Overlapping lines which are higher on the y-axis indicate lack of convergence as the bond dimension increases, which occurs in particular for larger values of $p$, likely where higher entanglement occurs in the circuit (although, an entanglement measure is not quantified in these simulations). On the other hand, lines which become substantially smaller on the y-axis as the bond dimension increases, for example in cases of small $p$ where the lines drop off the log-scale axis plots entirely, show clear numerical convergence of the Hamiltonian expectation value.  }
    \label{fig:PEPS_convergence}
\end{figure*}

\begin{figure*}[ht!]
    \centering
    \includegraphics[width=0.325\textwidth]{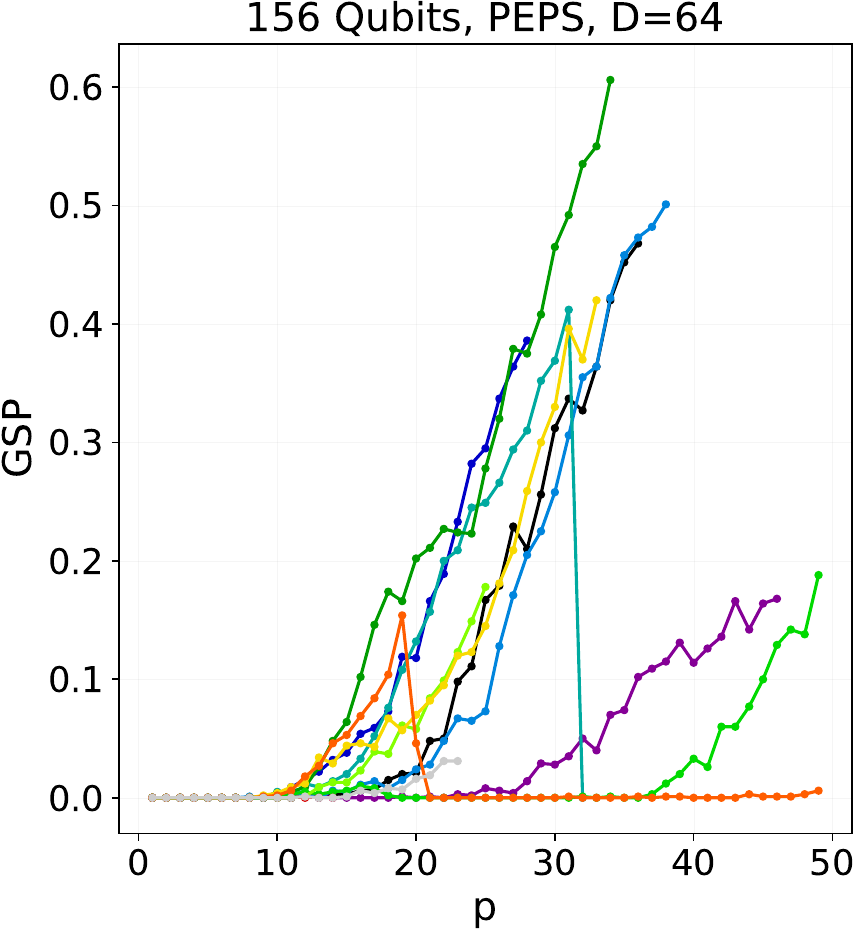}
    \includegraphics[width=0.325\textwidth]{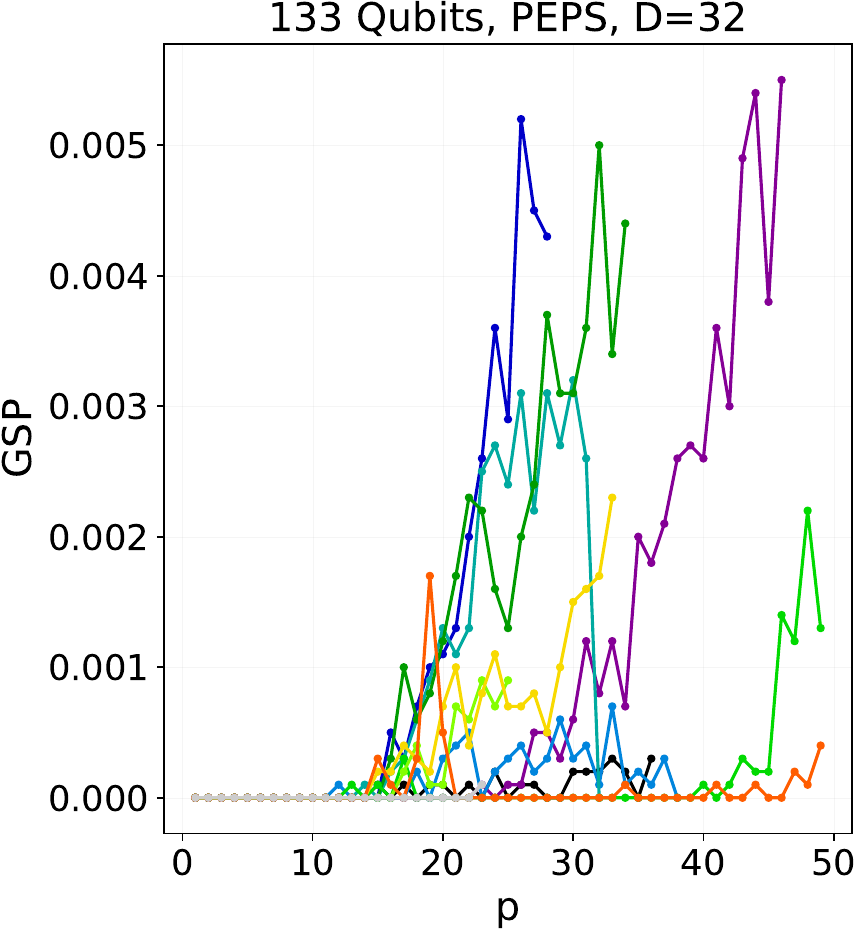}
    \includegraphics[width=0.325\textwidth]{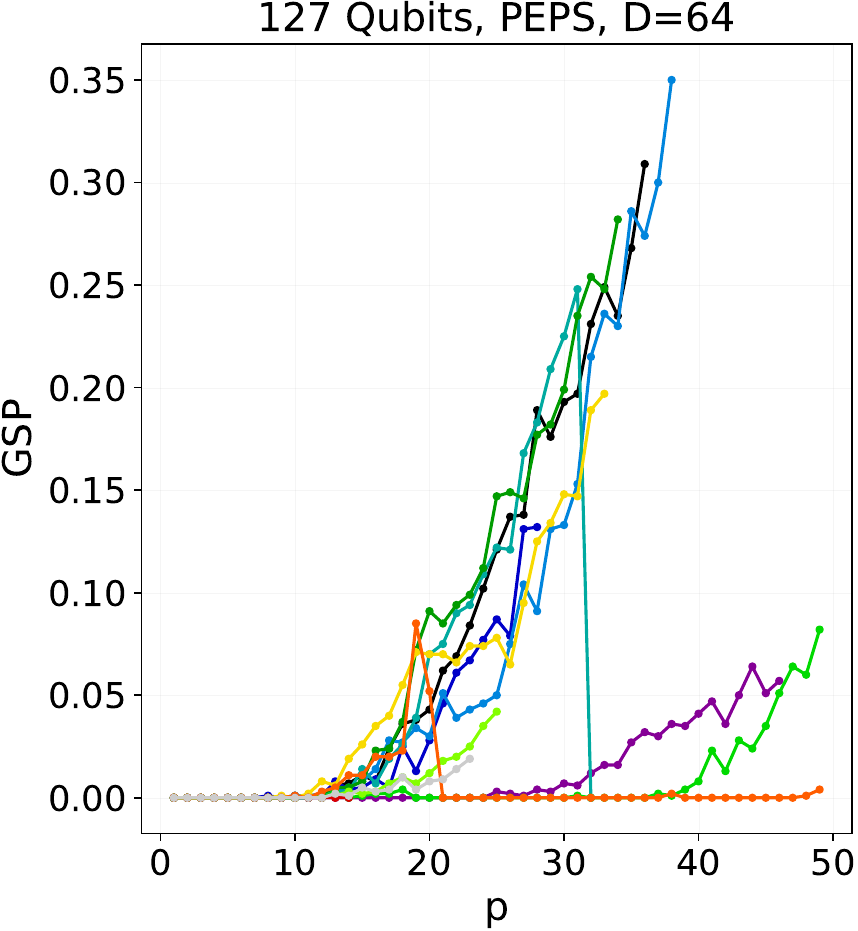}
    \includegraphics[width=1.0\textwidth]{figures/PEPS/27Q_legend.pdf}
    \caption{Ground-state energy QAOA sampling rate (probability of sampling the ground state) from $D=64$ and $D=32$ PEPS simulations, across the ensemble of $16$-qubit trained angles, on the $156$-qubit Ising model (left), $133$-qubit Ising model (middle) and the $127$-qubit Ising model (right). Interestingly, the $133$-qubit instance ground-state is rarely sampled, in comparison to the other two models. The ground-state sampling proportion is estimated using $1000$ samples, per parameter, ($10{,}000$ in the case of the $133$-qubit case, to reduce the effect of shot noise due to the low ground-state energy sampling rate) and each sub-plot y-axis is different for each Ising model. }
    \label{fig:PEPS_sampling_GSP}
\end{figure*}

Fig.~\ref{fig:PEPS_27Q_angles} shows similar trends of monotonic expectation value improvement as a function of $p$ from the $27$-qubit instance trained angles, as we would want from good QAOA evolution -- but with one exception of the entirely positive coefficient Ising model having a non-monotonic dip in performance. Comparing Fig.~\ref{fig:PEPS_16Q_angles} and Fig.~\ref{fig:PEPS_27Q_angles}, there is little difference between the angles that were trained either the slightly smaller ($16$ qubit) or larger ($27$ qubit) instances. 

\begin{figure*}[ht!]
    \centering
    \includegraphics[width=\textwidth]{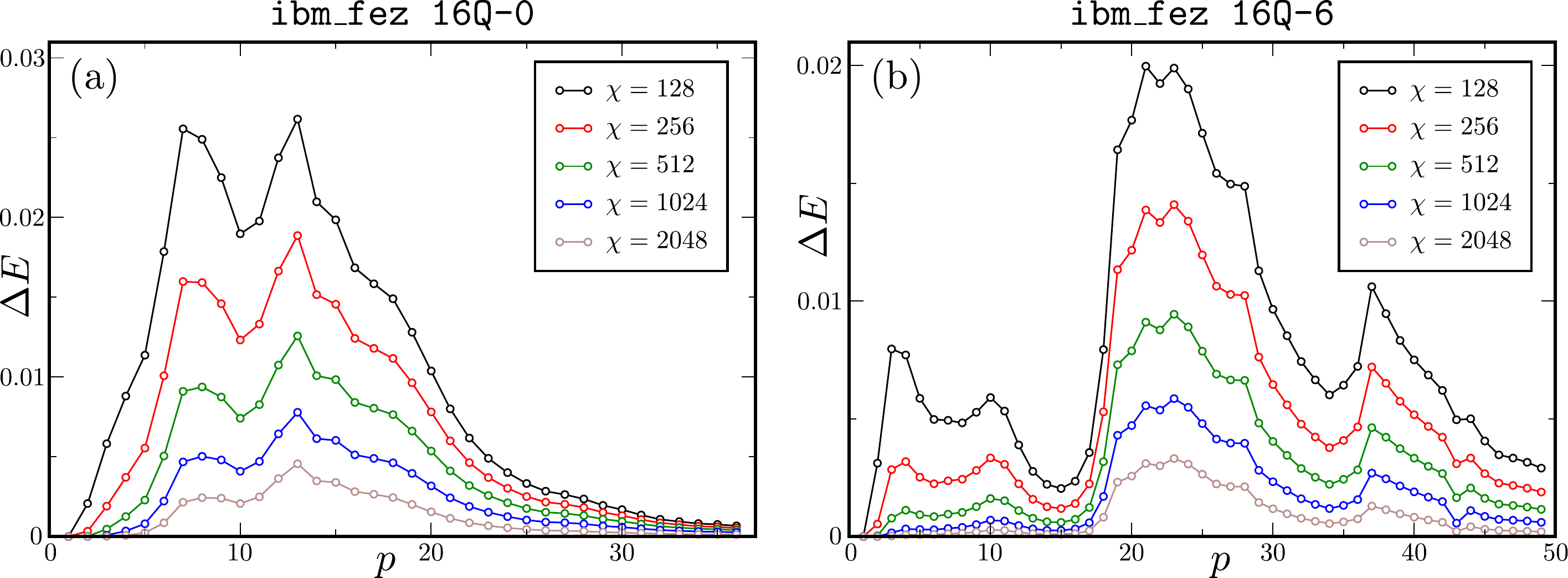}
	\caption{Discrepancy between the most accurate PEPS simulation at $D=128$ and multiple MPS simulations, for the largest QAOA circuit considered in this study of $156$ qubits (defined on the \texttt{ibm\_fez} hardware graph). The error in the energy $\Delta E$, defined in Eq.~\eqref{eq:DeltaE}, as a function of $p$ is shown for several values of MPS bond dimension $\chi$. The error $\Delta E$ decreases rapidly with increasing $\chi$ and drops below $0.005$ for hardest to simulate values of $p$. The results confirm the reliability of PEPS simulation and establish $D=128$ PEPS as our most accurate approximation to $\ket{\vec{\beta},\vec{\gamma}}$ across all values of $p$.}
	\label{fig:MPSvsPEPS}
\end{figure*}

Interestingly, one of the primary conclusions from all of these QAOA parameter transfer calculations is that overall, coefficient choice does not significantly impact the transfer of parameters (the only clear exception to this is that parameters from the positive coefficient case do not monotonically perform well on the other instances). Even very extreme coefficient choices, and random coefficient choices, at sufficiently large $p$ tend to work well. This suggests that these parameters could work well because the Hamiltonian connectivity structure was the same for all instances, however we do not evaluate parameter transfer across different connectivity graphs.

Fig.~\ref{fig:PEPS_convergence} plots the relative expectation value differences from PEPS simulations at different bond dimensions run on the largest Ising model we consider in this study ($156$ spins). This shows that the tensor network simulations clearly converge at small $p$ (typically $p \leq 7$) and typically fail to converge as strongly at larger $p$, with slower convergence occurring in the intermediate values of $p$ typically $p=10$ to $=20$. Moreover, the trend of PEPS convergence is quite distinct for each set of fixed QAOA angles. This empirical convergence evidence reflects what we would expect from highly optimized QAOA evolution~\cite{Dupont_2022}, similarly to quantum annealing, where in the middle of the evolution the system has high entanglement entropy, thus making it harder for the PEPS simulation to converge. 

Fig.~\ref{fig:complete_PEPS_sim_convergence} in Appendix~\ref{section:appendix_more_PEPS_simulations} shows the convergence properties of the remaining set of trained QAOA angles, run on the single $156$-qubit instance. Interestingly, this lack of full convergence at intermediate p values (typically this is approximately between $p=10$ and $p=30$) suggests that those specific QAOA angles result in circuits which are harder to approximate classically using tensor network methods. This convergence trend manifests across most of the angles, with the one exception of the all negative-coefficient angle set shown in Fig.~\ref{fig:complete_PEPS_sim_convergence}. However, the larger $p$ simulations, which use slightly different QAOA angles, are then closer to convergence. This suggests that there do exist particular QAOA parameter combinations for these circuits which are hard to classically simulate, however, unexpectedly the optimized angles, which typically approach the true ground-state energy, are easier to approximate classically. This could be related to the properties of the ground-states themselves which are approached in the high-$p$ regime, which can be expressed as a linear combination of many product states (i.e., due to high degeneracy of the Ising models).

Convergence of the PEPS simulations would resemble a clear progression downward on the y-axis of logarithm expectation value difference in Fig.~\ref{fig:PEPS_convergence}. Eventually at very large bond dimension, convergence results in the difference between bond dimensions (which are increasing as powers of 2) which have zero difference between each other. This effect is really only seen very strongly at low-p depth in Fig.~\ref{fig:PEPS_convergence}, and in some cases we see a strong lack of convergence at sufficiently high p, indicating those expectation values have higher uncertainty, even at $D=128$.

Lastly, a natural question in regards to quantum computation is the effect of sampling -- for example Fig.~\ref{fig:PEPS_16Q_angles} showed only the mean expectation value from the PEPS simulation, namely because all quantum computation requires measuring the state of the qubits at the end of the computation. Fig.~\ref{fig:PEPS_sampling_GSP} shows sampling from the PEPS simulation at $D=64$ using $1000$ samples (for each QAOA angle, at each $p$), where we quantify the sampled ground-state energy proportion. Interestingly, this shows that some of the parameter transfer angles work very well for ground-state sampling, typically well exceeding a proportion of $0.1$, while others do not perform well. This suggests in general for parameter transfer, using an ensemble of distinct QAOA angles is a good technique. However, the $133$-qubit instance ground-state energy is very rarely sampled. The gap between the best-performing angle and the ground-state energy of the $133$-qubit in Fig.~\ref{fig:PEPS_16Q_angles} is larger than the gap for the other two instances, and therefore the low ground-state sampling rate in Fig.~\ref{fig:PEPS_sampling_GSP} is consistent with the noiseless expectation values. This suggests that the low ground state (GS) sampling rate is an instance-dependent result, and the performance of these fixed angles likely has high variability across different random-coefficient problem instances. Fig.~\ref{fig:PEPS_sampling_GSP} reports only the QAOA angles originating from the training on $16$-qubit instances since the $27$-qubit instances were only trained up to $p=7$, and in that QAOA depth regime we verified with PEPS simulations that GS energy sampling rate (using $1000$ samples per $p$) is zero.

\subsection{Validating QAOA PEPS Simulations with MPS Simulations}
\label{section:results_MPS_simulations}

The MPS simulations are performed with $\chi = 2^k$, for $k = 7,\ldots,11$ and shown in Fig.~\ref{fig:MPSvsPEPS} for the $156$ qubit QAOA circuits. Panels (a) and (b) display the error in the energy $\Delta E$ as a function of $p$ for several values of MPS bond dimension $\chi$. Here, the error $\Delta E$ quantifies the difference between our most accurate PEPS energy approximation and the MPS approximation computed with a given $\chi$. It is defined as:
\begin{equation} \label{eq:DeltaE}
\Delta E = \left|\frac{E^\mathrm{PEPS}_{D=128} - E^\mathrm{MPS}_\chi}{E^\mathrm{PEPS}_{D=128}}\right| \ ,
\end{equation}
where $E^\mathrm{MPS}_\chi = \bra{\psi^\mathrm{MPS}_\chi} H_C \ket{\psi^\mathrm{MPS}_\chi}$ and $\ket{\psi^\mathrm{MPS}_\chi}$ is a bond dimension-$\chi$ MPS approximation to $\ket{\vec{\beta},\vec{\gamma}}$ in Eq.~\eqref{eq:QAOAstate}. Similarly, $E^\mathrm{PEPS}_D = \bra{\psi^\mathrm{PEPS}_D} H_C \ket{\psi^\mathrm{PEPS}_D}$, where $\ket{\psi^\mathrm{PEPS}_D}$ is PEPS approximation to $\ket{\vec{\beta},\vec{\gamma}}$ computed with a bond dimension $D$.

Panel (a) and (b) show the results obtained with angles from $16$-qubit random coefficient Ising model index $0$ and $6$. MPS simulations performed with $p \leq 2$ are exact with $\chi=2048$ and indeed $\Delta E = 0$ for those values of $p$. For larger $p$, the simulations are only approximate within the MPS framework and bond dimensions studied. Nevertheless, the errors $\Delta E$ are small and $\chi=2048$ simulations drop below $\Delta E = 0.005$ for all values of $p$ in the instances \texttt{16Q-0} and \texttt{16Q-6}. The plots show systematic and regular decrease in error as a function of $\chi$ for all values of $p$. This is a strong indication of high accuracy and dependability of our largest-$D$ PEPS simulations. MPS simulations closely recover the best PEPS results for some values of $p$ but struggle in the hardest region of intermediate $p$. Overall, MPS results presented here demonstrate that (i) large bond dimension PEPS simulations are very accurate and reliable and (ii) MPS, despite having to handle highly non-local interactions, provides good accuracy across a broad range of $p$.

\section{Discussion and Conclusion}
\label{section:conclusion}

With the overall goal of developing non-variational learning QAOA with the use of fixed parameters that work well on a general set of problem types, using parameter transfer by numerically learning on smaller problem instances is clearly a valuable approach to making QAOA not require significant resources for variational training. We have demonstrated this on a certain class of IBM NISQ computer hardware-compatible Ising models, using a combination of real NISQ computer hardware experiments and tensor network simulations. Notably, in the investigation of QAOA parameter transfer, we have experimentally demonstrated high-depth QAOA volumes of up to $1092$ and $1197$. Moreover, the ensemble of QAOA angles exhibit a general trend of improving expectation value when applied as transferred parameters, evaluated on on up to $156$-qubit instances.

However, parameter transfer does have limitations. Concretely, the following limitations exist for QAOA parameter transfer, primarily for the task of training on smaller problem instances and applying those parameters to larger problems. The most direct limitation is that once the QAOA training saturates the maximum approximation ratio of $1$, meaning no further improvement is possible, then the training can not supply good angles past that saturation point (for larger $p$) to use on any larger problem instances. This saturation is an issue specifically when we generally expect higher QAOA depth simulations to be required, to reach a certain threshold of accuracy (such as a ground-state sampling rate), as problem size increases. This limit necessitates the development of other techniques to be able to efficiently compute good QAOA angles, such as extrapolating to higher QAOA depths based on the already learned angles.

This previous limitation could be mitigated by finding fixed angle rules, or extrapolating QAOA angle trends, that heuristically work well, based on the QAOA training on smaller problem instances. However, such extrapolation would need to be carefully crafted for higher depth QAOA circuits. Further study of using QAOA angle extrapolation based on QAOA parameter transferred angles is warranted, and could result in very scalable high depth variational-learning free QAOA. Here, variational-free refers to the property that the fixed angles work reasonably well across different problem instances, even at high QAOA depth and on larger Ising model sizes -- meaning, one does not need to re-optimize the QAOA parameters for every problem instance. 

The transferred QAOA parameters do not always work well, in the sense they do not necessarily result in continual solution quality improvement as a function of $p$. Despite this, in the QAOA parameters we evaluated numerically there is a \emph{general} solution quality improvement when using all transferred angles, despite local blips of non-continual improvement. Additionally, QAOA parameter transfer does not always work well because of the variability that can result from training on specific problem instances -- we observe a relatively large distribution of QAOA angle performances across the different trained instances. This then suggests that parameter transfer can be applied in two distinct ways; either computing average angles from an ensemble of trained instances, or using an ensemble of trained angles.

\section*{Acknowledgments}
\label{sec:acknowledgments}

We thank Stefan Woerner for constructive discussions. This work was supported by the U.S. Department of Energy through the Los Alamos National Laboratory. Los Alamos National Laboratory is operated by Triad National Security, LLC, for the National Nuclear Security Administration of U.S. Department of Energy (Contract No. 89233218CNA000001). Research presented in this article was supported by the NNSA's Advanced Simulation and Computing Beyond Moore's Law Program at Los Alamos National Laboratory. This research used resources provided by the Darwin testbed at Los Alamos National Laboratory (LANL) which is funded by the Computational Systems and Software Environments subprogram of LANL's Advanced Simulation and Computing program (NNSA/DOE). This research used resources provided by the Los Alamos National Laboratory Institutional Computing Program, which is supported by the U.S. Department of Energy National Nuclear Security Administration under Contract No.~89233218CNA000001. 
We acknowledge the support of the National Science Center (NCN),
Poland, under projects No.~2020/38/E/ST3/00150 (M.M.R.) and No.~2022/47/D/ST2/03393 (P.C.). 
The research was also supported by a grant from the Priority Research Area DigiWorld under the Strategic Programme Excellence Initiative at Jagiellonian University (M.M.R., P.C.). 
We acknowledge the use of IBM Quantum services for this work. The views expressed are those of the authors, and do not reflect the official policy or position of IBM or the IBM Quantum team. \\ LANL report LA-UR-25-28746

\section*{Data and Code Availability}
Code and data publicly available at~\cite{marek_github, pelofske_2025_17110170}

\appendix

\section{Detailed QAOA Circuit Description}
\label{section:appendix_QAOA_circuit}

\begin{figure*}[ht!]
    \centering
    \includegraphics[width=0.93\linewidth]{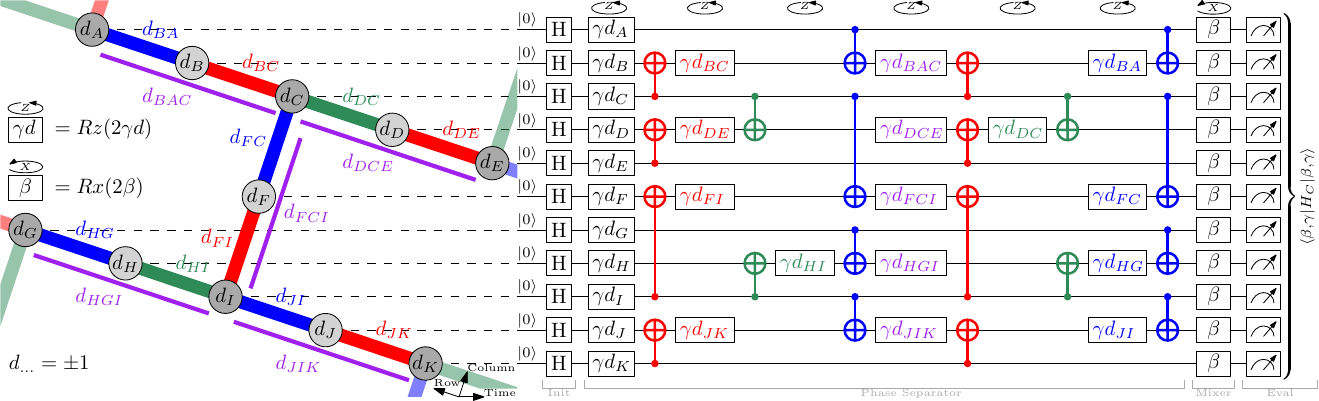}
    \caption{Heavy-hex defined Ising model with geometrically local cubic terms (left), corresponding $p=1$ QAOA circuit description (right). Geometrically local cubic terms are denoted by the short purple lines that are adjacent to lines of three qubits. Red, blue, and green edge coloring denotes an example unique 3-edge-coloring of the lattice. Light and dark grey node coloring denotes a heavy-hex lattice bipartition. The coefficient weights of the local fields and of the quadratic terms are denoted as $d_{x}$. This QAOA circuit illustration is from Refs.~\cite{pelofske2023short, pelofske2023qavsqaoa}. }
    \label{fig:QAOA_circuit}
\end{figure*}

\begin{figure*}[ht!]
    \centering
    \includegraphics[width=0.495\textwidth]{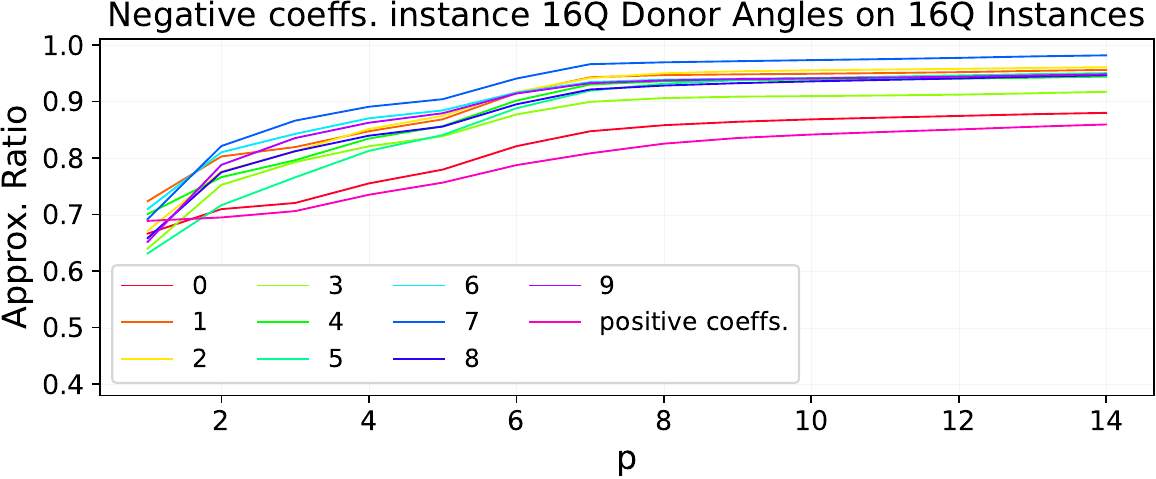}
    \includegraphics[width=0.495\textwidth]{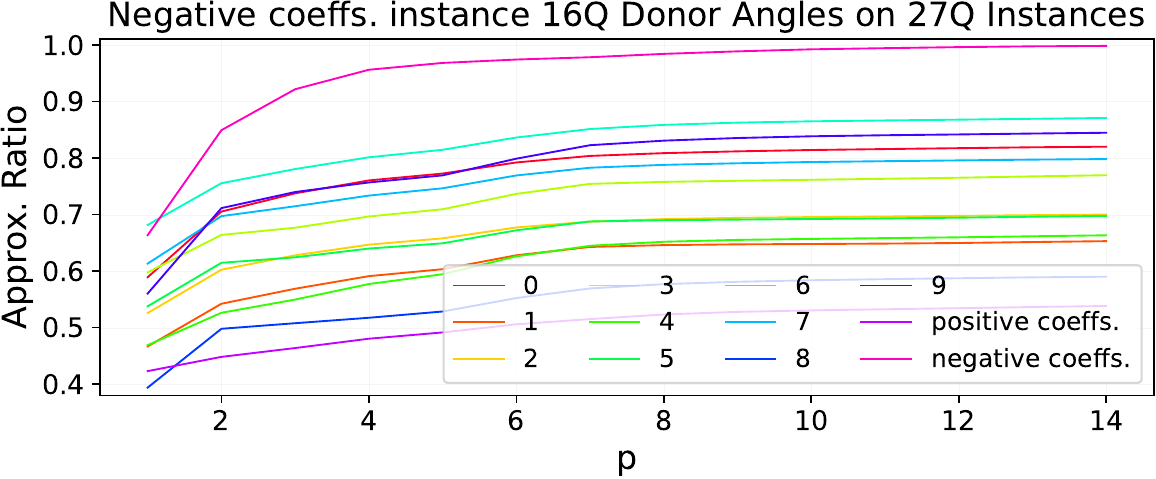}
    \caption{Numerical simulation of parameter transfer using the donor source graph of an entirely negative coefficient Ising model, applied to the other $11$ distinct $16-$qubit problem instances (left) and onto the $12$ separate $27-$qubit problem instances (right). Notably, this set of trained QAOA angles performed well in the sense that the angles result in continuous improvement as a function of $p$. However, the trained angles only go up to $p=14$ and on the $27$-qubit unseen instances none of the trained QAOA angles passed an approximation ratio of $0.9$.  }
    \label{fig:parameter_transfer_16_qubits_instance_negative}
\end{figure*}

\begin{figure*}[ht!]
    \centering
    \includegraphics[width=0.495\textwidth]{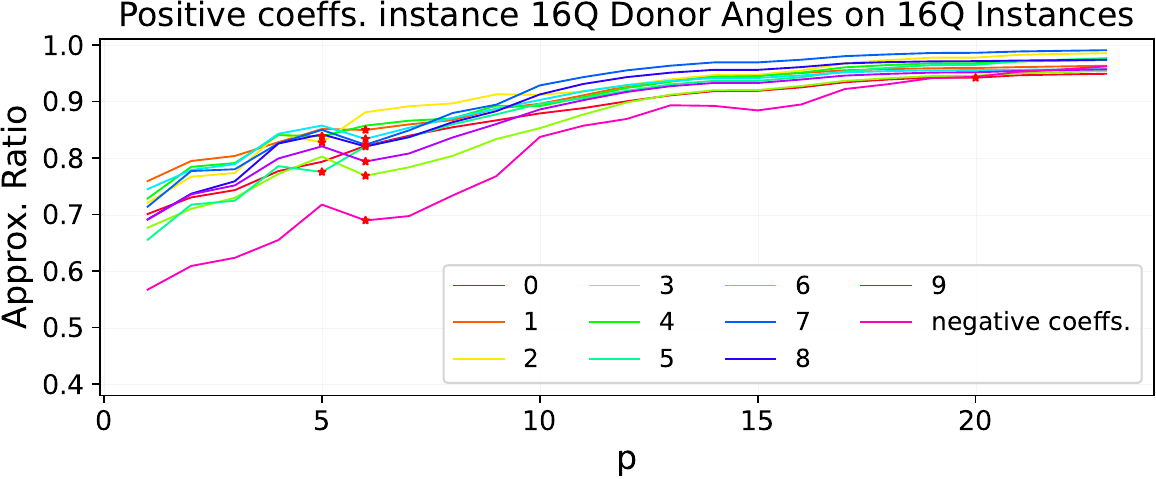}
    \includegraphics[width=0.495\textwidth]{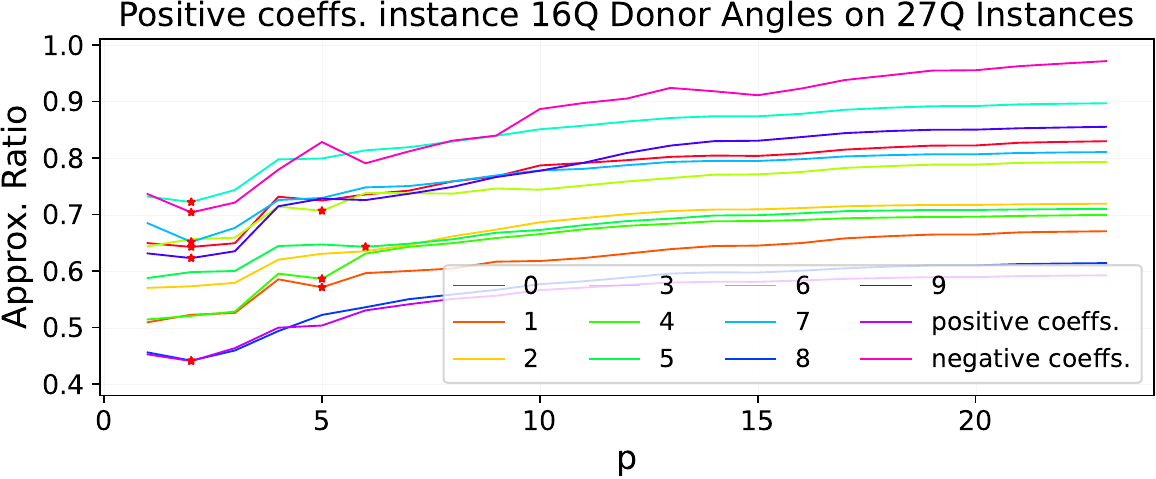}
    \caption{Numerical simulation of QAOA parameter transfer using the donor source graph of an entirely positive coefficient model, applied to the other $11$ distinct $16-$qubit problem instances (left) and onto the $12$ $27-$qubit problem instances (right). The first $p$ step where the parameter angles have a decrease in approximation ratio compared to the previous $p$ steps are marked with a red asterisk (if applicable). The red asterisks denote sub-optimal convergence behavior of the QAOA parameters (ideally the QAOA performance should consistently increase as a function of $p$).  }
    \label{fig:parameter_transfer_16_qubits_instance_positive}
\end{figure*}

\begin{figure*}[!ht]
    \centering
    \includegraphics[width=0.98\textwidth]{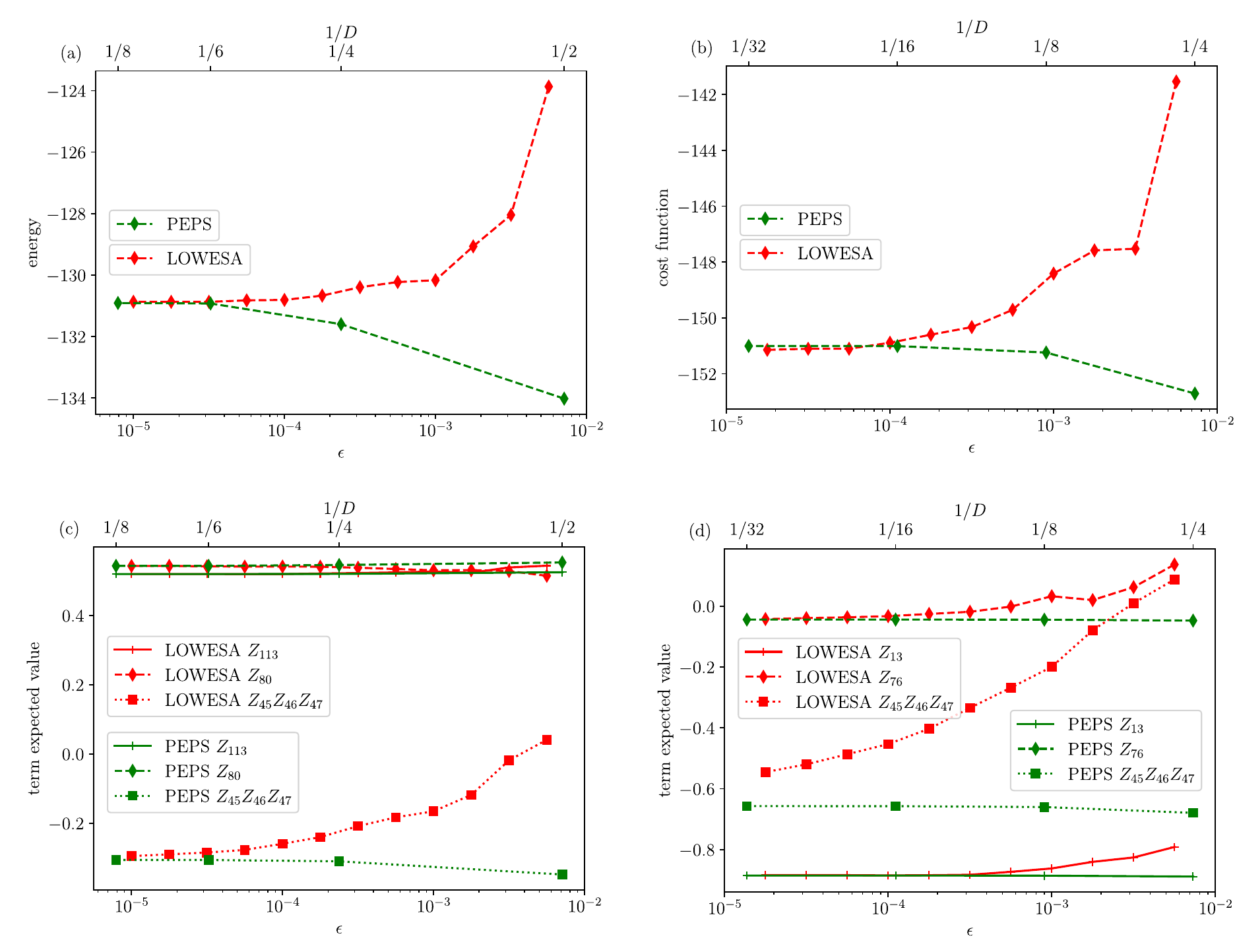}
    \caption{ A comparison of LOWESA and PEPS energy and Hamiltonian terms convergence. We plot here results for $p=3$ (a,c) and $p=5$ (b,d) QAOA angles trained on the 16-qubit instance 0 and applied to the 127-spin Ising model. In (a,b) we analyze convergence of the energy and in panels (c,d) we plot convergence of the expected values of selected Hamiltonian terms. The LOWESA results are plotted versus coefficient cutoff $\epsilon$ shown in the bottom x-axis, while the PEPS simulations are plotted versus an inverse bond dimension $1/D$ in the upper axis. We use a logarithmic scale for both $1/D$ and $\epsilon$.  }
    \label{fig:LOWESA}
\end{figure*}

\begin{figure*}[!ht]
    \centering
    \includegraphics[width=0.495\linewidth]{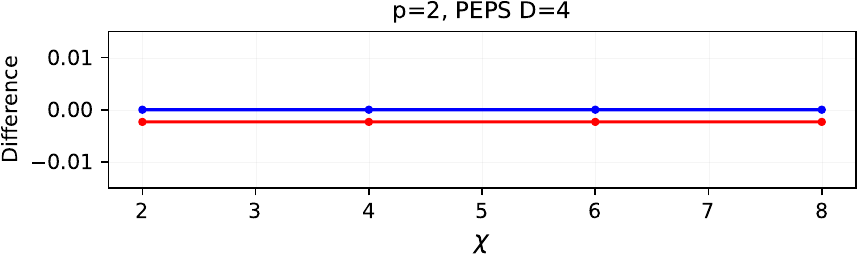}
    \includegraphics[width=0.495\linewidth]{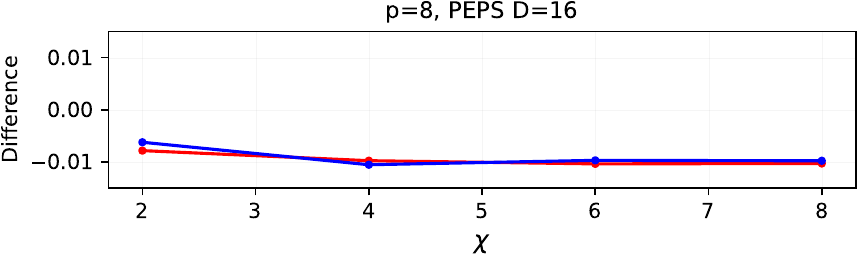}
    \includegraphics[width=0.495\linewidth]{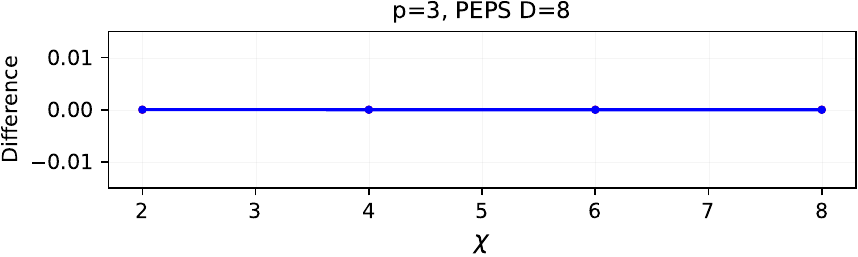}
    \includegraphics[width=0.495\linewidth]{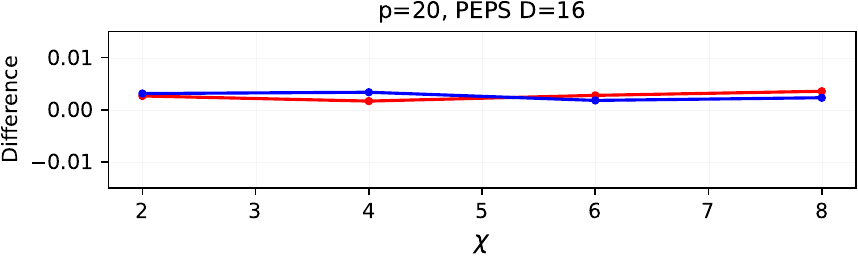}
    \includegraphics[width=0.495\linewidth]{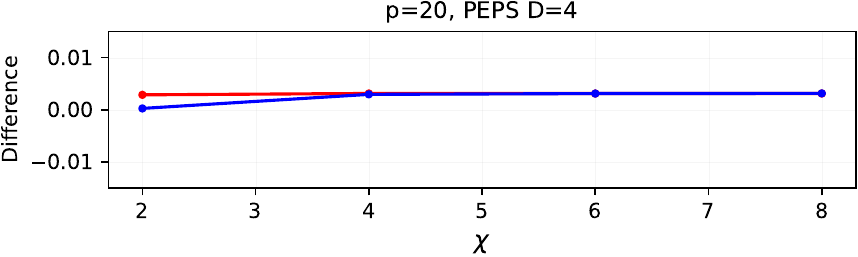}
    \includegraphics[width=0.495\linewidth]{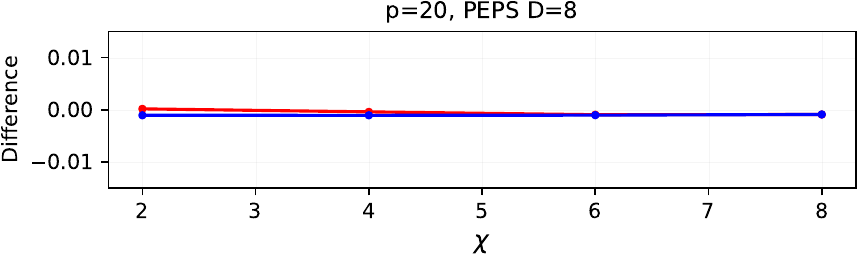}\\
    \includegraphics[width=0.32\linewidth]{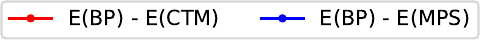}
    \caption{Comparison of the three different PEPS contraction methods contraction methods, for a few selected representative example QAOA angles, from the 16-qubit-0 instance applied to the $156$-spin Ising model, at various $p$ depths. The $D$ in the sub-figure title denotes the PEPS bond dimension. The y-axes are Hamiltonian expectation value (energy, no shot noise) relative difference with respect to the expectation value computed from belief propagation. The x-axes are the bond dimension $\chi$, which here refers to the refinement parameter for the MPS and CTM methods (belief propagation has no refinement parameter, so it is constant in all plots and is therefore used to compute relative error). The y-axis scale is consistent for all sub-plots, showing consistently high agreement between the three approximate tensor network contraction methods. }
    \label{fig:PEPS_tensor_network_contraction_comparison}
\end{figure*}

Fig.~\ref{fig:QAOA_circuit} gives an explicit description of the QAOA circuit formulation that is used in this study to sample this specific class of current IBM-processor defined two-qubit gate interaction graph. This QAOA circuit description can be extended to arbitrary $p$ by repeating this same $p=1$ structure. All QAOA simulations used the standard X mixer. All qubits are measured at the end of the simulation in the computational Z-basis. 

The QAOA circuits were all compiled with optimization using Qiskit~\cite{Qiskit, javadiabhari2024quantumcomputingqiskit} prior to submission to the QPUs -- not however that in terms of two qubit gate complexity, these circuits are already very heavily optimized, by hand compilation. Therefore, the primary function this served was to adapt the circuits to the correct gateset.

\section{All Positive and Negative Donor QAOA Angles Parameter Transfer on $27$ and $16$ Qubit Instances}
\label{section:appendix_all_positive_all_negative_param_transfer}

Fig.~\ref{fig:parameter_transfer_16_qubits_instance_negative} shows parameter transfer performance for another set of fixed trained QAOA angles, specifically trained on the $16$-qubit Ising model with only negative coefficients. Fig.~\ref{fig:parameter_transfer_16_qubits_instance_positive} shows the same thing but for the $16$-qubit Ising model with all positive coefficients. 

Figs.~\ref{fig:parameter_transfer_16_qubits_instance_0} and~\ref{fig:parameter_transfer_16_qubits_instance_negative} and~\ref{fig:parameter_transfer_16_qubits_instance_positive} all show that the different trained QAOA angles do transfer onto other problem instances. In particular at sufficiently large $p$, these parameters do eventually show continuing improvement in energy as a function of $p$. The general finding is that despite very different random coefficient $\{+1, -1\}$ distributions, the QAOA angles still can be effectively transferred.

\section{LOWESA Numerical Simulations}
\label{section:appendix_LOWESA}

We compare PEPS simulations with a LOWESA classical numerical method, which is capable of large circuit simulations. The LOWESA simulates the expected value of a Pauli observable by propagation in the Heisenberg picture~\cite{fontana2025classical,rudolph2023classical}. The propagated observable is represented as a combination of Pauli observables. Each non-Clifford gate splits a Pauli observable into multiple Paulis increasing computational cost. Thus, for large circuits with many non-Clifford gates, such as our QAOA circuits, a truncation scheme is necessary that discards terms with small contribution to the expected value. Here, we use a small coefficient truncation, which keeps only Pauli terms with coefficients larger than $\epsilon$~\cite{rudolph2025pauli}. We use LOWESA implementation from the \textit{PauliPropagation} package~\cite{rudolph2025pauli} and perform the truncation after each gate. We simulate $p=3$ and $p=5$ QAOA angles trained on the 16-qubit instance 0 and applied to a 127-qubit problem. We investigate the convergence of the energy and expectation values of individual Hamiltonian terms~\eqref{equation:problem_instance} versus $\epsilon$. We perform the LOWESA simulation for each Hamiltonian term independently and use $10^{-2.25} \ge \epsilon \ge 10^{-5}$ ($10^{-2.25} \ge \epsilon \ge 10^{-4.75}$) for $p=3$ ($p=5$). We compare the results with the PEPS convergence with increasing bond dimension $D$, using $D=2,4,6,8$ ($D=4,8,16,32$ for $p=3$ ($p=5$). We gather the results in Fig.~\ref{fig:LOWESA}. We find that the LOWESA and PEPS energies converge to the same values in a limit of small $\epsilon$ and large $D$, respectively. The convergence of the Hamiltonian terms varies strongly between the terms for LOWESA. In contrast, the term expectation values converge similarly quickly in the PEPS simulations. The LOWESA estimates of the terms expected values move towards the PEPS results when $\epsilon$ is decreased, with the convergence being faster for $p=3$. Because of this more regular convergence behavior, in main QAOA parameter transfer plots we exclusively report PEPS results. We remark that in general the LOWESA convergence depends on a truncation methods and that the optimal truncation scheme for QAOA circuits simulation is an interesting open research problem.

\section{PEPS Tensor Network Contraction Methods}
\label{section:appendix_PEPS_contraction_methods_comparison}

We evaluate three different methods for approximate PEPS tensor network contraction: i) BP, ii) corner transfer matrix renormalization group (CTM)~\cite{nishino1998,orus2009ctm,corboz2011} (where we adjust the procedure of~\cite{corboz2011} to the heavy-hex lattice), and iii) boundary matrix product state (MPS)~\cite{verstraete2008matrix}.
Both ii) and iii) have their own bond dimension, $\chi$, which controls the precision of the approximation. Empirically, we find that BP provides both numerically stable and accurate expectation values compared to both MPS and CTM, where the contraction error is typically sub-leading to the finite-$D$ PEPS representation error. This also corroborates BP-based simple update scheme for PEPS time evolution.

In Fig.~\ref{fig:PEPS_tensor_network_contraction_comparison} we show comparisons using these different approximate tensor network contraction methods. Here we show examples of PEPS expectation values on some subset of the QAOA angle simulations on the $156$-qubit hardware-compatible Ising model instance, some of which show absolute agreement among the three tensor network contraction methods (in particular at $p=1$) and others which show small disagreement between expectation values. In general we see high agreement between boundary MPS, CTM, and BP methods, and therefore we use BP for all PEPS QAOA simulations reported in this study (with the exception of a handful of examples, such as in Fig.~\ref{fig:PEPS_tensor_network_contraction_comparison}, where the agreement is not absolutely 0), because it is more computationally efficient than the other three methods.

\begin{figure*}[ht!]
    \centering
    \includegraphics[width=0.495\linewidth]{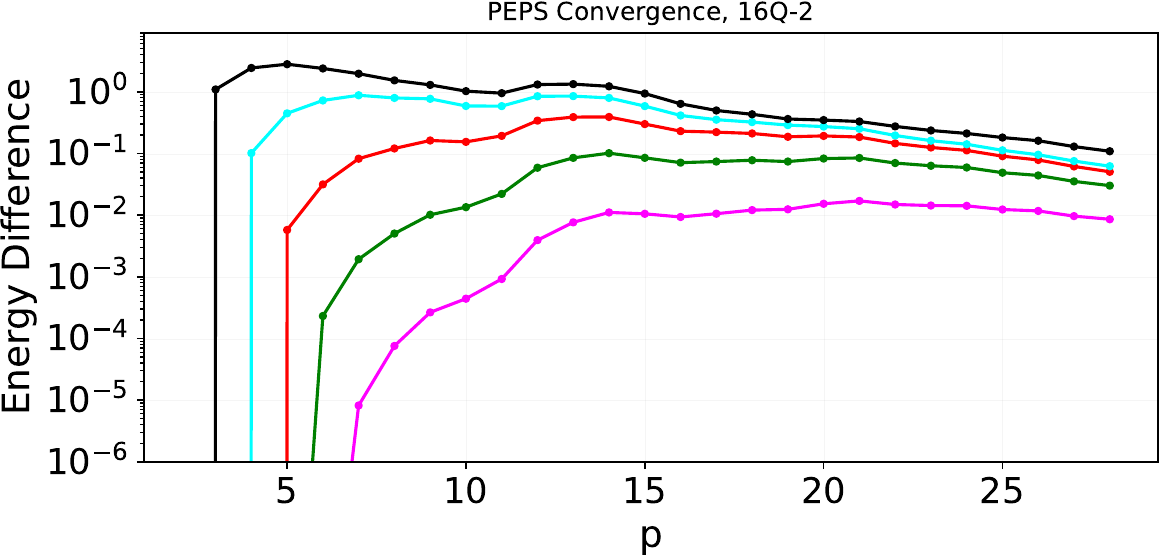}
    \includegraphics[width=0.495\linewidth]{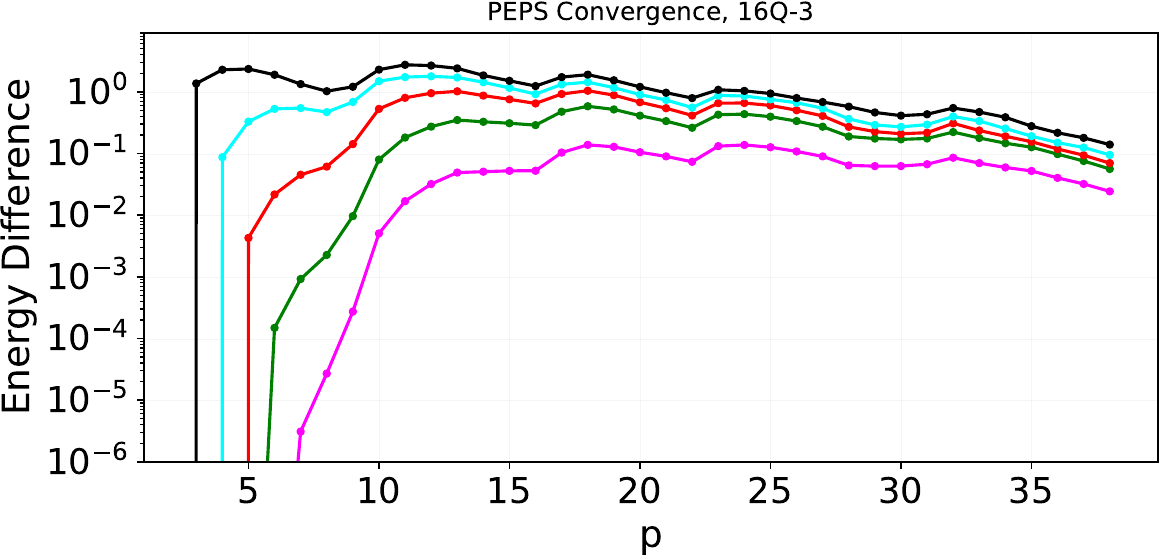}
    \includegraphics[width=0.495\linewidth]{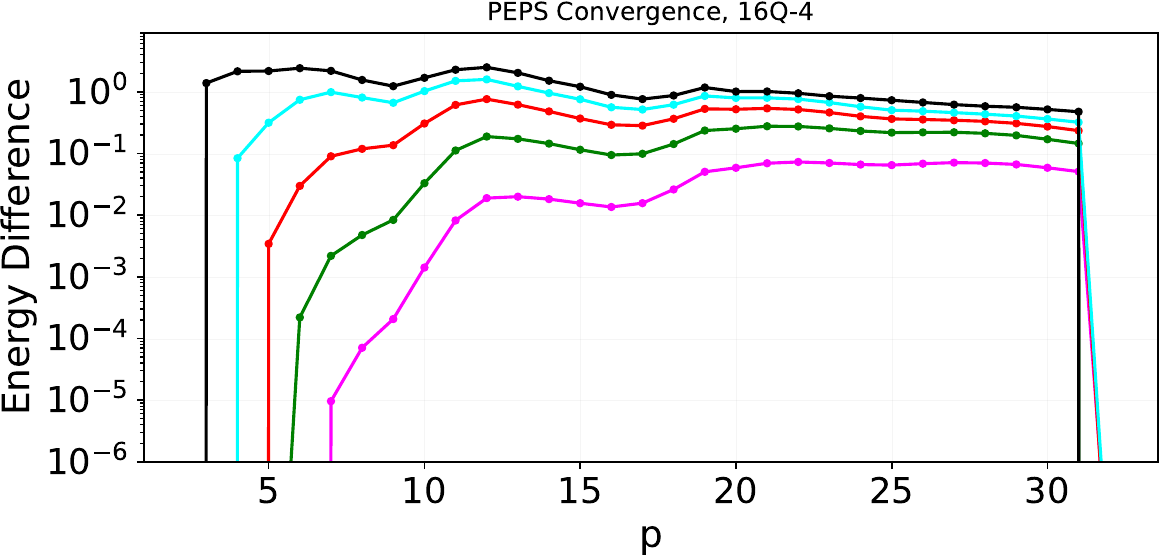}
    \includegraphics[width=0.495\linewidth]{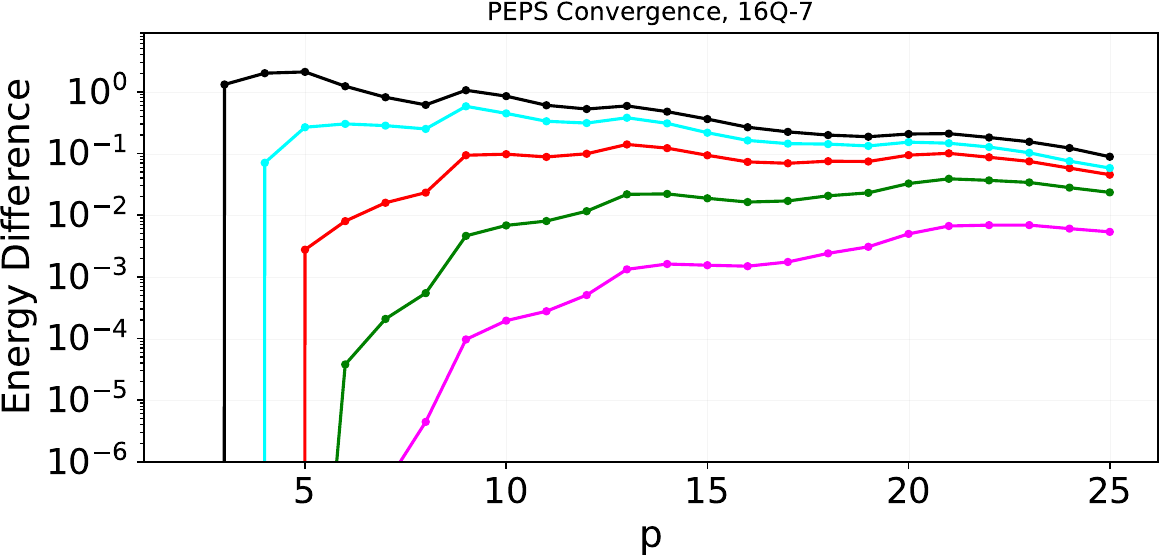}
    \includegraphics[width=0.495\linewidth]{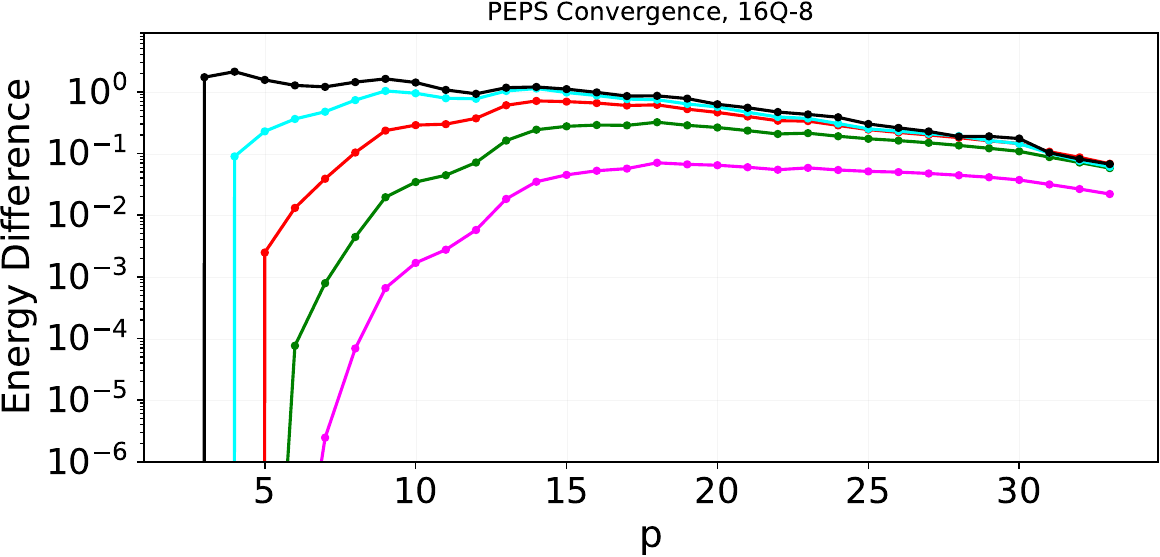}
    \includegraphics[width=0.495\linewidth]{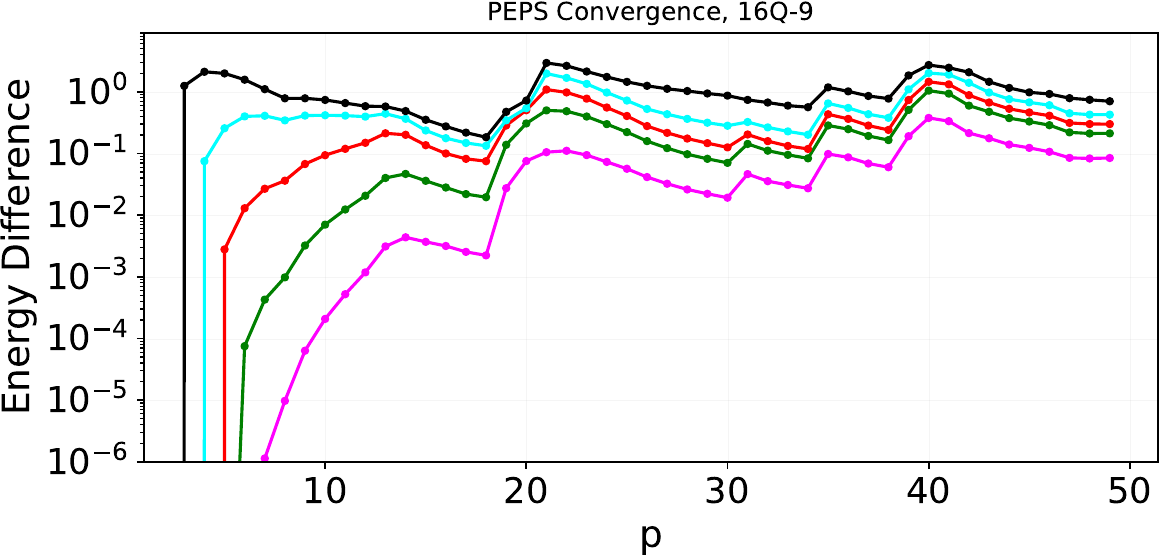}
    \includegraphics[width=0.495\linewidth]{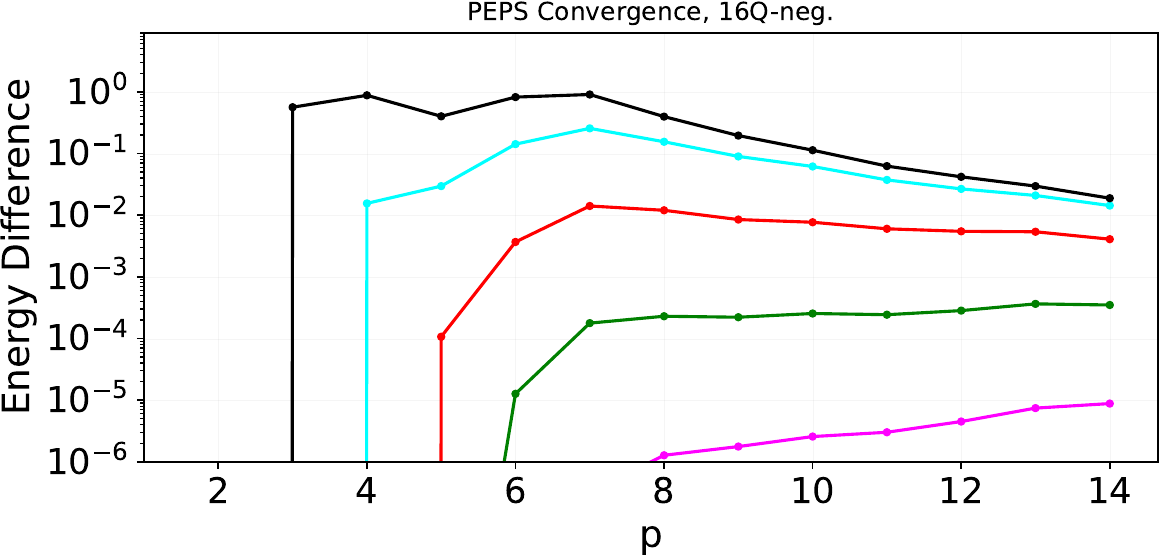}
    \includegraphics[width=0.495\linewidth]{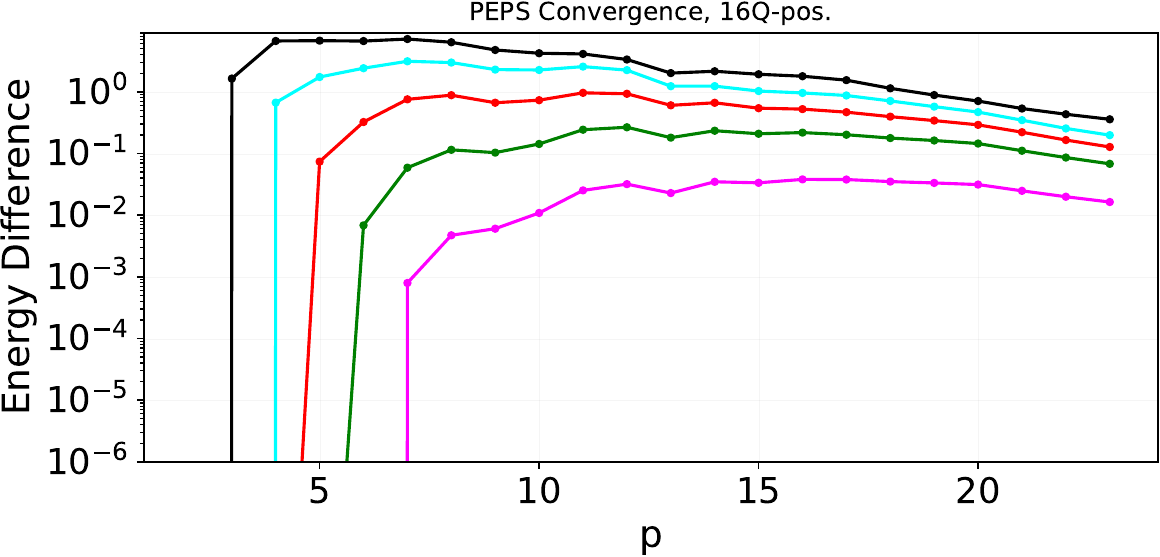}
    \includegraphics[width=0.99\textwidth]{figures/PEPS_convergence/legend.pdf}
    \caption{PEPS convergence plots for $156$-qubit QAOA circuits, using the other $8$ trained angle sets from $16$-qubit instances, continuing from Fig.~\ref{fig:PEPS_convergence}. }
    \label{fig:complete_PEPS_sim_convergence}
\end{figure*}

\section{Remaining PEPS Convergence Plots}
\label{section:appendix_more_PEPS_simulations}

Fig.~\ref{fig:complete_PEPS_sim_convergence} plots a complete set of PEPS convergence results for the remaining $8$ distinct sets of ``QAOA schedules'' trained on $16$-qubit problem instances. Notably, the convergence properties of each set of QAOA angles, shown in the form of different source training problem instances in each sub-plot, vary considerably. Ranging from a lack of strong convergence at high p, such as on the \emph{16Q-8} instance, to very strong convergence in the case of the all-negative coefficient Ising model case (suggesting a lack of high entanglement produced by that set of QAOA circuits). 

All of the convergence plots in Fig.~\ref{fig:complete_PEPS_sim_convergence}, as well as Fig.~\ref{fig:PEPS_convergence}, are normalized to use the same log-scale y-axis in the interval from $9$ to $10^{-6}$.

\clearpage

\bibliography{references}
\end{document}